\documentclass[aps,prx,reprint,superscriptaddress,floatfix]{revtex4-2}

\usepackage{dcolumn}  
\usepackage[T1]{fontenc}
\usepackage{silence}
\usepackage{fdsymbol}
\usepackage{patches}
\begin{document}

\title{Singular jets in compound drop impact}

\author{Zeyang Mou}
\affiliation{
State Key Laboratory for Strength and Vibration of Mechanical Structures,
International Center for Applied Mechanics, School of Aerospace,
Xi'an Jiaotong University, Xi'an 710049, P. R. China
}
\author{Zheng Zheng}
\affiliation{
State Key Laboratory for Strength and Vibration of Mechanical Structures,
International Center for Applied Mechanics, School of Aerospace,
Xi'an Jiaotong University, Xi'an 710049, P. R. China
}
\author{Zhen Jian}
\affiliation{
State Key Laboratory for Strength and Vibration of Mechanical Structures,
International Center for Applied Mechanics, School of Aerospace,
Xi'an Jiaotong University, Xi'an 710049, P. R. China
}
\author{Carlo Antonini}
\affiliation{
Department of Materials Science, University of Milano-Bicocca, Via R. Cozzi 55, 20125, Milano, Italy
}
\author{Christophe Josserand}
\affiliation{Laboratoire d’Hydrodynamique (LadHyX), UMR 7646 CNRS-Ecole Polytechnique, IP Paris, F-91128 Palaiseau CEDEX, France}
\author{Marie-Jean Thoraval}
\email[]{mjthoraval@xjtu.edu.cn}
\affiliation{
State Key Laboratory for Strength and Vibration of Mechanical Structures,
International Center for Applied Mechanics, School of Aerospace,
Xi'an Jiaotong University, Xi'an 710049, P. R. China
}

\date{\today}

\begin{abstract}
Compound drop impacting on a solid surface is of considerable importance in industrial applications, such as combustion, food industry, and drug encapsulation.
An intriguing phenomenon associated with this process is the occurrence of singular jets that are up to dozens of times faster than the impact velocity.
These jets break into micro-droplets, which can produce aerosols and affect the quality of printing technologies.
Here, we investigate experimentally and numerically the jetting process after a coaxial water-in-oil compound drop impacts on a glass substrate with different releasing heights and volumetric ratios.
After impact, the water core spreads and retracts,
giving rise to a vertical jet initially made of oil.
For certain values of the impacting velocity, high speed and very thin jets are observed, the so-called singular jets.
Depending on the volumetric ratio, one or two velocity peaks can be observed when varying the impact velocity, triggered by the contraction dynamics of a deep and cylindrical cavity.
The self-similar time–evolution of the collapse for the first singularity regime follows a $1/2$ power law in time, which can be derived from bubble pinch-off.
In contrast, the collapse at the second peak follows a $2/3$ power law, which can be accounted for by a balance between inertial and capillary forces.
\end{abstract}


\maketitle

\section{Introduction}
Compound drops, consisting of multiple immiscible fluids, are encountered in a wide range of industrial applications \cite{Blanken2021}, such as drug encapsulation \cite{Yuan2015, Han2016}, food industry \cite{Dordevic2015, Ray2016}, combustion \cite{Shinjo2016, Lyu2021}, and additive manufacturing of complex (bio)materials \cite{Visser2018, Kamperman2018, Li2019}.
The impact of these compound drops produces a rich variety of dynamics \cite{Blanken2021, Chiu2005, Gao2011, Liu2018Tran, Liu2018Ding, Antonini2019, Liu2019, Blanken2020, Zhang2020, Han2021}.
In the case of a coaxial water-in-oil compound drop impacting on a solid surface, Blanken \textit{et.al} \cite{Blanken2020} discovered the emergence of a very thin and fast oil jet during the rebound of the water core for some impact parameters.
Such singular jet could be detrimental in printing technologies, due to the contamination of droplets breaking up from the liquid jet and contaminating the substrate \cite{Eggers2008}, or the formation of aerosols \cite{Lhuissier2012, Joung2015, Huang2021}.
On the other hand, the atomization of the jetted liquid filament could also be harnessed to provide a simple micro-drop generator system from large drop impact, extending to compound drops the technique recently proposed by Lin \textit{et al.} \cite{Lin2021}.
It is therefore important to understand the origin of these jets and their controlling parameters.

The jets observed by Blanken \textit{et al.} \cite{Blanken2020} emerge from similar dynamics as the low velocity impact of a single phase drop on a non-wetting surface: the drop first spreads during the impact before recoiling by surface tension and eventually rebound \cite{Josserand2016}.
In that geometry, the singular jets emerge from the collapse of the air cavity on the axis of symmetry before rebound \cite{Bartolo2006, Yamamoto2018, Chen2017, Guo2020, Siddique2020, Mitra2021}.
Similar rebound and jetting dynamics can also be observed for the impact on other surfaces that enable the sliding of the drop during its contraction phase, with the lubrication provided either by the air cushioning \cite{Kolinski2014, DeRuiter2015, DeRuiter2015JFM, Lakshman2021}, the vapor layer produced by the liquid droplet on superheated surfaces \cite{Tran2012, Quere2013, Shirota2016, Bouillant2018, Lee2020}, the sublimation of an ice substrate \cite{Antonini2013}, or the liquid layer on lubricated surfaces \cite{Wong2011, Lee2014}.
In all cases, the "singular" jets, with the largest velocities, are observed near a topological transition of the collapsing interface, either to the rupture of the drop liquid into a toroidal shape \cite{Bartolo2006}, or to the entrapment of a bubble at the bottom of the cavity \cite{Bartolo2006, Chen2017, Guo2020}.
Interestingly, the singular jets observed in other interfacial flows, such as collapsing waves or bubble bursting, also appear near a transition leading to bubble entrapment \cite{Zeff2000, Duchemin2002, Michon2017, Thoroddsen2018, Yang2020}.
From a fundamental perspective, the formation of finite-time singularities in interfacial flows constitute an important feature appearing in a wide range of configurations \cite{Eggers2009, Eggers2015}.
These singular dynamics produce very thin and fast geometries evolving in a self-similar way from larger boundaries to the smallest scales, such as for bubble or drop pinch-off \cite{Day1998, Burton2005, Bergmann2006, Castrejon-Pita2015, Lagarde2018, Ruth2019, Pahlavan2019} and coalescence \cite{Paulsen2012, Hack2020} for instance.
The formation of singular jets has been observed previously in many configurations, such as the collapse of Faraday waves \cite{Hogrefe1998, Zeff2000, Brenner2000, KrishnaRaja2019, Basak2021},
cavitation bubbles \cite{Longuet-Higgins1983, Thoroddsen2009, Reuter2021},
collapsing cavities \cite{Ismail2018}, 
drop impact on a solid surface \cite{Bartolo2006, Yamamoto2018, Chen2017, Guo2020, Siddique2020, Mitra2021, Lin2021},
drop impact on a liquid pool \cite{Michon2017, Thoroddsen2018, Yang2020, VanRijn2021, BlancoRodriguez2021},
solid impact on a pool \cite{Lohse2004, Gekle2009, Truscott2014},
and bubble bursting at a free surface \cite{Duchemin2002, Lee2011, Ghabache2014, Krishnan2017, Ganan-Calvo2017, Ganan-Calvo2018, Lai2018, Brasz2018, Gordillo2019, BlancoRodriguez2020, BlancoRodriguez2021, Ganan-Calvo2021}.
However, only few studies have observed such singular jets in interfacial flows with multiple immiscible liquids \cite{Zhang2020, Yang2020, Dhuper2021}.
For a drop impacting on a pool of another immiscible liquid, Yang \textit{et al.} \cite{Yang2020} demonstrated that the additional complexity induced by multiple interfaces leads to a large variety of cavity shapes producing singular collapse.
The understanding of the high speed jets produced by the impact of compound drops could therefore extend our fundamental understanding of singularities to multiple interfaces flows.

In this work, we show experimentally and numerically that the impact of a water-in-oil compound drop on a hydrophilic solid surface can produce two distinct types of singular jets.
We demonstrate that both types of singular jets are associated with topological changes involving encapsulations.
Finally, we explain the critical role of the oil above the water core in the generation of the second type of singular jet.

\begin{figure*}[htp]
	\centering
	\includegraphics[width=\linewidth]{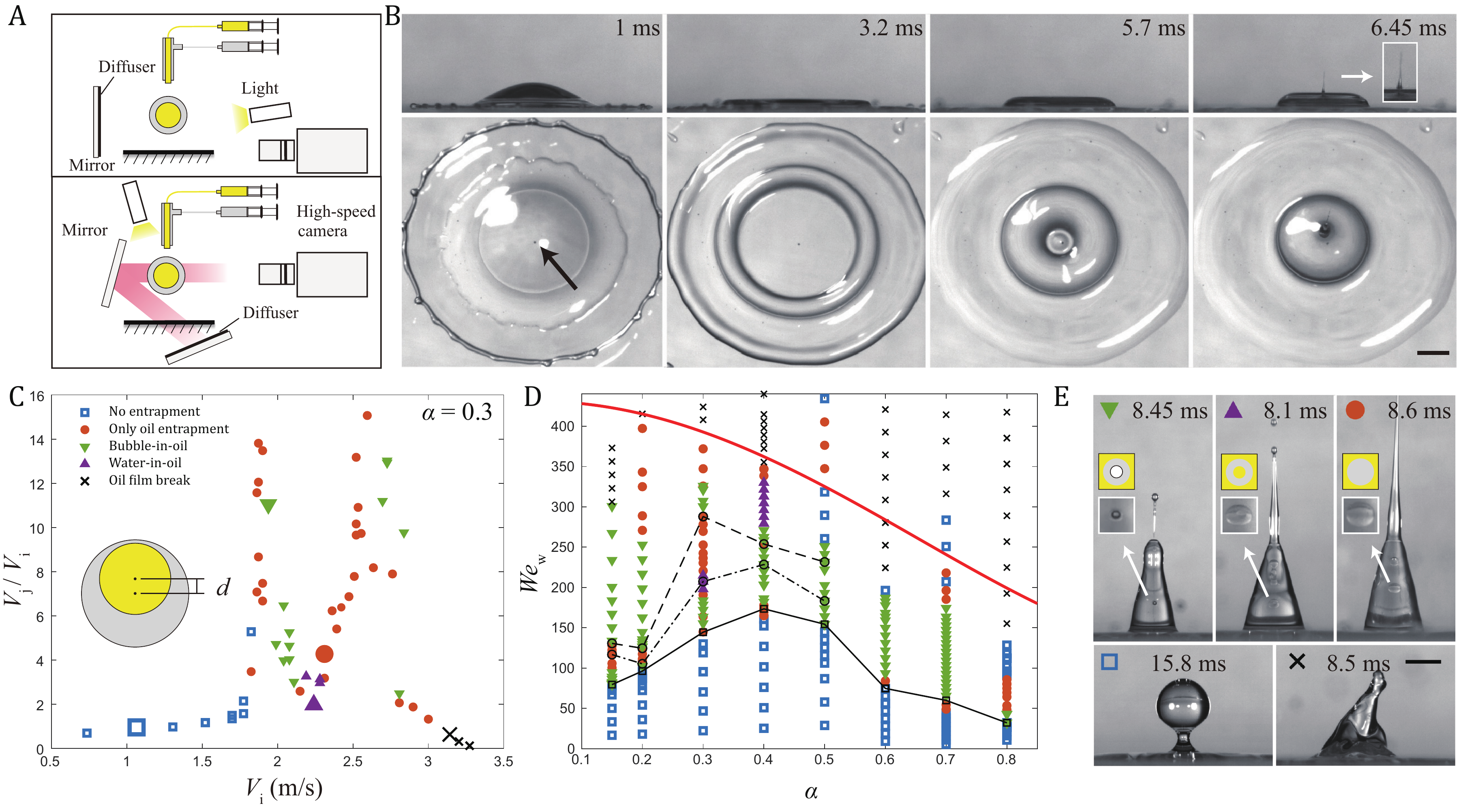}
	\vspace{\spaceBelowFigure}
	\phantomsubfloat{fig:ExperimentMap:A}
    \phantomsubfloat{fig:ExperimentMap:B}
    \phantomsubfloat{fig:ExperimentMap:C}
    \phantomsubfloat{fig:ExperimentMap:D}
    \phantomsubfloat{fig:ExperimentMap:E}
    \vspace{-2\baselineskip}
    \caption{Experimental results of impact dynamic and jet velocity.
    	(\textit{A}) Schematic of the experimental setup for side-view and top-view.
		(\textit{B}) Sequences of experimental snapshots from side-view and top-view corresponding to the first jetting velocity peak at $V_{\textnormal{i}} = \SI{1.87}{\m\per\s}$, $We_{\textnormal{w}} = 144$, $\alpha = 0.3$ (Movie S1 \& S2). 
	    $t = 0$ represents the moment when the drop first contacts the surface.
	    (\textit{C}) Maximal jet velocity $V_{\textnormal{j}}$ normalized by impact velocity $V_{\textnormal{i}}$ versus $V_{\textnormal{i}}$ together with entrapment types at $\alpha = 0.3$.
		(\textit{D}) Encapsulation phase diagram for various volume ratio $\alpha$.
		The black solid line, dash-dot line and the dashed line indicate the critical $We_{\textnormal{w}}$ of the first velocity peak, the valley value and the second peak respectively.
		The red solid line is obtained from Blanken \textit{et al.} \cite{Blanken2020} predicting the limit from which the oil film ruptures due to the core sinking to the bottom.
		(\textit{E}) Experimental snapshots to show encapsulation, represented as the magnified symbols in (\textit{C}) (Movie S4-S8). 
		The scale bar in the images corresponds to \SI{1}{\mm}.
		}
    \vspace{\spaceBelowCaption}
    \label{fig:ExperimentMap}
\end{figure*} 

\section{Results and Discussion}

\subsection{Experiments: two jetting velocity peaks and multiple types of encapsulations}

We use a similar experimental setup as Blanken \textit{et al.} \cite{Blanken2020} to investigate the dynamics of a core-shell compound drop impacting on a glass substrate (see details in \textit{Materials and Methods}), with a focus on the vertical jet.
The target surface is both hydrophilic and oleophilic.
We combine side-view and top-view imaging to capture the evolution of the interfaces during impact and rebound (\cref{fig:ExperimentMap:A}).
The compound drop is generated from a coaxial needle connected to two syringe pumps, producing a water inner core in a \SI{5}{cSt} silicone oil outer shell.
The water core of diameter $D_{\textnormal{w}}$ is initially located at the top of the compound drop of outer diameter $D_{\textnormal{o}}$, because of the nearly simultaneous pinch-off of core and shell from the nozzle.
During the fall of the compound drop, the water core gradually sinks to the bottom of the oil shell, due to the air drag and its larger density.
The eccentricity $d$ of the compound drop is defined as the distance between the vertical coordinates of the geometrical centers of the outer and inner drops, $d = z_{\textnormal{w}} - z_{\textnormal{o}}$, with the nondimensional eccentricity $d^* = 2d / (D_{\textnormal{o}} - D_{\textnormal{w}})$ (inset of \cref{fig:ExperimentMap:C}).
This offset is difficult to control experimentally, but can be estimated theoretically as a function of falling height \cite{Blanken2020}, as detailed in \textit{Materials and Methods}.

A typical impact dynamics is presented in \cref{fig:ExperimentMap:B} from both side and top views.
The water core always remains wrapped by oil, due to the positive spreading parameter of the oil on water: $S = \sigma_{\textnormal{w}} - \sigma_{\textnormal{ow}} - \sigma_{\textnormal{o}} > 0$.
The early contact between the drop and the surface entraps a thin air disk contracting into a micro-bubble at the center of the drop \cite{Thoroddsen2005, Li2015}, as indicated by the black arrow in \cref{fig:ExperimentMap:B}.
Both the oil shell and the water core first spread on the solid surface, with the oil splashing droplets radially.
While the wetting oil remains at its maximum spreading diameter, the water forms a toroidal rim and recoils back by surface tension towards the center.
This contraction is enabled by the lubrication of the oil layer below the core, preventing the contact between the water core and the solid surface for impact velocities below a critical value \cite{Blanken2020}.
The contraction dynamics generates waves propagating inwards and focusing at the center, as illustrated at $t = \SI{5.7}{\ms}$ of \cref{fig:ExperimentMap:B}.
When $t = \SI{6.45}{\ms}$, a high-speed jet consisting only of oil is shot out vertically just after the collapse of the air cavity.
Finally, the core liquid wrapped with a thin oil layer rebounds \cite{Blanken2020}.

As we focus on the jet generated by the contraction dynamics of the water core covered by an oil layer, we characterize the impact dynamics with the Weber and Reynolds numbers based on the water core size $D_{\textnormal{w}}$, water density $\rho_{\textnormal{w}}$ and viscosity $\mu_{\textnormal{w}}$, and the oil-water interfacial tension $\sigma_{\textnormal{ow}}$:
\begin{equation*}
We_{\textnormal{w}} = \frac{\rho_{\textnormal{w}} D_{\textnormal{w}} V_{\textnormal{i}}^2}{\sigma_{\textnormal{ow}}},
\qquad
Re_{\textnormal{w}} = \frac{\rho_{\textnormal{w}} D_{\textnormal{w}} V_{\textnormal{i}}}{\mu_{\textnormal{w}}},
\end{equation*}
with $V_{\textnormal{i}}$ the impact velocity of the drop.
As the Reynolds number of the water core remains $Re_{\textnormal{w}} > 1900$ at the first velocity peak, we neglect viscosity effects and use only the $We_{\textnormal{w}}$ to characterize the impact dynamics in the remaining of this study.
We systematically vary the volumetric ratio of water $\alpha=\left(D_{\textnormal{w}}/D_{\textnormal{o}}\right)^3$ and the impact velocity $V_{\textnormal{i}}$ to study the formation of the jet.
The jet velocity $V_{\textnormal{j}}$ is calculated in the experiments when it first becomes visible from side-view imaging, emerging from the top surface of the rim.
Its evolution with $V_{\textnormal{i}}$ is presented in \cref{fig:ExperimentMap:C} for $\alpha = 0.3$, showing two separate velocity peaks.
The first velocity peak reaches $V_{\textnormal{j}}/V_{\textnormal{i}} = 13.8$ ($V_{\textnormal{j}} = \SI{25.9}{\m\per\s}$) at $We_{\textnormal{w}} = 144$ ($V_{\textnormal{i}} = \SI{1.87}{\m\per\s}$)
and the second peak reaches a slightly larger non-dimensional value of $V_{\textnormal{j}}/V_{\textnormal{i}} = 15$ ($V_{\textnormal{j}} = \SI{39.1}{\m\per\s}$) at $We_{\textnormal{w}} = 288$ ($V_{\textnormal{i}} = \SI{2.59}{\m\per\s}$).
Even with larger magnification, the diameters of the jet and droplets produced are smaller than the pixel size at both velocity peaks, showing that they are smaller than \SI{3.2}{\um} (see \textit{SI Appendix} Fig.~S2).
The highest jet speed observed in our experiments is \SI{46.3}{\m\per\s} ($V_{\textnormal{j}}/V_{\textnormal{i}}=23.1$) at the first velocity peak ($We_{\textnormal{w}} = 173$, $V_{\textnormal{i}} = \SI{2}{\m\per\s}$) for $\alpha=0.4$,
while the maximal nondimensional jet velocity is $V_{\textnormal{j}}/V_{\textnormal{i}}=51$ when $We_{\textnormal{w}} = 32$ ($V_{\textnormal{i}} = \SI{0.77}{\m\per\s}$) with $\alpha=0.8$ (see \textit{SI Appendix} Fig.~S4).
In comparison, two jetting velocity peaks were also previously observed experimentally for a single water drop of diameter $\SI{2}{mm}$ impacting a hydrophobic surface \cite{Bartolo2006}, but they occurred at much lower impact velocity (first peak at $V_{\textnormal{i}}=\SI{0.45}{\m\per\s}$, $We=5.6$ and second peak at $V_{\textnormal{i}}=\SI{0.68}{\m\per\s}$, $We=12.8$), creating water jets of smaller velocities $\sim$ \SI{15}{\m\per\s} \cite{Bartolo2006,Chen2017} than the ones observed in our experiments.
Moreover, similarly high speed jets were also observed in the collapse of Faraday waves \cite{Zeff2000, KrishnaRaja2019} and bubble collapse induced by drop impact on a deep pool \cite{Yang2020, Thoroddsen2018} reaching up to $\sim$ \SI{50}{\m\per\s}, while bubble bursting at water interface showed smaller jet velocities ($\sim$ \SI{12}{\m\per\s}) \cite{Ghabache2014, Ghabache2016, Spiel1995}.

From side-view imaging, we identify multiple types of entrapments inside the water drop (\cref{fig:ExperimentMap:E}), ascribed to the closure of the air cavity: only oil entrapment, bubble-in-oil entrapment, and water-in-oil entrapment.
The correlation between the different types of entrapments and the jetting velocity is presented with different symbols in \cref{fig:ExperimentMap:C} for $\alpha=0.3$.
The velocity peaks coincide with a transition from oil entrapment when $V_{\textnormal{j}}$ rises with $V_{\textnormal{i}}$, to bubble-in-oil entrapment with $V_{\textnormal{j}}$ decreasing.
For all volume ratios (see \cref{fig:ExperimentMap:D} and Fig.~S4 in \textit{SI Appendix}), they also always correspond to a transition to bubble entrapment when increasing $V_{\textnormal{i}}$.
This suggests that the jetting mechanism is due to the focusing of the outer oil interface at the center, driven by the contraction dynamics of the water core.
This is consistent with the observation that the leading part of the high-speed jet is composed only of oil, as observed in \cref{fig:ExperimentMap:B,fig:ExperimentMap:E}.
As the impact velocity increases, the oil-air interface becomes more and more vertical when it approaches the center, in a similar way as for the impact of a water drop on a superhydrophobic surface.
The singular jets appear at the transition condition when the cavity collapses with vertical walls. 
For larger impact velocity cases entrapping a bubble, the focusing effect is reduced, leading to a decrease of the jetting velocity.
It should be mentioned here that the bubble entrapment itself is due to a singularity dynamics corresponding to the air column break-up leading to high-speed jet.
Therefore each velocity peak can be understood as a singularity separating a regular flow from a singular dynamics (entrapping a bubble).

However, for a core-shell compound drop, two interfaces are participating in the cavity collapse and encapsulations.
The overturn of the water-oil interface leads to the entrapment of oil in the water drop, while the overturn of both the water-oil and oil-air interfaces lead to the bubble-in-oil entrapment.
Oil encapsulation is thus a transition between no entrapment and bubble entrapment when increasing $V_{\textnormal{i}}$.
When approaching the lowest jet velocity conditions between the two peaks in \cref{fig:ExperimentMap:C}, water-in-oil entrapment cases are observed (purple triangle in \cref{fig:ExperimentMap:E}).
All the different types of entrapments observed here were also observed in studies of drop impact on a pool of immiscible liquid \cite{Jain2019, Yang2020}, suggesting that these collapsing oil covered water cavities share some common features.
Especially, the water-in-oil entrapment is reminiscent of the encapsulation observed by Yang \textit{et al.} \cite{Yang2020} in their Fig.~4(b), where they studied the impact of a PP1 drop on a water pool.
It suggests that the pinch-off of a protrusion created by the convergence of the capillary waves near the center could also be responsible for this type of encapsulation here (similar to numerical result of Movie S3).

At large impact velocity, the water core approaches to the bottom of the compound drop ($d^* \sim -1$). The lubricating oil film beneath the core ruptures and the water core contacts the hydrophilic substrate directly during spreading, which is clearly captured from top-view images (\textit{SI Appendix} Fig.~S1\textit{B}). From side-view observation, the contact angle between water column and the substrate is nearly $0 \degree$, shown as black cross in \cref{fig:ExperimentMap:E} \cite{Blanken2020}.

From experimental observations, both jet velocity peaks appear very similar, with similar amplitude and similar topological transitions from oil entrapment to bubble entrapment.
In order to gain a more detailed insight in the formation of these singular jets, we also perform numerical simulations.
\begin{figure*}[ht]
	\centering
	\vspace{\spaceAboveFigure}
	\includegraphics[width=\linewidth]{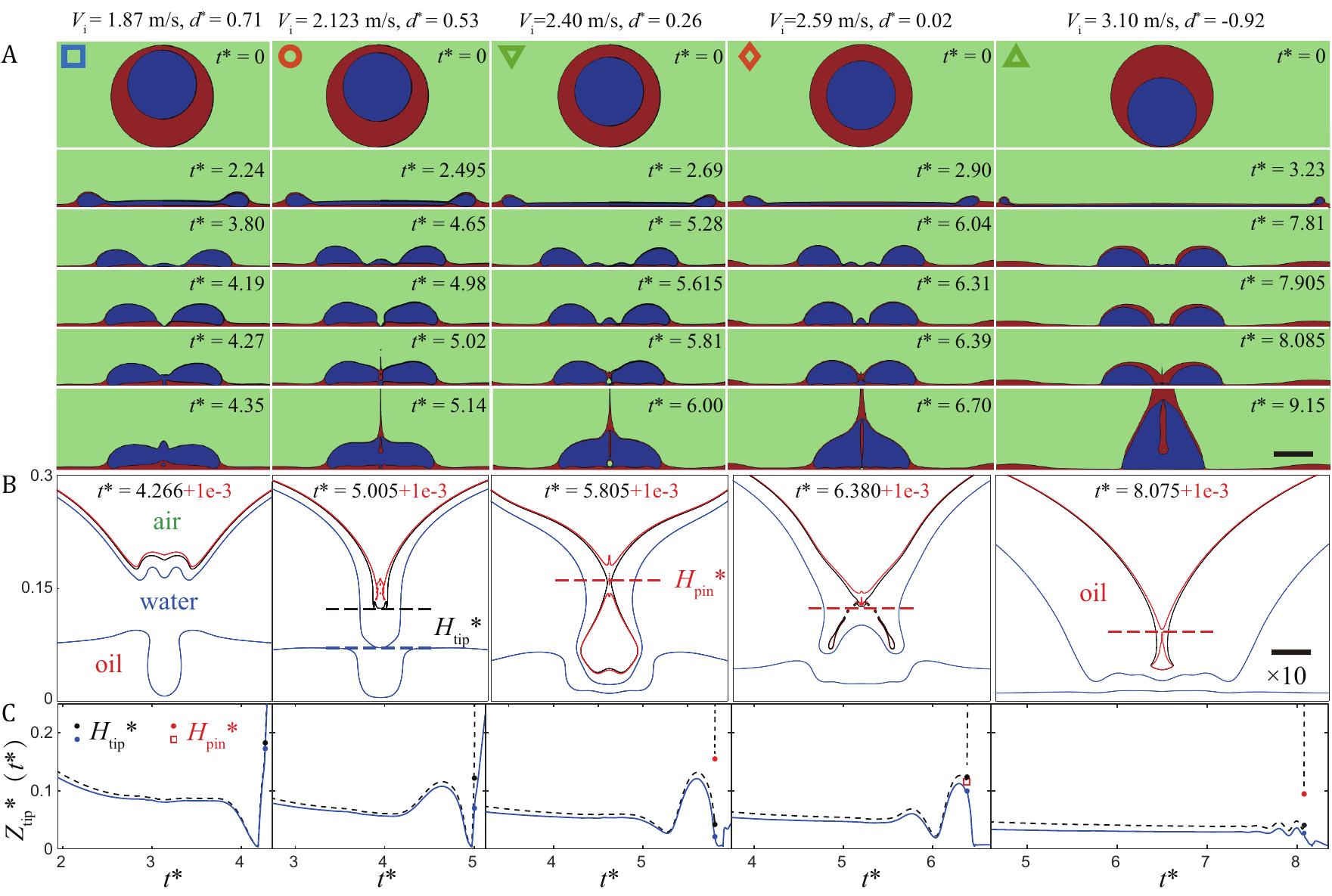}
	\phantomsubfloat{fig:TimeEvolutionSim:A}
    \phantomsubfloat{fig:TimeEvolutionSim:B}
    \phantomsubfloat{fig:TimeEvolutionSim:C}
    \vspace{-2\baselineskip}
    \caption{Simulations to reproduce experiments when $\alpha=0.3$ in \cref{fig:ExperimentMap:C}, corresponding to the five magnified symbols in \cref{fig:Simulation:A}.
	(\textit{A}) Numerical time sequences (Movie S9-S13).
	The first row represents the initial configuration of compound drop derived from the theory of Blanken \textit{et al.} \cite{Blanken2020}.
	The subsequent images depict the geometry when the water core reaches its maximal spreading, capillary waves propagate and converge, a jet is emitted out, and finally the water-oil interface collapses to entrap single or complex inclusions.
	The encapsulations are thus defined as no entrapment (column 1), only oil entrapment (column 2), and bubble-in-oil entrapment (column 3-5). The bubble encapsulation is then subdivided to central bubble (column 3 \& 5) and toroidal bubble (column 4), based on whether the bubble is continuous at the axis of symmetry once it is sealed.
	(\textit{B}) Cavity geometry captured at 10 times zoom-in compared to (\textit{A}), showing the air-oil interface (black line), oil-water interface (blue line), and the air-oil interface after the non-dimensional numerical time interval $\Delta t^* = \num{1e-3}$ (red line) to demonstrate the occurrence of jets. The $V_{\textnormal{j}}$ is calculated during this $\Delta t^*$ when there is no bubble entrapment (column 1-2). While for the cases that pinches-off a bubble (column 3-5), the air cavity alters from open to close over the period, but $V_{\textnormal{j}}$ is calculated after the cavity closure.
	(\textit{C}) Time evolution of the vertical coordinate of the tip $Z_\textnormal{tip}^*(t^*)$ (normalized by $D_\textnormal{w}$) along the axis of axial symmetry for two interfaces. The dashed black line and the solid blue line correspond to the outer and inner interfaces respectively.
    Here $H_\textnormal{tip}^*=Z_\textnormal{tip}^*$ at the time in (\textit{B}), $H_\textnormal{pin}^*$ is the pinch-off height defined in (\textit{B}). They are both a constant at a fixed $V_{\textnormal{i}}$.
	}
	\vspace{\spaceBelowCaption}
	\label{fig:TimeEvolutionSim}
\end{figure*}

\subsection{Simulations}

We use the open source code \textit{Basilisk} \cite{Popinet2021, Popinet2003, Popinet2009, Popinet2018} to perform axisymmetric numerical simulations of the compound drop impacts (see \textit{Materials and Methods}).

\begin{figure}[ht]
	\centering
	\includegraphics[width=\linewidth]{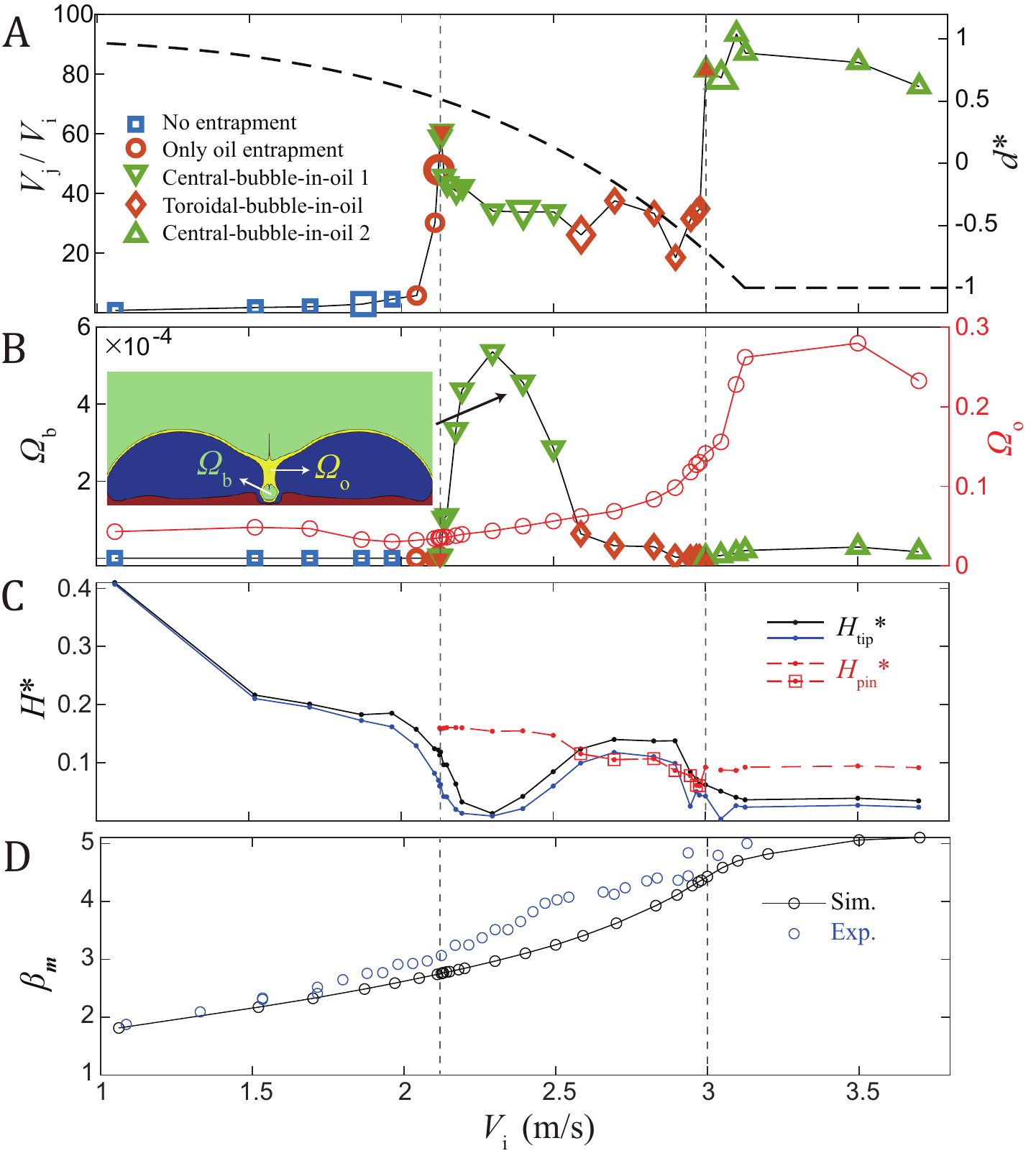}
	\phantomsubfloat{fig:Simulation:A}
    \phantomsubfloat{fig:Simulation:B}
    \phantomsubfloat{fig:Simulation:C}
    \phantomsubfloat{fig:Simulation:D}
    \vspace{-2\baselineskip}
    \caption{Numerical results for $\alpha=0.3$ with similar conditions as the experiments in \cref{fig:ExperimentMap:C}, varying both $V_{\textnormal{i}}$ and $d^*$.
	   (\textit{A}) Left axis: nondimensional jet velocity with modified definition of encapsulations. Right axis: The long dashed curve is corresponding to the eccentricity $d^*$. 
	   The full symbols overlapped by the two dashed lines indicate the transition of encapsulations, where two types of entrapments at the both sides occur in the certain cases.
	   (\textit{B}) Nondimensional volume (normalized by the volume of water core) of bubble encapsulation $\Omega_\textnormal{b} $and oil on top $\Omega_\textnormal{o}$ at the instant of cavity collapse.
	   (\textit{C}) Tip height $H_\textnormal{tip}^*$ of two interfaces and pinch-off height $H_\textnormal{pin}^*$ 
	   defined in \cref{fig:TimeEvolutionSim:B} and \cref{fig:TimeEvolutionSim:C}. The red dots indicate the vertical height of the minimal radius of cavity just before ($\Delta t^* = \num{1e-3}$) the cavity closure. The squares correspond to the positions of closure to entrap toroidal bubbles, which are not located on the axis of symmetry.
	   (\textit{D}) Maximal spreading ratio $\beta_{\textnormal{m}}$ of the water core.}
	\vspace{\spaceBelowCaption}
	\label{fig:Simulation}
\end{figure}

We select impact conditions similar to the experiments presented in \cref{fig:ExperimentMap:C} with volume ratio $\alpha=0.3$, varying $V_{\textnormal{i}}$ and taking into account the variations of the eccentricity $d^*$ (see the derivation in \textit{Materials and Methods}).
\Cref{fig:TimeEvolutionSim:A} shows time evolution sequences of the simulations for five typical impact conditions, while the systematic quantification of the jetting velocity and the geometry of the impact is presented in \cref{fig:Simulation}.
The simulations correctly reproduce the formation of a high-speed jet at the time of collapse of the water rim, together with the different types of entrapments as the impact velocity increases.
We observe in \cref{fig:Simulation:A} that $V_{\textnormal{j}}/V_{\textnormal{i}}$ also presents two peaks when varying  $V_{\textnormal{i}}$, in agreement with the experimental observations.
The jet velocities in the simulations are significantly higher than in the experiments, with $V_{\textnormal{j}}/V_{\textnormal{i}}=60.6$ ($V_{\textnormal{j}}=\SI{129.1}{\m\per\s}$) at the first peak and $V_{\textnormal{j}}/V_{\textnormal{i}}=93.3$ ($V_{\textnormal{j}}=\SI{289.2}{\m\per\s}$) for the second peak.
These higher values could be explained by slight differences between the experimental geometry of the drop at impact compared to the perfectly axisymmetric conditions in the simulations.
Moreover, in the experiments, the jets are tracked from side-view imaging after they pass above the water rim, where their speed is already reduced by air drag.
The jetting velocity peaks observed in the simulations are strongly correlated with the different types of entrapments, as shown in \cref{fig:Simulation:A}, in a similar way as for the experimental results in \cref{fig:ExperimentMap:C}.
Both singular jets appear at a topological transition, between oil entrapment and bubble entrapment for the first peak, and between toroidal bubble and central bubble entrapment for the second peak.

\begin{figure*}[ht]
	\centering
	\includegraphics[width=\linewidth]{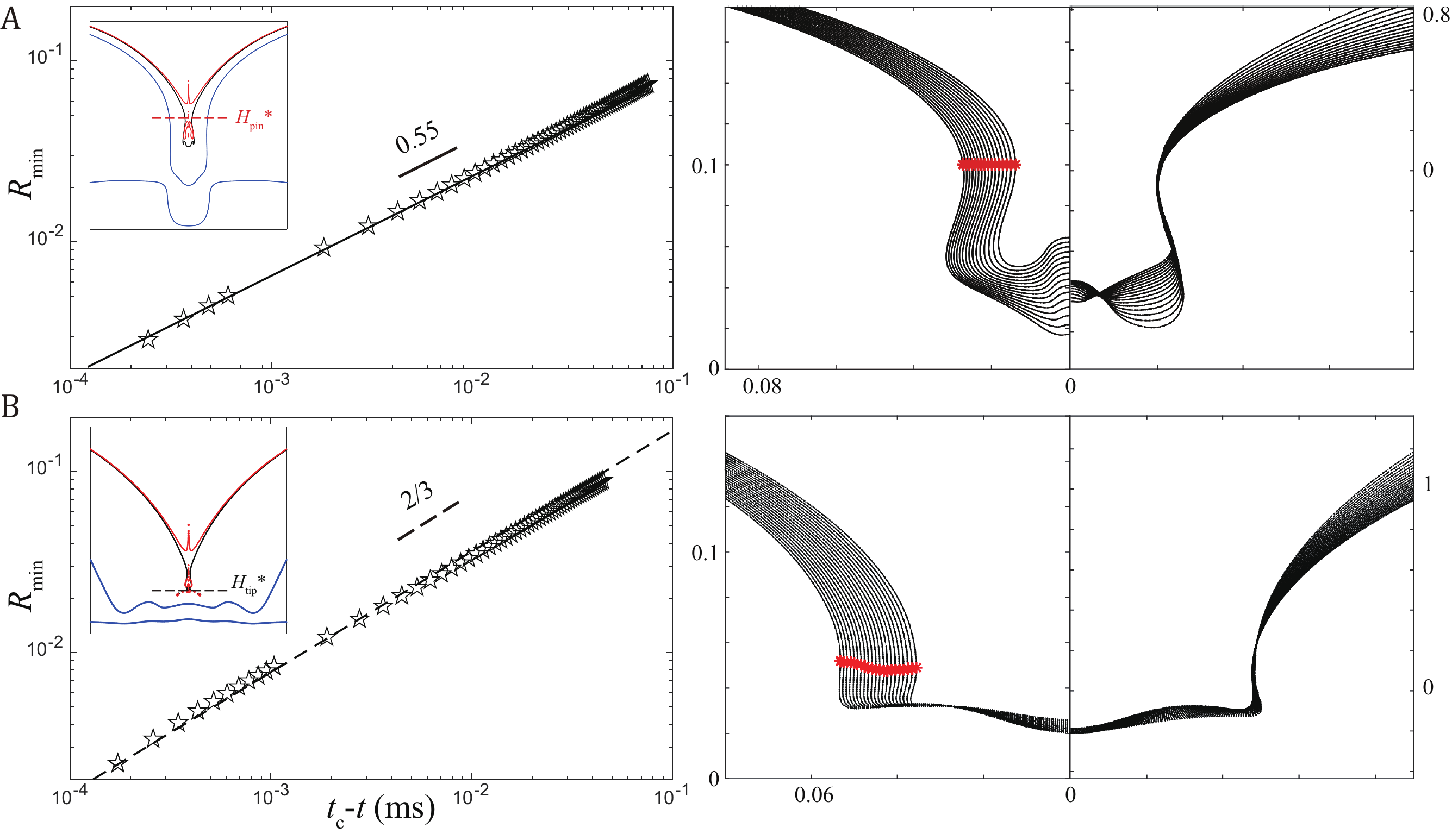}
	\phantomsubfloat{fig:PowerLaw:A}
    \phantomsubfloat{fig:PowerLaw:B}
    \vspace{-2\baselineskip}
	\caption{Air cavity collapse versus time prior to singularity at two jet velocity peaks from numerical results of $\alpha=0.3$ in \cref{fig:Simulation:A}.
	Here $R_{\textnormal{min}}$ is the nondimensional minimum radius of the air cavity (normalized by $D_\textnormal{w}$) tracked at red stars until the cavity eventually reverses back or pinch-off ($t = t_{\textnormal{c}}$). 
	We set the non-dimensional time interval between profiles $t^* = \num{1e-4}$ in our simulations to define the singularity time $t_{\textnormal{c}}$ precisely (corresponding to the red line in the inset sketches).
	(\textit{A}) $V_{\textnormal{i}} = \SI{2.13}{\m\per\s}$, represented by the orange triangle of green border overlapped by the first dashed line in \cref{fig:Simulation:A}.
	Solid line: $R_{\textnormal{min}} = 0.29(t_{\textnormal{c}}-t)^{0.55}$.
	The right image is an assembly of unrescaled and rescaled profiles. These profiles are rescaled radially as $x/(t_{\textnormal{c}}-t)^{0.55}$, vertically as $(y-H_\textnormal{pin}^*)/(t_{\textnormal{c}}-t)^{0.55}$, where $H_\textnormal{pin}^*$ is the height of pinch-off point located at the waist of the cavity. The successive profiles are plotted from $t_{\textnormal{c}}-t=\SI{0.0261}{ms}$ to $\SI{0.0079}{ms}$.
	(\textit{B}) $V_{\textnormal{i}} = \SI{3.0}{\m\per\s}$, represented by the orange triangle of green border overlapped by the last dashed line in \cref{fig:Simulation:A} (Movie S6).
	The profiles are rescaled radially as $x/(t_{\textnormal{c}}-t)^{2/3}$, vertically as $(y-H_\textnormal{tip}^*)/(t_{\textnormal{c}}-t)^{2/3}$, where $H_\textnormal{tip}^*$ is the height of cavity base that just precedes the jet emergence.
	dashed line: $R_{\textnormal{min}} = 0.78(t_{\textnormal{c}}-t)^{2/3}$.
    The profiles start at $t_{\textnormal{c}}-t=\SI{0.0372}{ms}$ and end at $\SI{0.0199}{ms}$.
	}
	\vspace{\spaceBelowCaption}
	\label{fig:PowerLaw}
\end{figure*}

Beyond the replication of the experimental results, the numerical simulations also give access to the detailed cavity evolution that is hidden by the water rim in the experimental side-view images.
We observe that the contraction dynamics of the water rim induces capillary waves which propagate inward towards the axis of symmetry, as visible in particular in the third and the forth rows of \cref{fig:TimeEvolutionSim:A}.
The focusing of the waves determines the final geometry of the cavity at the time the water rim collapses at the center, as shown in \cref{fig:TimeEvolutionSim:B}.
As the waves reach the center, they induce oscillations in time of the top oil and water interfaces heights $Z_{\textnormal{tip}}^*(t^*)$ on the axis of symmetry, plotted on \cref{fig:TimeEvolutionSim:C}.

At low impact velocity (first column in \cref{fig:TimeEvolutionSim}), the contracting rim only produces a small depression on the interfaces ahead of it.
When this circle of lower height reaches the center, it induces a smooth reversal of the cavity, followed by another acceleration of $Z_{\textnormal{tip}}^*(t^*)$ to a larger velocity when the water rim reaches the axis of symmetry.
This corresponds to the water rim reaching the center during the rising phase of $Z_{\textnormal{tip}}^*(t^*)$.
As $V_{\textnormal{i}}$ increases (second column in \cref{fig:TimeEvolutionSim}), stronger waves are generated ahead of the rim, leading to the rise of the two interfaces at the center followed by a fast decrease of $Z_{\textnormal{tip}}^*(t^*)$ ahead of the arrival of the water rim.
This induces the steepening of the air cavity, responsible for the increasing jet velocity.
Further increasing $V_{\textnormal{i}}$ produces a stronger upward and downward motion on the axis of symmetry,
forcing the rim to close at a higher location $H_\textnormal{pin}^*$, thus entrapping a central bubble, as shown in the third column of \cref{fig:TimeEvolutionSim}.
The singular jet at the first peak corresponds to the intermediate configuration when the downward motion of the wave generates a vertical cavity at the time of the rim collapse.
Therefore, it is controlled by the interplay between the focusing waves on the axis of symmetry and the collapsing rim.

With increasing $V_{\textnormal{i}}$, both the waves and the rim take more time to reach the center, as illustrated in the time evolution curves of \cref{fig:TimeEvolutionSim:C}.
However, the phase shift between them decreases with $V_{\textnormal{i}}$, with the rim collapsing earlier relative to the wave.
This phase shift is directly correlated with the variation of the bubble volume $\Omega_\textnormal{b}$ represented in \cref{fig:Simulation:B}.
The maximum bubble entrapment corresponds to the cavity collapsing at the time of $Z_{\textnormal{tip}}^*(t^*)$ reaching its minimum, while the pinch-off height $H_\textnormal{pin}^*$ remains nearly constant in this first bubble entrapment regime.
As the phase shift further decreases, the collapsing cavity reaches the center at an earlier time when $Z_{\textnormal{tip}}^*(t^*)$ is decreasing, leading to a larger $H_\textnormal{tip}^*$.
When $H_\textnormal{tip}^*$ reaches the same height as $H_\textnormal{pin}^*$ at $V_{\textnormal{i}} = \SI{2.59}{\m\per\s}$, this corresponds to the collapse of the rim onto a central protrusion, leading to the entrapment of a toroidal bubble (fourth column of \cref{fig:TimeEvolutionSim}).
A similar toroidal geometry was also reported in Fig.~5(b) of the experimental study of Yang \textit{et al.} \cite{Yang2020}, but the asymmetry of the contraction dynamics led there to the shedding of separate micro-bubbles rather than a toroidal bubble.
Asymmetric instabilities of the sharp toroidal film in our experiments could also explain that we did not observe any bubble entrapment in this regime.

Finally, for larger $V_{\textnormal{i}}$, the toroidal entrapment disappears and is replaced by a central bubble entrapped at the bottom of the air cavity after the second jet velocity peak (last column of \cref{fig:TimeEvolutionSim}).
It corresponds to the closure of the cavity at a smaller pinch-off height $H_\textnormal{pin}^*$ (\cref{fig:TimeEvolutionSim:B,fig:Simulation:C}).
Consequently, the volume of the entrapped bubble $\Omega_\textnormal{b}$ is much smaller after the second peak than after the first peak (\cref{fig:Simulation:B}, see the comparison with experiments results in Fig.~S3 in \textit{SI Appendix}).
To understand this transition at the second peak, we notice that the geometry of this smaller bubble entrapment results from a very different collapse of the cavity, as shown in the last column of \cref{fig:TimeEvolutionSim:C}.
It appears at large impact velocities, where the water core spreads to a maximal diameter larger than $4$ times its initial diameter $D_\textnormal{w}$ ($\beta_\textnormal{m} > 4$, \cref{fig:Simulation:D}).
Furthermore, a thicker layer of oil (in red) is pushed ahead of the contracting water rim (in blue) during the retraction dynamics, leading to the decoupling of the air-oil and oil-water interfaces.
This layer suppresses the main wave ahead of the water rim, leaving only small waves propagating towards the axis of symmetry.
The rim therefore pushes the oil interface onto a nearly flat liquid surface ahead of it, leading to a conical geometry of the interface at the time of collapse.
This transition can be quantified by calculating the oil volume $\Omega_\textnormal{o}$ on top of the water core at the moment of cavity closure (red circle and line in \cref{fig:Simulation:B}), which indeed increases significantly around the second jet velocity peak.
At larger impact velocities, the jetting velocity remains very high (\cref{fig:Simulation:A}), in contrast with the sharp decrease observed experimentally in \cref{fig:ExperimentMap:C}.
This decrease observed experimentally can be explained by the development of an azimuthal instability around the water rim, breaking the axisymmetry necessary to produce singular jets (see Fig.~S1\textit{A}).
A more similar plateau behavior of the jet velocity after the second velocity peak can be observed experimentally for $\alpha = 0.15$ or $0.2$ (see Fig.~S4).
The second velocity peak therefore appears as a stable regime characterized by the suppression of focusing waves, in contrast with the first peak which is only a transition due to the oscillations produced by the focusing waves.

In order to further characterize the singular jets corresponding to the two velocity peaks, we will now investigate in details the final stage of the interfaces collapse near these transition regions.

\subsection{Two singularity regimes and scaling law}

As mentioned in the introduction, the fast and thin jets observed here are reminiscent of other singular jets observed in other physical configurations.
To understand which type of jet is observed in this study, we characterize the collapsing dynamics of the cavity before the formation of the jet (\cref{fig:PowerLaw}).

At the first jet velocity peak (\cref{fig:PowerLaw:A}), the minimum radius of the cavity follows a $(t_{\textnormal{c}}-t)^{0.55}$ power law.
The rescaled profiles on the right indicate a single-point collapse, while the profiles are not self-similar very near or far away from the axis of symmetry.
This collapsing dynamics is similar to the inertial collapse  an infinite cylinder, which can be described through the two-dimensional Rayleigh-Plesset equation, leading to a $(t_{\textnormal{c}}-t)^{1/2}$ scaling law for the cylinder collapse~\cite{Plesset1977}.
Such cylindrical self-similar dynamics have been observed in particular for bubble pinch-off and drop impact on super-hydrophobic surface~\cite{Burton2005, Bergmann2006, Bartolo2006, Lohse2004, Chen2017, Ruth2019}.
The larger observed exponent $0.55$ compared to the inertial power law of $0.5$ is consistent with bubble pinch-off experimental measurements \cite{Yang2020, Bergmann2006, Thoroddsen2007, Eggers2007}, where they obtained an exponent varying from $0.55$ to $0.57$.
This variation of the exponent could in fact witness the logarithmic correction deduced from the detailed asymptotic~\cite{Eggers2007, Gordillo2007}. 

In contrast, at the second velocity peak (\cref{fig:PowerLaw:B}), the collapsing cavity follows a $(t_{\textnormal{c}}-t)^{2/3}$ power law.
The profiles superimpose well after being rescaled, even near the axis of symmetry.
They form a conical cavity akin to surface shape of other self-similar flows observed for collapsing cavities \cite{Zeff2000, Duchemin2002}.
Such singular dynamics corresponds to the inertia-capillary collapse of a cavity that exhibits a corner like self-similar shape~\cite{Keller1983}.

The analysis of the collapsing cavity at the two jet velocity peaks therefore also reveal that they follow different power laws, suggesting that they differ not only in their geometry, but also fundamentally in their singular nature.

\section{Conclusion and outlooks}

In conclusion, we have investigated experimentally and numerically the fast jets produced by the impact of a compound drop onto a solid surface.
We discovered that the rebound of the water core covered by the oil layer can produce two fundamentally different types of singular jets.
At lower impact velocities, the singular jet is formed by the interaction between the collapsing rim and the focusing waves on the axis of symmetry.
In that case, the cavity collapse follows a $0.55$ power law similar to the inertial collapse of a cylindrical cavity, such as for bubble pinch-off.
In contrast, a different type of singular jet is formed at higher impact velocities.
It emerges from the collapse of a conical cavity on a flat interface.
It is characterized by a $2/3$ power law for the collapsing cavity, characteristic of an inertial-capillary collapse such as observed in Faraday waves of bubble bursting.

The emergence of these robust singular jets provides a promising way to design micro-droplet generators, limiting the use of chemical by the presence of the water core.
Furthermore, these dynamics are observed on a hydrophilic surface, and therefore do not need any complex treatment of the solid surface to design this setup.
Finally, the additional complexity of these singular flow illustrate the rich dynamics of multicomponent fluids, relevant in many industrial processes.

\section{Materials and Methods}

\subsection{Experiments}

The inner liquid is water of density $\rho_{\textnormal{w}} = \SI{998}{\kg\per\cubic\m}$,
viscosity $\mu_{\textnormal{w}}=\SI{0.89}{mPa \cdot s}$,
surface tension $\sigma_{\textnormal{w}} = \SI{72}{\mN\per\m}$.
The outer shell is silicone oil of density $\rho_{\textnormal{o}} = \SI{913}{\kg\per\cubic\m}$, viscosity $\mu_{\textnormal{o}}=\SI{4.57}{mPa \cdot s}$, surface tension $\sigma_{\textnormal{o}} = \SI{20}{\mN\per\m}$.
The interfacial tension between them is $\sigma_{\textnormal{ow}} = \SI{42}{\mN\per\m}$.
The solid surface is hydrophilic glass substrate with static contact angle smaller than $5\degree$ for both water and oil.
We use ultra-high-speed cameras (Photron FASTCAM-SAZ) equipped with a long-distance microscope (LEICA Z16 APO) to capture the spreading process and jetting dynamics from side-view and top-view (\cref{fig:ExperimentMap:B}).
We use frame rates varying from \SIrange{20 000}{100 000}{\fps} to capture the slower dynamics (spread and retraction) and faster process (jetting) separately, and increase magnification to capture the thread coming from drop if necessary (Fig.~S2).
All the experiments are conducted in a dust-free clean environment with constant temperature of \SI{22}{\celsius} and humidity of \SI{45}{\percent}. 

\subsection{Theoretical model to calculate eccentricity}

In our study, we adopt the theory of Blanken \textit{et al.} \cite{Blanken2020} to calculate the relative position $d^*$ between water core and oil shell.
During the fall, the air drag reduces the falling velocity of the compound drop in comparison with free fall.
Taking into account the buoyancy effect (due to the density difference between water and oil) and the drag force (which is modeled by a Stokes drag), the water core moves upward relative to the outer shell.
The mathematical formulation is derived in \cite{Blanken2020} and is directly implemented in our simulations:
\begin{align}
\begin{cases}
\frac{dV_{\textnormal{i}}}{dt}   &= - g \left( 1-\frac{ V_{\textnormal{i}}^2 }{ V_{\textnormal{T}}^2 } \right) \\
\frac{dV_{\textnormal{rel}}}{dt} &= - \frac{ \rho_{\textnormal{w}} - \rho_{\textnormal{o}} }{ \rho_{\textnormal{w}} } \left( g + \frac{ dV_{\textnormal{i}} }{ dt } \right) - \frac{ 18 \mu_{\textnormal{o}} }{ \rho_{\textnormal{w}} D_{\textnormal{w}}^2 } V_{rel}
\end{cases}
\label{model_nathan}
\end{align}
where $V_{\textnormal{i}}$ is the impact velocity of the compound drop, $V_{\textnormal{T}}$ is the terminal velocity, which is $V_{\textnormal{T}} = \SI{6.9}{\m\per\s}$, and $V_{rel}=\textnormal{d}d/\textnormal{d}t$.
By solving numerically \cref{model_nathan}, the evolution of the vertical position of the water core versus impact velocity can be deduced and is shown on the right axis of \cref{fig:Simulation:A}. 

\subsection{Simulations}

Numerical simulations are carried out with the open-source code \textit{Basilisk} which solves the dimensionless incompressible Navier-Stokes equations using a finite-volume method on an adaptive tree-structured mesh. The non-dimensionalization is done with respect to the physical properties of water ($\rho_{\textnormal{w}}, \mu_{\textnormal{w}}$), the diameter of outer drop ($D_{\textnormal{o}}$) and impact velocity ($V_{\textnormal{i}}$) of the compound drop. The interfaces are tracked by the Volume-Of-Fluid (VOF) method, and in our case, two different interface tracers ($f_{\textnormal{wo}}$ and $f_{\textnormal{oa}}$) are defined for tracking the water-oil and oil-air  interfaces \cite{Ramirez-Soto2020, Wei2021}. The properties of the tracers are defined as follow:
\begin{align}
f_{\textnormal{wo}} &=
\begin{cases}
    0 \rightarrow $ oil \& air phase$  \\
    1 \rightarrow $ water phase$  
\end{cases}
\label{f1}\\
f_{\textnormal{oa}} &=
\begin{cases}
    0 \rightarrow $ air phase$  \\
    1 \rightarrow $ water \& oil phase$  
\end{cases}
\label{f2}
\end{align}
Consequently, the density and viscosity fields (as a function of two tracers) are computed based on the VOF fraction $f_{\textnormal{wo}}$ and $f_{\textnormal{oa}}$:
\begin{eqnarray}
\footnotesize
\begin{cases}
\rho(f_{\textnormal{wo}},f_{\textnormal{oa}})=(f_{\textnormal{wo}}f_{\textnormal{oa}} )\rho_{\textnormal{w}} + f_{\textnormal{oa}}(1-f_{\textnormal{wo}})\rho_{\textnormal{o}} + (1-f_{\textnormal{oa}})\rho_{\textnormal{a}}\\
\mu(f_{\textnormal{wo}},f_{\textnormal{oa}})=(f_{\textnormal{wo}}f_{\textnormal{oa}} )\mu_{\textnormal{w}} + f_{\textnormal{oa}}(1-f_{\textnormal{wo}})\mu_{\textnormal{o}} + (1-f_{\textnormal{oa}})\mu_{\textnormal{a}}
\label{func}
\end{cases}
\end{eqnarray}
where $f_{\textnormal{oa}}$ and $f_{\textnormal{wo}} \in \lbrack0,1\rbrack$.

The contact angles of the oil on the substrate is $0\degree$, and water in oil on the substrate is $180\degree$ to avoid water-substrate contact.
However, the water core is continuously wrapped by oil shell throughout the simulation (see numerical snapshots from \cref{fig:TimeEvolutionSim:A}) due to the relationship of their interfacial tension, so the contact angle of water in oil hardly play a role.
A no-slip (Dirichlet) boundary condition is applied to the solid surface while the Neumann boundary condition is used for other boundaries of the domain, ensuring normal free outflow conditions. 
The computational domain size $L$ is kept large compared to the outer drop diameter $D_{\textnormal{o}}$ ($L = 3.86 D_{\textnormal{o}}$) to eliminate boundary effects and to cover all the spreading and rebound dynamics of the drop.
We consider an axisymmetric problem and neglect the gravity effects in all simulations.
We removed in our simulations the small air bubbles formed from a thin air film entrapped beneath the drop once the drop contacts the surface \cite{Josserand2016}.

\begin{acknowledgments}
The work was supported by the Natural Science Foundation of China (Grant Nos.
12072258, 11850410439, and 11702210)
and the Project No. B18040.
M.-J.T. was supported by the Cyrus Tang Foundation through the Tang Scholar program.
This research also received support from the HPC Platform of Xi‘an Jiaotong University.
\end{acknowledgments}

\bibliographystyle{jabbrv_apsrev4-2}
\bibliography{References}

\begin{thebibliography}{99}%
\makeatletter
\providecommand \@ifxundefined [1]{%
 \@ifx{#1\undefined}
}%
\providecommand \@ifnum [1]{%
 \ifnum #1\expandafter \@firstoftwo
 \else \expandafter \@secondoftwo
 \fi
}%
\providecommand \@ifx [1]{%
 \ifx #1\expandafter \@firstoftwo
 \else \expandafter \@secondoftwo
 \fi
}%
\providecommand \natexlab [1]{#1}%
\providecommand \enquote  [1]{``#1''}%
\providecommand \bibnamefont  [1]{#1}%
\providecommand \bibfnamefont [1]{#1}%
\providecommand \citenamefont [1]{#1}%
\providecommand \href@noop [0]{\@secondoftwo}%
\providecommand \href [0]{\begingroup \@sanitize@url \@href}%
\providecommand \@href[1]{\@@startlink{#1}\@@href}%
\providecommand \@@href[1]{\endgroup#1\@@endlink}%
\providecommand \@sanitize@url [0]{\catcode `\\12\catcode `\$12\catcode
  `\&12\catcode `\#12\catcode `\^12\catcode `\_12\catcode `\%12\relax}%
\providecommand \@@startlink[1]{}%
\providecommand \@@endlink[0]{}%
\providecommand \url  [0]{\begingroup\@sanitize@url \@url }%
\providecommand \@url [1]{\endgroup\@href {#1}{\urlprefix }}%
\providecommand \urlprefix  [0]{URL }%
\providecommand \Eprint [0]{\href }%
\providecommand \doibase [0]{https://doi.org/}%
\providecommand \selectlanguage [0]{\@gobble}%
\providecommand \bibinfo  [0]{\@secondoftwo}%
\providecommand \bibfield  [0]{\@secondoftwo}%
\providecommand \translation [1]{[#1]}%
\providecommand \BibitemOpen [0]{}%
\providecommand \bibitemStop [0]{}%
\providecommand \bibitemNoStop [0]{.\EOS\space}%
\providecommand \EOS [0]{\spacefactor3000\relax}%
\providecommand \BibitemShut  [1]{\csname bibitem#1\endcsname}%
\let\auto@bib@innerbib\@empty
\bibitem [{\citenamefont {Blanken}\ \emph {et~al.}(2021)\citenamefont
  {Blanken}, \citenamefont {Saleem}, \citenamefont {Thoraval},\ and\
  \citenamefont {Antonini}}]{Blanken2021}%
  \BibitemOpen
  \bibfield  {author} {\bibinfo {author} {\bibfnamefont {N.}~\bibnamefont
  {Blanken}}, \bibinfo {author} {\bibfnamefont {M.~S.}\ \bibnamefont {Saleem}},
  \bibinfo {author} {\bibfnamefont {M.-J.}\ \bibnamefont {Thoraval}},\ and\
  \bibinfo {author} {\bibfnamefont {C.}~\bibnamefont {Antonini}},\ }\bibfield
  {title} {\bibinfo {title} {\emph {{Impact of compound drops: a
  perspective}}},\ }\href {https://doi.org/10.1016/j.cocis.2020.09.002}
  {\bibfield  {journal} {\bibinfo  {journal} {\protect\JournalTitle{Current
  Opinion in Colloid {\&} Interface Science}}\ }\textbf {\bibinfo {volume}
  {51}},\ \bibinfo {pages} {101389} (\bibinfo {year} {2021})}\BibitemShut
  {NoStop}%
\bibitem [{\citenamefont {Yuan}\ \emph {et~al.}(2015)\citenamefont {Yuan},
  \citenamefont {Lei}, \citenamefont {Liu}, \citenamefont {Tong}, \citenamefont
  {Si},\ and\ \citenamefont {Xu}}]{Yuan2015}%
  \BibitemOpen
  \bibfield  {author} {\bibinfo {author} {\bibfnamefont {S.}~\bibnamefont
  {Yuan}}, \bibinfo {author} {\bibfnamefont {F.}~\bibnamefont {Lei}}, \bibinfo
  {author} {\bibfnamefont {Z.}~\bibnamefont {Liu}}, \bibinfo {author}
  {\bibfnamefont {Q.}~\bibnamefont {Tong}}, \bibinfo {author} {\bibfnamefont
  {T.}~\bibnamefont {Si}},\ and\ \bibinfo {author} {\bibfnamefont {R.~X.}\
  \bibnamefont {Xu}},\ }\bibfield  {title} {\bibinfo {title} {\emph {{Coaxial
  Electrospray of Curcumin-Loaded Microparticles for Sustained Drug
  Release}}},\ }\href {https://doi.org/10.1371/journal.pone.0132609} {\bibfield
   {journal} {\bibinfo  {journal} {\protect\JournalTitle{PLoS ONE}}\ }\textbf
  {\bibinfo {volume} {10}},\ \bibinfo {pages} {e0132609} (\bibinfo {year}
  {2015})}\BibitemShut {NoStop}%
\bibitem [{\citenamefont {Han}\ \emph {et~al.}(2016)\citenamefont {Han},
  \citenamefont {Thurecht}, \citenamefont {Whittaker},\ and\ \citenamefont
  {Smith}}]{Han2016}%
  \BibitemOpen
  \bibfield  {author} {\bibinfo {author} {\bibfnamefont {F.~Y.}\ \bibnamefont
  {Han}}, \bibinfo {author} {\bibfnamefont {K.~J.}\ \bibnamefont {Thurecht}},
  \bibinfo {author} {\bibfnamefont {A.~K.}\ \bibnamefont {Whittaker}},\ and\
  \bibinfo {author} {\bibfnamefont {M.~T.}\ \bibnamefont {Smith}},\ }\bibfield
  {title} {\bibinfo {title} {\emph {{Bioerodable PLGA-Based Microparticles for
  Producing Sustained-Release Drug Formulations and Strategies for Improving
  Drug Loading}}},\ }\href {https://doi.org/10.3389/fphar.2016.00185}
  {\bibfield  {journal} {\bibinfo  {journal} {\protect\JournalTitle{Frontiers
  in Pharmacology}}\ }\textbf {\bibinfo {volume} {7}},\ \bibinfo {pages} {185}
  (\bibinfo {year} {2016})}\BibitemShut {NoStop}%
\bibitem [{\citenamefont {Đorđevi{\'{c}}}\ \emph {et~al.}(2015)\citenamefont
  {Đorđevi{\'{c}}}, \citenamefont {Balan{\v{c}}}, \citenamefont
  {Bel{\v{s}}{\v{c}}ak-Cvitanovi{\'{c}}}, \citenamefont {Levi{\'{c}}},
  \citenamefont {Trifkovi{\'{c}}}, \citenamefont {Kalu{\v{s}}evi{\'{c}}},
  \citenamefont {Kosti{\'{c}}}, \citenamefont {Komes}, \citenamefont
  {Bugarski},\ and\ \citenamefont {Nedovi{\'{c}}}}]{Dordevic2015}%
  \BibitemOpen
  \bibfield  {author} {\bibinfo {author} {\bibfnamefont {V.}~\bibnamefont
  {Đorđevi{\'{c}}}}, \bibinfo {author} {\bibfnamefont {B.}~\bibnamefont
  {Balan{\v{c}}}}, \bibinfo {author} {\bibfnamefont {A.}~\bibnamefont
  {Bel{\v{s}}{\v{c}}ak-Cvitanovi{\'{c}}}}, \bibinfo {author} {\bibfnamefont
  {S.}~\bibnamefont {Levi{\'{c}}}}, \bibinfo {author} {\bibfnamefont
  {K.}~\bibnamefont {Trifkovi{\'{c}}}}, \bibinfo {author} {\bibfnamefont
  {A.}~\bibnamefont {Kalu{\v{s}}evi{\'{c}}}}, \bibinfo {author} {\bibfnamefont
  {I.}~\bibnamefont {Kosti{\'{c}}}}, \bibinfo {author} {\bibfnamefont
  {D.}~\bibnamefont {Komes}}, \bibinfo {author} {\bibfnamefont
  {B.}~\bibnamefont {Bugarski}},\ and\ \bibinfo {author} {\bibfnamefont
  {V.}~\bibnamefont {Nedovi{\'{c}}}},\ }\bibfield  {title} {\bibinfo {title}
  {\emph {{Trends in Encapsulation Technologies for Delivery of Food Bioactive
  Compounds}}},\ }\href {https://doi.org/10.1007/s12393-014-9106-7} {\bibfield
  {journal} {\bibinfo  {journal} {\protect\JournalTitle{Food Engineering
  Reviews}}\ }\textbf {\bibinfo {volume} {7}},\ \bibinfo {pages} {452}
  (\bibinfo {year} {2015})}\BibitemShut {NoStop}%
\bibitem [{\citenamefont {Ray}\ \emph {et~al.}(2016)\citenamefont {Ray},
  \citenamefont {Raychaudhuri},\ and\ \citenamefont {Chakraborty}}]{Ray2016}%
  \BibitemOpen
  \bibfield  {author} {\bibinfo {author} {\bibfnamefont {S.}~\bibnamefont
  {Ray}}, \bibinfo {author} {\bibfnamefont {U.}~\bibnamefont {Raychaudhuri}},\
  and\ \bibinfo {author} {\bibfnamefont {R.}~\bibnamefont {Chakraborty}},\
  }\bibfield  {title} {\bibinfo {title} {\emph {{An overview of encapsulation
  of active compounds used in food products by drying technology}}},\ }\href
  {https://doi.org/10.1016/j.fbio.2015.12.009} {\bibfield  {journal} {\bibinfo
  {journal} {\protect\JournalTitle{Food Bioscience}}\ }\textbf {\bibinfo
  {volume} {13}},\ \bibinfo {pages} {76} (\bibinfo {year} {2016})}\BibitemShut
  {NoStop}%
\bibitem [{\citenamefont {Shinjo}\ \emph {et~al.}(2016)\citenamefont {Shinjo},
  \citenamefont {Xia}, \citenamefont {Ganippa},\ and\ \citenamefont
  {Megaritis}}]{Shinjo2016}%
  \BibitemOpen
  \bibfield  {author} {\bibinfo {author} {\bibfnamefont {J.}~\bibnamefont
  {Shinjo}}, \bibinfo {author} {\bibfnamefont {J.}~\bibnamefont {Xia}},
  \bibinfo {author} {\bibfnamefont {L.~C.}\ \bibnamefont {Ganippa}},\ and\
  \bibinfo {author} {\bibfnamefont {A.}~\bibnamefont {Megaritis}},\ }\bibfield
  {title} {\bibinfo {title} {\emph {{Puffing-enhanced fuel/air mixing of an
  evaporating n-decane/ethanol emulsion droplet and a droplet group under
  convective heating}}},\ }\href {https://doi.org/10.1017/jfm.2016.130}
  {\bibfield  {journal} {\bibinfo  {journal} {\protect\JournalTitle{Journal of
  Fluid Mechanics}}\ }\textbf {\bibinfo {volume} {793}},\ \bibinfo {pages}
  {444} (\bibinfo {year} {2016})}\BibitemShut {NoStop}%
\bibitem [{\citenamefont {Lyu}\ \emph {et~al.}(2021)\citenamefont {Lyu},
  \citenamefont {Tan}, \citenamefont {Wakata}, \citenamefont {Yang},
  \citenamefont {Law}, \citenamefont {Lohse},\ and\ \citenamefont
  {Sun}}]{Lyu2021}%
  \BibitemOpen
  \bibfield  {author} {\bibinfo {author} {\bibfnamefont {S.}~\bibnamefont
  {Lyu}}, \bibinfo {author} {\bibfnamefont {H.}~\bibnamefont {Tan}}, \bibinfo
  {author} {\bibfnamefont {Y.}~\bibnamefont {Wakata}}, \bibinfo {author}
  {\bibfnamefont {X.}~\bibnamefont {Yang}}, \bibinfo {author} {\bibfnamefont
  {C.~K.}\ \bibnamefont {Law}}, \bibinfo {author} {\bibfnamefont
  {D.}~\bibnamefont {Lohse}},\ and\ \bibinfo {author} {\bibfnamefont
  {C.}~\bibnamefont {Sun}},\ }\bibfield  {title} {\bibinfo {title} {\emph {{On
  explosive boiling of a multicomponent Leidenfrost drop}}},\ }\href
  {https://doi.org/10.1073/pnas.2016107118} {\bibfield  {journal} {\bibinfo
  {journal} {\protect\JournalTitle{Proceedings of the National Academy of
  Sciences}}\ }\textbf {\bibinfo {volume} {118}},\ \bibinfo {pages}
  {e2016107118} (\bibinfo {year} {2021})}\BibitemShut {NoStop}%
\bibitem [{\citenamefont {Visser}\ \emph {et~al.}(2018)\citenamefont {Visser},
  \citenamefont {Kamperman}, \citenamefont {Karbaat}, \citenamefont {Lohse},\
  and\ \citenamefont {Karperien}}]{Visser2018}%
  \BibitemOpen
  \bibfield  {author} {\bibinfo {author} {\bibfnamefont {C.~W.}\ \bibnamefont
  {Visser}}, \bibinfo {author} {\bibfnamefont {T.}~\bibnamefont {Kamperman}},
  \bibinfo {author} {\bibfnamefont {L.~P.}\ \bibnamefont {Karbaat}}, \bibinfo
  {author} {\bibfnamefont {D.}~\bibnamefont {Lohse}},\ and\ \bibinfo {author}
  {\bibfnamefont {M.}~\bibnamefont {Karperien}},\ }\bibfield  {title} {\bibinfo
  {title} {\emph {{In-air microfluidics enables rapid fabrication of emulsions,
  suspensions, and 3D modular (bio)materials}}},\ }\href
  {https://doi.org/10.1126/sciadv.aao1175} {\bibfield  {journal} {\bibinfo
  {journal} {\protect\JournalTitle{Science Advances}}\ }\textbf {\bibinfo
  {volume} {4}},\ \bibinfo {pages} {eaao1175} (\bibinfo {year}
  {2018})}\BibitemShut {NoStop}%
\bibitem [{\citenamefont {Kamperman}\ \emph {et~al.}(2018)\citenamefont
  {Kamperman}, \citenamefont {Trikalitis}, \citenamefont {Karperien},
  \citenamefont {Visser},\ and\ \citenamefont {Leijten}}]{Kamperman2018}%
  \BibitemOpen
  \bibfield  {author} {\bibinfo {author} {\bibfnamefont {T.}~\bibnamefont
  {Kamperman}}, \bibinfo {author} {\bibfnamefont {V.~D.}\ \bibnamefont
  {Trikalitis}}, \bibinfo {author} {\bibfnamefont {M.}~\bibnamefont
  {Karperien}}, \bibinfo {author} {\bibfnamefont {C.~W.}\ \bibnamefont
  {Visser}},\ and\ \bibinfo {author} {\bibfnamefont {J.}~\bibnamefont
  {Leijten}},\ }\bibfield  {title} {\bibinfo {title} {\emph
  {{Ultrahigh-Throughput Production of Monodisperse and Multifunctional Janus
  Microparticles Using in-Air Microfluidics}}},\ }\href
  {https://doi.org/10.1021/acsami.8b05227} {\bibfield  {journal} {\bibinfo
  {journal} {\protect\JournalTitle{ACS Applied Materials {\&} Interfaces}}\
  }\textbf {\bibinfo {volume} {10}},\ \bibinfo {pages} {23433} (\bibinfo {year}
  {2018})}\BibitemShut {NoStop}%
\bibitem [{\citenamefont {Li}\ \emph {et~al.}(2019)\citenamefont {Li},
  \citenamefont {Zhang}, \citenamefont {Yi}, \citenamefont {Huang},
  \citenamefont {Lv},\ and\ \citenamefont {Duan}}]{Li2019}%
  \BibitemOpen
  \bibfield  {author} {\bibinfo {author} {\bibfnamefont {X.}~\bibnamefont
  {Li}}, \bibinfo {author} {\bibfnamefont {J.~M.}\ \bibnamefont {Zhang}},
  \bibinfo {author} {\bibfnamefont {X.}~\bibnamefont {Yi}}, \bibinfo {author}
  {\bibfnamefont {Z.}~\bibnamefont {Huang}}, \bibinfo {author} {\bibfnamefont
  {P.}~\bibnamefont {Lv}},\ and\ \bibinfo {author} {\bibfnamefont
  {H.}~\bibnamefont {Duan}},\ }\bibfield  {title} {\bibinfo {title} {\emph
  {{Multimaterial Microfluidic 3D Printing of Textured Composites with Liquid
  Inclusions}}},\ }\href {https://doi.org/10.1002/advs.201800730} {\bibfield
  {journal} {\bibinfo  {journal} {\protect\JournalTitle{Advanced Science}}\
  }\textbf {\bibinfo {volume} {6}},\ \bibinfo {pages} {1800730} (\bibinfo
  {year} {2019})}\BibitemShut {NoStop}%
\bibitem [{\citenamefont {Chiu}\ and\ \citenamefont {Lin}(2005)}]{Chiu2005}%
  \BibitemOpen
  \bibfield  {author} {\bibinfo {author} {\bibfnamefont {S.-L.}\ \bibnamefont
  {Chiu}}\ and\ \bibinfo {author} {\bibfnamefont {T.-H.}\ \bibnamefont {Lin}},\
  }\bibfield  {title} {\bibinfo {title} {\emph {{Experiment on the dynamics of
  a compound drop impinging on a hot surface}}},\ }\href
  {https://doi.org/10.1063/1.2139101} {\bibfield  {journal} {\bibinfo
  {journal} {\protect\JournalTitle{Physics of Fluids}}\ }\textbf {\bibinfo
  {volume} {17}},\ \bibinfo {pages} {122103} (\bibinfo {year}
  {2005})}\BibitemShut {NoStop}%
\bibitem [{\citenamefont {Gao}\ and\ \citenamefont {Feng}(2011)}]{Gao2011}%
  \BibitemOpen
  \bibfield  {author} {\bibinfo {author} {\bibfnamefont {P.}~\bibnamefont
  {Gao}}\ and\ \bibinfo {author} {\bibfnamefont {J.~J.}\ \bibnamefont {Feng}},\
  }\bibfield  {title} {\bibinfo {title} {\emph {{Spreading and breakup of a
  compound drop on a partially wetting substrate}}},\ }\href
  {https://doi.org/10.1017/jfm.2011.235} {\bibfield  {journal} {\bibinfo
  {journal} {\protect\JournalTitle{Journal of Fluid Mechanics}}\ }\textbf
  {\bibinfo {volume} {682}},\ \bibinfo {pages} {415} (\bibinfo {year}
  {2011})}\BibitemShut {NoStop}%
\bibitem [{\citenamefont {Liu}\ and\ \citenamefont {Tran}(2018)}]{Liu2018Tran}%
  \BibitemOpen
  \bibfield  {author} {\bibinfo {author} {\bibfnamefont {D.}~\bibnamefont
  {Liu}}\ and\ \bibinfo {author} {\bibfnamefont {T.}~\bibnamefont {Tran}},\
  }\bibfield  {title} {\bibinfo {title} {\emph {{Emergence of two lamellas
  during impact of compound droplets}}},\ }\href
  {https://doi.org/10.1063/1.5026821} {\bibfield  {journal} {\bibinfo
  {journal} {\protect\JournalTitle{Applied Physics Letters}}\ }\textbf
  {\bibinfo {volume} {112}},\ \bibinfo {pages} {203702} (\bibinfo {year}
  {2018})}\BibitemShut {NoStop}%
\bibitem [{\citenamefont {Liu}\ \emph {et~al.}(2018)\citenamefont {Liu},
  \citenamefont {Zhang}, \citenamefont {Gao}, \citenamefont {Lu},\ and\
  \citenamefont {Ding}}]{Liu2018Ding}%
  \BibitemOpen
  \bibfield  {author} {\bibinfo {author} {\bibfnamefont {H.-R.}\ \bibnamefont
  {Liu}}, \bibinfo {author} {\bibfnamefont {C.-Y.}\ \bibnamefont {Zhang}},
  \bibinfo {author} {\bibfnamefont {P.}~\bibnamefont {Gao}}, \bibinfo {author}
  {\bibfnamefont {X.-Y.}\ \bibnamefont {Lu}},\ and\ \bibinfo {author}
  {\bibfnamefont {H.}~\bibnamefont {Ding}},\ }\bibfield  {title} {\bibinfo
  {title} {\emph {{On the maximal spreading of impacting compound drops}}},\
  }\href {https://doi.org/10.1017/jfm.2018.702} {\bibfield  {journal} {\bibinfo
   {journal} {\protect\JournalTitle{Journal of Fluid Mechanics}}\ }\textbf
  {\bibinfo {volume} {854}},\ \bibinfo {pages} {R6} (\bibinfo {year}
  {2018})}\BibitemShut {NoStop}%
\bibitem [{\citenamefont {Antonini}\ \emph {et~al.}(2019)\citenamefont
  {Antonini}, \citenamefont {Wu}, \citenamefont {Zimmermann}, \citenamefont
  {Kherbeche}, \citenamefont {Thoraval}, \citenamefont {Nystr{\"{o}}m},\ and\
  \citenamefont {Geiger}}]{Antonini2019}%
  \BibitemOpen
  \bibfield  {author} {\bibinfo {author} {\bibfnamefont {C.}~\bibnamefont
  {Antonini}}, \bibinfo {author} {\bibfnamefont {T.}~\bibnamefont {Wu}},
  \bibinfo {author} {\bibfnamefont {T.}~\bibnamefont {Zimmermann}}, \bibinfo
  {author} {\bibfnamefont {A.}~\bibnamefont {Kherbeche}}, \bibinfo {author}
  {\bibfnamefont {M.-J.}\ \bibnamefont {Thoraval}}, \bibinfo {author}
  {\bibfnamefont {G.}~\bibnamefont {Nystr{\"{o}}m}},\ and\ \bibinfo {author}
  {\bibfnamefont {T.}~\bibnamefont {Geiger}},\ }\bibfield  {title} {\bibinfo
  {title} {\emph {{Ultra-Porous Nanocellulose Foams: A Facile and Scalable
  Fabrication Approach}}},\ }\href {https://doi.org/10.3390/nano9081142}
  {\bibfield  {journal} {\bibinfo  {journal}
  {\protect\JournalTitle{Nanomaterials}}\ }\textbf {\bibinfo {volume} {9}},\
  \bibinfo {pages} {1142} (\bibinfo {year} {2019})}\BibitemShut {NoStop}%
\bibitem [{\citenamefont {Liu}\ and\ \citenamefont {Tran}(2019)}]{Liu2019}%
  \BibitemOpen
  \bibfield  {author} {\bibinfo {author} {\bibfnamefont {D.}~\bibnamefont
  {Liu}}\ and\ \bibinfo {author} {\bibfnamefont {T.}~\bibnamefont {Tran}},\
  }\bibfield  {title} {\bibinfo {title} {\emph {{The ejecting lamella of
  impacting compound droplets}}},\ }\href {https://doi.org/10.1063/1.5097370}
  {\bibfield  {journal} {\bibinfo  {journal} {\protect\JournalTitle{Applied
  Physics Letters}}\ }\textbf {\bibinfo {volume} {115}},\ \bibinfo {pages}
  {073702} (\bibinfo {year} {2019})}\BibitemShut {NoStop}%
\bibitem [{\citenamefont {Blanken}\ \emph {et~al.}(2020)\citenamefont
  {Blanken}, \citenamefont {Saleem}, \citenamefont {Antonini},\ and\
  \citenamefont {Thoraval}}]{Blanken2020}%
  \BibitemOpen
  \bibfield  {author} {\bibinfo {author} {\bibfnamefont {N.}~\bibnamefont
  {Blanken}}, \bibinfo {author} {\bibfnamefont {M.~S.}\ \bibnamefont {Saleem}},
  \bibinfo {author} {\bibfnamefont {C.}~\bibnamefont {Antonini}},\ and\
  \bibinfo {author} {\bibfnamefont {M.-J.}\ \bibnamefont {Thoraval}},\
  }\bibfield  {title} {\bibinfo {title} {\emph {{Rebound of self-lubricating
  compound drops}}},\ }\href {https://doi.org/10.1126/sciadv.aay3499}
  {\bibfield  {journal} {\bibinfo  {journal} {\protect\JournalTitle{Science
  Advances}}\ }\textbf {\bibinfo {volume} {6}},\ \bibinfo {pages} {eaay3499}
  (\bibinfo {year} {2020})}\BibitemShut {NoStop}%
\bibitem [{\citenamefont {Zhang}\ \emph {et~al.}(2020)\citenamefont {Zhang},
  \citenamefont {Li},\ and\ \citenamefont {Thoroddsen}}]{Zhang2020}%
  \BibitemOpen
  \bibfield  {author} {\bibinfo {author} {\bibfnamefont {J.~M.}\ \bibnamefont
  {Zhang}}, \bibinfo {author} {\bibfnamefont {E.~Q.}\ \bibnamefont {Li}},\ and\
  \bibinfo {author} {\bibfnamefont {S.~T.}\ \bibnamefont {Thoroddsen}},\
  }\bibfield  {title} {\bibinfo {title} {\emph {{Fine radial jetting during the
  impact of compound drops}}},\ }\href {https://doi.org/10.1017/jfm.2019.885}
  {\bibfield  {journal} {\bibinfo  {journal} {\protect\JournalTitle{Journal of
  Fluid Mechanics}}\ }\textbf {\bibinfo {volume} {883}},\ \bibinfo {pages}
  {A46} (\bibinfo {year} {2020})}\BibitemShut {NoStop}%
\bibitem [{\citenamefont {Han}\ \emph {et~al.}(2021)\citenamefont {Han},
  \citenamefont {Li}, \citenamefont {Zhao}, \citenamefont {Li}, \citenamefont
  {Tang},\ and\ \citenamefont {Wang}}]{Han2021}%
  \BibitemOpen
  \bibfield  {author} {\bibinfo {author} {\bibfnamefont {X.}~\bibnamefont
  {Han}}, \bibinfo {author} {\bibfnamefont {W.}~\bibnamefont {Li}}, \bibinfo
  {author} {\bibfnamefont {H.}~\bibnamefont {Zhao}}, \bibinfo {author}
  {\bibfnamefont {J.}~\bibnamefont {Li}}, \bibinfo {author} {\bibfnamefont
  {X.}~\bibnamefont {Tang}},\ and\ \bibinfo {author} {\bibfnamefont
  {L.}~\bibnamefont {Wang}},\ }\bibfield  {title} {\bibinfo {title} {\emph
  {{Slippery damper of an overlay for arresting and manipulating droplets on
  nonwetting surfaces}}},\ }\href {https://doi.org/10.1038/s41467-021-23511-3}
  {\bibfield  {journal} {\bibinfo  {journal} {\protect\JournalTitle{Nature
  Communications}}\ }\textbf {\bibinfo {volume} {12}},\ \bibinfo {pages} {3154}
  (\bibinfo {year} {2021})}\BibitemShut {NoStop}%
\bibitem [{\citenamefont {Eggers}\ and\ \citenamefont
  {Villermaux}(2008)}]{Eggers2008}%
  \BibitemOpen
  \bibfield  {author} {\bibinfo {author} {\bibfnamefont {J.}~\bibnamefont
  {Eggers}}\ and\ \bibinfo {author} {\bibfnamefont {E.}~\bibnamefont
  {Villermaux}},\ }\bibfield  {title} {\bibinfo {title} {\emph {{Physics of
  liquid jets}}},\ }\href {https://doi.org/10.1088/0034-4885/71/3/036601}
  {\bibfield  {journal} {\bibinfo  {journal} {\protect\JournalTitle{Reports on
  Progress in Physics}}\ }\textbf {\bibinfo {volume} {71}},\ \bibinfo {pages}
  {036601} (\bibinfo {year} {2008})}\BibitemShut {NoStop}%
\bibitem [{\citenamefont {Lhuissier}\ and\ \citenamefont
  {Villermaux}(2012)}]{Lhuissier2012}%
  \BibitemOpen
  \bibfield  {author} {\bibinfo {author} {\bibfnamefont {H.}~\bibnamefont
  {Lhuissier}}\ and\ \bibinfo {author} {\bibfnamefont {E.}~\bibnamefont
  {Villermaux}},\ }\bibfield  {title} {\bibinfo {title} {\emph {{Bursting
  bubble aerosols}}},\ }\href {https://doi.org/10.1017/jfm.2011.418} {\bibfield
   {journal} {\bibinfo  {journal} {\protect\JournalTitle{Journal of Fluid
  Mechanics}}\ }\textbf {\bibinfo {volume} {696}},\ \bibinfo {pages} {5}
  (\bibinfo {year} {2012})}\BibitemShut {NoStop}%
\bibitem [{\citenamefont {Joung}\ and\ \citenamefont {Buie}(2015)}]{Joung2015}%
  \BibitemOpen
  \bibfield  {author} {\bibinfo {author} {\bibfnamefont {Y.~S.}\ \bibnamefont
  {Joung}}\ and\ \bibinfo {author} {\bibfnamefont {C.~R.}\ \bibnamefont
  {Buie}},\ }\bibfield  {title} {\bibinfo {title} {\emph {{Aerosol generation
  by raindrop impact on soil}}},\ }\href {https://doi.org/10.1038/ncomms7083}
  {\bibfield  {journal} {\bibinfo  {journal} {\protect\JournalTitle{Nature
  Communications}}\ }\textbf {\bibinfo {volume} {6}},\ \bibinfo {pages} {6083}
  (\bibinfo {year} {2015})}\BibitemShut {NoStop}%
\bibitem [{\citenamefont {Huang}\ \emph {et~al.}(2021)\citenamefont {Huang},
  \citenamefont {Mahrt}, \citenamefont {Xu}, \citenamefont {Shiraiwa},
  \citenamefont {Zuend},\ and\ \citenamefont {Bertram}}]{Huang2021}%
  \BibitemOpen
  \bibfield  {author} {\bibinfo {author} {\bibfnamefont {Y.}~\bibnamefont
  {Huang}}, \bibinfo {author} {\bibfnamefont {F.}~\bibnamefont {Mahrt}},
  \bibinfo {author} {\bibfnamefont {S.}~\bibnamefont {Xu}}, \bibinfo {author}
  {\bibfnamefont {M.}~\bibnamefont {Shiraiwa}}, \bibinfo {author}
  {\bibfnamefont {A.}~\bibnamefont {Zuend}},\ and\ \bibinfo {author}
  {\bibfnamefont {A.~K.}\ \bibnamefont {Bertram}},\ }\bibfield  {title}
  {\bibinfo {title} {\emph {{Coexistence of three liquid phases in individual
  atmospheric aerosol particles}}},\ }\href
  {https://doi.org/10.1073/pnas.2102512118} {\bibfield  {journal} {\bibinfo
  {journal} {\protect\JournalTitle{Proceedings of the National Academy of
  Sciences}}\ }\textbf {\bibinfo {volume} {118}},\ \bibinfo {pages}
  {e2102512118} (\bibinfo {year} {2021})}\BibitemShut {NoStop}%
\bibitem [{\citenamefont {Lin}\ \emph {et~al.}(2021)\citenamefont {Lin},
  \citenamefont {Wang}, \citenamefont {Zhang}, \citenamefont {Jin},
  \citenamefont {Li}, \citenamefont {Bonaccurso}, \citenamefont {You},
  \citenamefont {Deng},\ and\ \citenamefont {Chen}}]{Lin2021}%
  \BibitemOpen
  \bibfield  {author} {\bibinfo {author} {\bibfnamefont {S.}~\bibnamefont
  {Lin}}, \bibinfo {author} {\bibfnamefont {D.}~\bibnamefont {Wang}}, \bibinfo
  {author} {\bibfnamefont {L.}~\bibnamefont {Zhang}}, \bibinfo {author}
  {\bibfnamefont {Y.}~\bibnamefont {Jin}}, \bibinfo {author} {\bibfnamefont
  {Z.}~\bibnamefont {Li}}, \bibinfo {author} {\bibfnamefont {E.}~\bibnamefont
  {Bonaccurso}}, \bibinfo {author} {\bibfnamefont {Z.}~\bibnamefont {You}},
  \bibinfo {author} {\bibfnamefont {X.}~\bibnamefont {Deng}},\ and\ \bibinfo
  {author} {\bibfnamefont {L.}~\bibnamefont {Chen}},\ }\bibfield  {title}
  {\bibinfo {title} {\emph {{Macrodrop‐Impact‐Mediated Fluid
  Microdispensing}}},\ }\href {https://doi.org/10.1002/advs.202101331}
  {\bibfield  {journal} {\bibinfo  {journal} {\protect\JournalTitle{Advanced
  Science}}\ }\textbf {\bibinfo {volume} {8}},\ \bibinfo {pages} {2101331}
  (\bibinfo {year} {2021})}\BibitemShut {NoStop}%
\bibitem [{\citenamefont {Josserand}\ and\ \citenamefont
  {Thoroddsen}(2016)}]{Josserand2016}%
  \BibitemOpen
  \bibfield  {author} {\bibinfo {author} {\bibfnamefont {C.}~\bibnamefont
  {Josserand}}\ and\ \bibinfo {author} {\bibfnamefont {S.~T.}\ \bibnamefont
  {Thoroddsen}},\ }\bibfield  {title} {\bibinfo {title} {\emph {{Drop Impact on
  a Solid Surface}}},\ }\href
  {https://doi.org/10.1146/annurev-fluid-122414-034401} {\bibfield  {journal}
  {\bibinfo  {journal} {\protect\JournalTitle{Annual Review of Fluid
  Mechanics}}\ }\textbf {\bibinfo {volume} {48}},\ \bibinfo {pages} {365}
  (\bibinfo {year} {2016})}\BibitemShut {NoStop}%
\bibitem [{\citenamefont {Bartolo}\ \emph {et~al.}(2006)\citenamefont
  {Bartolo}, \citenamefont {Josserand},\ and\ \citenamefont
  {Bonn}}]{Bartolo2006}%
  \BibitemOpen
  \bibfield  {author} {\bibinfo {author} {\bibfnamefont {D.}~\bibnamefont
  {Bartolo}}, \bibinfo {author} {\bibfnamefont {C.}~\bibnamefont {Josserand}},\
  and\ \bibinfo {author} {\bibfnamefont {D.}~\bibnamefont {Bonn}},\ }\bibfield
  {title} {\bibinfo {title} {\emph {{Singular Jets and Bubbles in Drop
  Impact}}},\ }\href {https://doi.org/10.1103/PhysRevLett.96.124501} {\bibfield
   {journal} {\bibinfo  {journal} {\protect\JournalTitle{Physical Review
  Letters}}\ }\textbf {\bibinfo {volume} {96}},\ \bibinfo {pages} {124501}
  (\bibinfo {year} {2006})}\BibitemShut {NoStop}%
\bibitem [{\citenamefont {Yamamoto}\ \emph {et~al.}(2018)\citenamefont
  {Yamamoto}, \citenamefont {Motosuke},\ and\ \citenamefont
  {Ogata}}]{Yamamoto2018}%
  \BibitemOpen
  \bibfield  {author} {\bibinfo {author} {\bibfnamefont {K.}~\bibnamefont
  {Yamamoto}}, \bibinfo {author} {\bibfnamefont {M.}~\bibnamefont {Motosuke}},\
  and\ \bibinfo {author} {\bibfnamefont {S.}~\bibnamefont {Ogata}},\ }\bibfield
   {title} {\bibinfo {title} {\emph {{Initiation of the Worthington jet on the
  droplet impact}}},\ }\href {https://doi.org/10.1063/1.5020085} {\bibfield
  {journal} {\bibinfo  {journal} {\protect\JournalTitle{Applied Physics
  Letters}}\ }\textbf {\bibinfo {volume} {112}},\ \bibinfo {pages} {093701}
  (\bibinfo {year} {2018})}\BibitemShut {NoStop}%
\bibitem [{\citenamefont {Chen}\ \emph {et~al.}(2017)\citenamefont {Chen},
  \citenamefont {Li}, \citenamefont {Li},\ and\ \citenamefont
  {Zhang}}]{Chen2017}%
  \BibitemOpen
  \bibfield  {author} {\bibinfo {author} {\bibfnamefont {L.}~\bibnamefont
  {Chen}}, \bibinfo {author} {\bibfnamefont {L.}~\bibnamefont {Li}}, \bibinfo
  {author} {\bibfnamefont {Z.}~\bibnamefont {Li}},\ and\ \bibinfo {author}
  {\bibfnamefont {K.}~\bibnamefont {Zhang}},\ }\bibfield  {title} {\bibinfo
  {title} {\emph {{Submillimeter-Sized Bubble Entrapment and a High-Speed Jet
  Emission during Droplet Impact on Solid Surfaces}}},\ }\href
  {https://doi.org/10.1021/acs.langmuir.7b01506} {\bibfield  {journal}
  {\bibinfo  {journal} {\protect\JournalTitle{Langmuir}}\ }\textbf {\bibinfo
  {volume} {33}},\ \bibinfo {pages} {7225} (\bibinfo {year}
  {2017})}\BibitemShut {NoStop}%
\bibitem [{\citenamefont {Guo}\ \emph {et~al.}(2020)\citenamefont {Guo},
  \citenamefont {Zou}, \citenamefont {Lin}, \citenamefont {Zhao}, \citenamefont
  {Deng},\ and\ \citenamefont {Chen}}]{Guo2020}%
  \BibitemOpen
  \bibfield  {author} {\bibinfo {author} {\bibfnamefont {J.}~\bibnamefont
  {Guo}}, \bibinfo {author} {\bibfnamefont {S.}~\bibnamefont {Zou}}, \bibinfo
  {author} {\bibfnamefont {S.}~\bibnamefont {Lin}}, \bibinfo {author}
  {\bibfnamefont {B.}~\bibnamefont {Zhao}}, \bibinfo {author} {\bibfnamefont
  {X.}~\bibnamefont {Deng}},\ and\ \bibinfo {author} {\bibfnamefont
  {L.}~\bibnamefont {Chen}},\ }\bibfield  {title} {\bibinfo {title} {\emph
  {{Oblique droplet impact on superhydrophobic surfaces: Jets and bubbles}}},\
  }\href {https://doi.org/10.1063/5.0033729} {\bibfield  {journal} {\bibinfo
  {journal} {\protect\JournalTitle{Physics of Fluids}}\ }\textbf {\bibinfo
  {volume} {32}},\ \bibinfo {pages} {122112} (\bibinfo {year}
  {2020})}\BibitemShut {NoStop}%
\bibitem [{\citenamefont {Siddique}\ \emph {et~al.}(2020)\citenamefont
  {Siddique}, \citenamefont {Trimble}, \citenamefont {Zhao}, \citenamefont
  {Weislogel},\ and\ \citenamefont {Tan}}]{Siddique2020}%
  \BibitemOpen
  \bibfield  {author} {\bibinfo {author} {\bibfnamefont {A.~U.}\ \bibnamefont
  {Siddique}}, \bibinfo {author} {\bibfnamefont {M.}~\bibnamefont {Trimble}},
  \bibinfo {author} {\bibfnamefont {F.}~\bibnamefont {Zhao}}, \bibinfo {author}
  {\bibfnamefont {M.~M.}\ \bibnamefont {Weislogel}},\ and\ \bibinfo {author}
  {\bibfnamefont {H.}~\bibnamefont {Tan}},\ }\bibfield  {title} {\bibinfo
  {title} {\emph {{Jet ejection following drop impact on micropillared
  hydrophilic substrates}}},\ }\href
  {https://doi.org/10.1103/physrevfluids.5.063606} {\bibfield  {journal}
  {\bibinfo  {journal} {\protect\JournalTitle{Physical Review Fluids}}\
  }\textbf {\bibinfo {volume} {5}},\ \bibinfo {pages} {63606} (\bibinfo {year}
  {2020})}\BibitemShut {NoStop}%
\bibitem [{\citenamefont {Mitra}\ \emph {et~al.}(2021)\citenamefont {Mitra},
  \citenamefont {Vo},\ and\ \citenamefont {Tran}}]{Mitra2021}%
  \BibitemOpen
  \bibfield  {author} {\bibinfo {author} {\bibfnamefont {S.}~\bibnamefont
  {Mitra}}, \bibinfo {author} {\bibfnamefont {Q.}~\bibnamefont {Vo}},\ and\
  \bibinfo {author} {\bibfnamefont {T.}~\bibnamefont {Tran}},\ }\bibfield
  {title} {\bibinfo {title} {\emph {{Bouncing-to-wetting transition of water
  droplets impacting soft solids}}},\ }\href
  {https://doi.org/10.1039/D1SM00339A} {\bibfield  {journal} {\bibinfo
  {journal} {\protect\JournalTitle{Soft Matter}}\ }\textbf {\bibinfo {volume}
  {17}},\ \bibinfo {pages} {5969} (\bibinfo {year} {2021})}\BibitemShut
  {NoStop}%
\bibitem [{\citenamefont {Kolinski}\ \emph {et~al.}(2014)\citenamefont
  {Kolinski}, \citenamefont {Mahadevan},\ and\ \citenamefont
  {Rubinstein}}]{Kolinski2014}%
  \BibitemOpen
  \bibfield  {author} {\bibinfo {author} {\bibfnamefont {J.~M.}\ \bibnamefont
  {Kolinski}}, \bibinfo {author} {\bibfnamefont {L.}~\bibnamefont
  {Mahadevan}},\ and\ \bibinfo {author} {\bibfnamefont {S.~M.}\ \bibnamefont
  {Rubinstein}},\ }\bibfield  {title} {\bibinfo {title} {\emph {{Drops can
  bounce from perfectly hydrophilic surfaces}}},\ }\href
  {https://doi.org/10.1209/0295-5075/108/24001} {\bibfield  {journal} {\bibinfo
   {journal} {\protect\JournalTitle{EPL (Europhysics Letters)}}\ }\textbf
  {\bibinfo {volume} {108}},\ \bibinfo {pages} {24001} (\bibinfo {year}
  {2014})}\BibitemShut {NoStop}%
\bibitem [{\citenamefont {de~Ruiter}\ \emph
  {et~al.}(2015{\natexlab{a}})\citenamefont {de~Ruiter}, \citenamefont
  {Lagraauw}, \citenamefont {van~den Ende},\ and\ \citenamefont
  {Mugele}}]{DeRuiter2015}%
  \BibitemOpen
  \bibfield  {author} {\bibinfo {author} {\bibfnamefont {J.}~\bibnamefont
  {de~Ruiter}}, \bibinfo {author} {\bibfnamefont {R.}~\bibnamefont {Lagraauw}},
  \bibinfo {author} {\bibfnamefont {D.}~\bibnamefont {van~den Ende}},\ and\
  \bibinfo {author} {\bibfnamefont {F.}~\bibnamefont {Mugele}},\ }\bibfield
  {title} {\bibinfo {title} {\emph {{Wettability-independent bouncing on flat
  surfaces mediated by thin air films}}},\ }\href
  {https://doi.org/10.1038/nphys3145} {\bibfield  {journal} {\bibinfo
  {journal} {\protect\JournalTitle{Nature Physics}}\ }\textbf {\bibinfo
  {volume} {11}},\ \bibinfo {pages} {48} (\bibinfo {year}
  {2015}{\natexlab{a}})}\BibitemShut {NoStop}%
\bibitem [{\citenamefont {de~Ruiter}\ \emph
  {et~al.}(2015{\natexlab{b}})\citenamefont {de~Ruiter}, \citenamefont
  {Lagraauw}, \citenamefont {Mugele},\ and\ \citenamefont {van~den
  Ende}}]{DeRuiter2015JFM}%
  \BibitemOpen
  \bibfield  {author} {\bibinfo {author} {\bibfnamefont {J.}~\bibnamefont
  {de~Ruiter}}, \bibinfo {author} {\bibfnamefont {R.}~\bibnamefont {Lagraauw}},
  \bibinfo {author} {\bibfnamefont {F.}~\bibnamefont {Mugele}},\ and\ \bibinfo
  {author} {\bibfnamefont {D.}~\bibnamefont {van~den Ende}},\ }\bibfield
  {title} {\bibinfo {title} {\emph {{Bouncing on thin air: how squeeze forces
  in the air film during non-wetting droplet bouncing lead to momentum transfer
  and dissipation}}},\ }\href {https://doi.org/10.1017/jfm.2015.310} {\bibfield
   {journal} {\bibinfo  {journal} {\protect\JournalTitle{Journal of Fluid
  Mechanics}}\ }\textbf {\bibinfo {volume} {776}},\ \bibinfo {pages} {531}
  (\bibinfo {year} {2015}{\natexlab{b}})}\BibitemShut {NoStop}%
\bibitem [{\citenamefont {Lakshman}\ \emph {et~al.}(2021)\citenamefont
  {Lakshman}, \citenamefont {Tewes}, \citenamefont {Harth}, \citenamefont
  {Snoeijer},\ and\ \citenamefont {Lohse}}]{Lakshman2021}%
  \BibitemOpen
  \bibfield  {author} {\bibinfo {author} {\bibfnamefont {S.}~\bibnamefont
  {Lakshman}}, \bibinfo {author} {\bibfnamefont {W.}~\bibnamefont {Tewes}},
  \bibinfo {author} {\bibfnamefont {K.}~\bibnamefont {Harth}}, \bibinfo
  {author} {\bibfnamefont {J.~H.}\ \bibnamefont {Snoeijer}},\ and\ \bibinfo
  {author} {\bibfnamefont {D.}~\bibnamefont {Lohse}},\ }\bibfield  {title}
  {\bibinfo {title} {\emph {{Deformation and relaxation of viscous thin films
  under bouncing drops}}},\ }\href {https://doi.org/10.1017/jfm.2021.378}
  {\bibfield  {journal} {\bibinfo  {journal} {\protect\JournalTitle{Journal of
  Fluid Mechanics}}\ }\textbf {\bibinfo {volume} {920}},\ \bibinfo {pages} {A3}
  (\bibinfo {year} {2021})}\BibitemShut {NoStop}%
\bibitem [{\citenamefont {Tran}\ \emph {et~al.}(2012)\citenamefont {Tran},
  \citenamefont {Staat}, \citenamefont {Prosperetti}, \citenamefont {Sun},\
  and\ \citenamefont {Lohse}}]{Tran2012}%
  \BibitemOpen
  \bibfield  {author} {\bibinfo {author} {\bibfnamefont {T.}~\bibnamefont
  {Tran}}, \bibinfo {author} {\bibfnamefont {H.~J.~J.}\ \bibnamefont {Staat}},
  \bibinfo {author} {\bibfnamefont {A.}~\bibnamefont {Prosperetti}}, \bibinfo
  {author} {\bibfnamefont {C.}~\bibnamefont {Sun}},\ and\ \bibinfo {author}
  {\bibfnamefont {D.}~\bibnamefont {Lohse}},\ }\bibfield  {title} {\bibinfo
  {title} {\emph {{Drop Impact on Superheated Surfaces}}},\ }\href
  {https://doi.org/10.1103/PhysRevLett.108.036101} {\bibfield  {journal}
  {\bibinfo  {journal} {\protect\JournalTitle{Physical Review Letters}}\
  }\textbf {\bibinfo {volume} {108}},\ \bibinfo {pages} {036101} (\bibinfo
  {year} {2012})}\BibitemShut {NoStop}%
\bibitem [{\citenamefont {Qu{\'{e}}r{\'{e}}}(2013)}]{Quere2013}%
  \BibitemOpen
  \bibfield  {author} {\bibinfo {author} {\bibfnamefont {D.}~\bibnamefont
  {Qu{\'{e}}r{\'{e}}}},\ }\bibfield  {title} {\bibinfo {title} {\emph
  {{Leidenfrost Dynamics}}},\ }\href
  {https://doi.org/10.1146/annurev-fluid-011212-140709} {\bibfield  {journal}
  {\bibinfo  {journal} {\protect\JournalTitle{Annual Review of Fluid
  Mechanics}}\ }\textbf {\bibinfo {volume} {45}},\ \bibinfo {pages} {197}
  (\bibinfo {year} {2013})}\BibitemShut {NoStop}%
\bibitem [{\citenamefont {Shirota}\ \emph {et~al.}(2016)\citenamefont
  {Shirota}, \citenamefont {van Limbeek}, \citenamefont {Sun}, \citenamefont
  {Prosperetti},\ and\ \citenamefont {Lohse}}]{Shirota2016}%
  \BibitemOpen
  \bibfield  {author} {\bibinfo {author} {\bibfnamefont {M.}~\bibnamefont
  {Shirota}}, \bibinfo {author} {\bibfnamefont {M.~A.~J.}\ \bibnamefont {van
  Limbeek}}, \bibinfo {author} {\bibfnamefont {C.}~\bibnamefont {Sun}},
  \bibinfo {author} {\bibfnamefont {A.}~\bibnamefont {Prosperetti}},\ and\
  \bibinfo {author} {\bibfnamefont {D.}~\bibnamefont {Lohse}},\ }\bibfield
  {title} {\bibinfo {title} {\emph {{Dynamic Leidenfrost Effect: Relevant Time
  and Length Scales}}},\ }\href
  {https://doi.org/10.1103/PhysRevLett.116.064501} {\bibfield  {journal}
  {\bibinfo  {journal} {\protect\JournalTitle{Physical Review Letters}}\
  }\textbf {\bibinfo {volume} {116}},\ \bibinfo {pages} {064501} (\bibinfo
  {year} {2016})}\BibitemShut {NoStop}%
\bibitem [{\citenamefont {Bouillant}\ \emph {et~al.}(2018)\citenamefont
  {Bouillant}, \citenamefont {Mouterde}, \citenamefont {Bourrianne},
  \citenamefont {Lagarde}, \citenamefont {Clanet},\ and\ \citenamefont
  {Qu{\'{e}}r{\'{e}}}}]{Bouillant2018}%
  \BibitemOpen
  \bibfield  {author} {\bibinfo {author} {\bibfnamefont {A.}~\bibnamefont
  {Bouillant}}, \bibinfo {author} {\bibfnamefont {T.}~\bibnamefont {Mouterde}},
  \bibinfo {author} {\bibfnamefont {P.}~\bibnamefont {Bourrianne}}, \bibinfo
  {author} {\bibfnamefont {A.}~\bibnamefont {Lagarde}}, \bibinfo {author}
  {\bibfnamefont {C.}~\bibnamefont {Clanet}},\ and\ \bibinfo {author}
  {\bibfnamefont {D.}~\bibnamefont {Qu{\'{e}}r{\'{e}}}},\ }\bibfield  {title}
  {\bibinfo {title} {\emph {{Leidenfrost wheels}}},\ }\href
  {https://doi.org/10.1038/s41567-018-0275-9} {\bibfield  {journal} {\bibinfo
  {journal} {\protect\JournalTitle{Nature Physics}}\ }\textbf {\bibinfo
  {volume} {14}},\ \bibinfo {pages} {1188} (\bibinfo {year}
  {2018})}\BibitemShut {NoStop}%
\bibitem [{\citenamefont {Lee}\ \emph {et~al.}(2020)\citenamefont {Lee},
  \citenamefont {Rump}, \citenamefont {Harth}, \citenamefont {Kim},
  \citenamefont {Lohse}, \citenamefont {Fezzaa},\ and\ \citenamefont
  {Je}}]{Lee2020}%
  \BibitemOpen
  \bibfield  {author} {\bibinfo {author} {\bibfnamefont {S.-H.}\ \bibnamefont
  {Lee}}, \bibinfo {author} {\bibfnamefont {M.}~\bibnamefont {Rump}}, \bibinfo
  {author} {\bibfnamefont {K.}~\bibnamefont {Harth}}, \bibinfo {author}
  {\bibfnamefont {M.}~\bibnamefont {Kim}}, \bibinfo {author} {\bibfnamefont
  {D.}~\bibnamefont {Lohse}}, \bibinfo {author} {\bibfnamefont
  {K.}~\bibnamefont {Fezzaa}},\ and\ \bibinfo {author} {\bibfnamefont {J.~H.}\
  \bibnamefont {Je}},\ }\bibfield  {title} {\bibinfo {title} {\emph {{Downward
  jetting of a dynamic Leidenfrost drop}}},\ }\href
  {https://doi.org/10.1103/PhysRevFluids.5.074802} {\bibfield  {journal}
  {\bibinfo  {journal} {\protect\JournalTitle{Physical Review Fluids}}\
  }\textbf {\bibinfo {volume} {5}},\ \bibinfo {pages} {74802} (\bibinfo {year}
  {2020})}\BibitemShut {NoStop}%
\bibitem [{\citenamefont {Antonini}\ \emph {et~al.}(2013)\citenamefont
  {Antonini}, \citenamefont {Bernagozzi}, \citenamefont {Jung}, \citenamefont
  {Poulikakos},\ and\ \citenamefont {Marengo}}]{Antonini2013}%
  \BibitemOpen
  \bibfield  {author} {\bibinfo {author} {\bibfnamefont {C.}~\bibnamefont
  {Antonini}}, \bibinfo {author} {\bibfnamefont {I.}~\bibnamefont
  {Bernagozzi}}, \bibinfo {author} {\bibfnamefont {S.}~\bibnamefont {Jung}},
  \bibinfo {author} {\bibfnamefont {D.}~\bibnamefont {Poulikakos}},\ and\
  \bibinfo {author} {\bibfnamefont {M.}~\bibnamefont {Marengo}},\ }\bibfield
  {title} {\bibinfo {title} {\emph {{Water Drops Dancing on Ice: How
  Sublimation Leads to Drop Rebound}}},\ }\href
  {https://doi.org/10.1103/PhysRevLett.111.014501} {\bibfield  {journal}
  {\bibinfo  {journal} {\protect\JournalTitle{Physical Review Letters}}\
  }\textbf {\bibinfo {volume} {111}},\ \bibinfo {pages} {014501} (\bibinfo
  {year} {2013})}\BibitemShut {NoStop}%
\bibitem [{\citenamefont {Wong}\ \emph {et~al.}(2011)\citenamefont {Wong},
  \citenamefont {Kang}, \citenamefont {Tang}, \citenamefont {Smythe},
  \citenamefont {Hatton}, \citenamefont {Grinthal},\ and\ \citenamefont
  {Aizenberg}}]{Wong2011}%
  \BibitemOpen
  \bibfield  {author} {\bibinfo {author} {\bibfnamefont {T.-S.}\ \bibnamefont
  {Wong}}, \bibinfo {author} {\bibfnamefont {S.~H.}\ \bibnamefont {Kang}},
  \bibinfo {author} {\bibfnamefont {S.~K.~Y.}\ \bibnamefont {Tang}}, \bibinfo
  {author} {\bibfnamefont {E.~J.}\ \bibnamefont {Smythe}}, \bibinfo {author}
  {\bibfnamefont {B.~D.}\ \bibnamefont {Hatton}}, \bibinfo {author}
  {\bibfnamefont {A.}~\bibnamefont {Grinthal}},\ and\ \bibinfo {author}
  {\bibfnamefont {J.}~\bibnamefont {Aizenberg}},\ }\bibfield  {title} {\bibinfo
  {title} {\emph {{Bioinspired self-repairing slippery surfaces with
  pressure-stable omniphobicity}}},\ }\href
  {https://doi.org/10.1038/nature10447} {\bibfield  {journal} {\bibinfo
  {journal} {\protect\JournalTitle{Nature}}\ }\textbf {\bibinfo {volume}
  {477}},\ \bibinfo {pages} {443} (\bibinfo {year} {2011})}\BibitemShut
  {NoStop}%
\bibitem [{\citenamefont {Lee}\ \emph {et~al.}(2014)\citenamefont {Lee},
  \citenamefont {Kim},\ and\ \citenamefont {Nam}}]{Lee2014}%
  \BibitemOpen
  \bibfield  {author} {\bibinfo {author} {\bibfnamefont {C.}~\bibnamefont
  {Lee}}, \bibinfo {author} {\bibfnamefont {H.}~\bibnamefont {Kim}},\ and\
  \bibinfo {author} {\bibfnamefont {Y.}~\bibnamefont {Nam}},\ }\bibfield
  {title} {\bibinfo {title} {\emph {{Drop impact dynamics on oil-infused
  nanostructured surfaces}}},\ }\href {https://doi.org/10.1021/la501341x}
  {\bibfield  {journal} {\bibinfo  {journal} {\protect\JournalTitle{Langmuir}}\
  }\textbf {\bibinfo {volume} {30}},\ \bibinfo {pages} {8400} (\bibinfo {year}
  {2014})}\BibitemShut {NoStop}%
\bibitem [{\citenamefont {Zeff}\ \emph {et~al.}(2000)\citenamefont {Zeff},
  \citenamefont {Kleber}, \citenamefont {Fineberg},\ and\ \citenamefont
  {Lathrop}}]{Zeff2000}%
  \BibitemOpen
  \bibfield  {author} {\bibinfo {author} {\bibfnamefont {B.~W.}\ \bibnamefont
  {Zeff}}, \bibinfo {author} {\bibfnamefont {B.}~\bibnamefont {Kleber}},
  \bibinfo {author} {\bibfnamefont {J.}~\bibnamefont {Fineberg}},\ and\
  \bibinfo {author} {\bibfnamefont {D.~P.}\ \bibnamefont {Lathrop}},\
  }\bibfield  {title} {\bibinfo {title} {\emph {{Singularity dynamics in
  curvature collapse and jet eruption on a fluid surface}}},\ }\href
  {https://doi.org/10.1038/35000151} {\bibfield  {journal} {\bibinfo  {journal}
  {\protect\JournalTitle{Nature}}\ }\textbf {\bibinfo {volume} {403}},\
  \bibinfo {pages} {401} (\bibinfo {year} {2000})}\BibitemShut {NoStop}%
\bibitem [{\citenamefont {Duchemin}\ \emph {et~al.}(2002)\citenamefont
  {Duchemin}, \citenamefont {Popinet}, \citenamefont {Josserand},\ and\
  \citenamefont {Zaleski}}]{Duchemin2002}%
  \BibitemOpen
  \bibfield  {author} {\bibinfo {author} {\bibfnamefont {L.}~\bibnamefont
  {Duchemin}}, \bibinfo {author} {\bibfnamefont {S.}~\bibnamefont {Popinet}},
  \bibinfo {author} {\bibfnamefont {C.}~\bibnamefont {Josserand}},\ and\
  \bibinfo {author} {\bibfnamefont {S.}~\bibnamefont {Zaleski}},\ }\bibfield
  {title} {\bibinfo {title} {\emph {{Jet formation in bubbles bursting at a
  free surface}}},\ }\href {https://doi.org/10.1063/1.1494072} {\bibfield
  {journal} {\bibinfo  {journal} {\protect\JournalTitle{Physics of Fluids}}\
  }\textbf {\bibinfo {volume} {14}},\ \bibinfo {pages} {3000} (\bibinfo {year}
  {2002})}\BibitemShut {NoStop}%
\bibitem [{\citenamefont {Michon}\ \emph {et~al.}(2017)\citenamefont {Michon},
  \citenamefont {Josserand},\ and\ \citenamefont {S{\'{e}}on}}]{Michon2017}%
  \BibitemOpen
  \bibfield  {author} {\bibinfo {author} {\bibfnamefont {G.-J.}\ \bibnamefont
  {Michon}}, \bibinfo {author} {\bibfnamefont {C.}~\bibnamefont {Josserand}},\
  and\ \bibinfo {author} {\bibfnamefont {T.}~\bibnamefont {S{\'{e}}on}},\
  }\bibfield  {title} {\bibinfo {title} {\emph {{Jet dynamics post drop impact
  on a deep pool}}},\ }\href {https://doi.org/10.1103/PhysRevFluids.2.023601}
  {\bibfield  {journal} {\bibinfo  {journal} {\protect\JournalTitle{Physical
  Review Fluids}}\ }\textbf {\bibinfo {volume} {2}},\ \bibinfo {pages} {023601}
  (\bibinfo {year} {2017})}\BibitemShut {NoStop}%
\bibitem [{\citenamefont {Thoroddsen}\ \emph {et~al.}(2018)\citenamefont
  {Thoroddsen}, \citenamefont {Takehara}, \citenamefont {Nguyen},\ and\
  \citenamefont {Etoh}}]{Thoroddsen2018}%
  \BibitemOpen
  \bibfield  {author} {\bibinfo {author} {\bibfnamefont {S.~T.}\ \bibnamefont
  {Thoroddsen}}, \bibinfo {author} {\bibfnamefont {K.}~\bibnamefont
  {Takehara}}, \bibinfo {author} {\bibfnamefont {H.~D.}\ \bibnamefont
  {Nguyen}},\ and\ \bibinfo {author} {\bibfnamefont {T.~G.}\ \bibnamefont
  {Etoh}},\ }\bibfield  {title} {\bibinfo {title} {\emph {{Singular jets during
  the collapse of drop-impact craters}}},\ }\href
  {https://doi.org/10.1017/jfm.2018.435} {\bibfield  {journal} {\bibinfo
  {journal} {\protect\JournalTitle{Journal of Fluid Mechanics}}\ }\textbf
  {\bibinfo {volume} {848}},\ \bibinfo {pages} {R3} (\bibinfo {year}
  {2018})}\BibitemShut {NoStop}%
\bibitem [{\citenamefont {Yang}\ \emph {et~al.}(2020)\citenamefont {Yang},
  \citenamefont {Tian},\ and\ \citenamefont {Thoroddsen}}]{Yang2020}%
  \BibitemOpen
  \bibfield  {author} {\bibinfo {author} {\bibfnamefont {Z.~Q.}\ \bibnamefont
  {Yang}}, \bibinfo {author} {\bibfnamefont {Y.~S.}\ \bibnamefont {Tian}},\
  and\ \bibinfo {author} {\bibfnamefont {S.~T.}\ \bibnamefont {Thoroddsen}},\
  }\bibfield  {title} {\bibinfo {title} {\emph {{Multitude of dimple shapes can
  produce singular jets during the collapse of immiscible drop-impact
  craters}}},\ }\href {https://doi.org/10.1017/jfm.2020.694} {\bibfield
  {journal} {\bibinfo  {journal} {\protect\JournalTitle{Journal of Fluid
  Mechanics}}\ }\textbf {\bibinfo {volume} {904}},\ \bibinfo {pages} {A19}
  (\bibinfo {year} {2020})}\BibitemShut {NoStop}%
\bibitem [{\citenamefont {Eggers}\ and\ \citenamefont
  {Fontelos}(2009)}]{Eggers2009}%
  \BibitemOpen
  \bibfield  {author} {\bibinfo {author} {\bibfnamefont {J.}~\bibnamefont
  {Eggers}}\ and\ \bibinfo {author} {\bibfnamefont {M.~A.}\ \bibnamefont
  {Fontelos}},\ }\bibfield  {title} {\bibinfo {title} {\emph {{The role of
  self-similarity in singularities of partial differential equations}}},\
  }\href {https://doi.org/10.1088/0951-7715/22/1/R01} {\bibfield  {journal}
  {\bibinfo  {journal} {\protect\JournalTitle{Nonlinearity}}\ }\textbf
  {\bibinfo {volume} {22}},\ \bibinfo {pages} {R1} (\bibinfo {year}
  {2009})}\BibitemShut {NoStop}%
\bibitem [{\citenamefont {Eggers}\ and\ \citenamefont
  {Fontelos}(2015)}]{Eggers2015}%
  \BibitemOpen
  \bibfield  {author} {\bibinfo {author} {\bibfnamefont {J.}~\bibnamefont
  {Eggers}}\ and\ \bibinfo {author} {\bibfnamefont {M.~A.}\ \bibnamefont
  {Fontelos}},\ }\href {https://www.cambridge.org/9781107485495
  https://1lib.us/book/3434368/5c2b2f?id=3434368{\&}secret=5c2b2f} {\emph
  {\bibinfo {title} {{Singularities: Formation, Structure, and Propagation}}}}\
  (\bibinfo  {publisher} {Cambridge University Press},\ \bibinfo {address}
  {Cambridge, United Kingdom},\ \bibinfo {year} {2015})\BibitemShut {NoStop}%
\bibitem [{\citenamefont {Day}\ \emph {et~al.}(1998)\citenamefont {Day},
  \citenamefont {Hinch},\ and\ \citenamefont {Lister}}]{Day1998}%
  \BibitemOpen
  \bibfield  {author} {\bibinfo {author} {\bibfnamefont {R.~F.}\ \bibnamefont
  {Day}}, \bibinfo {author} {\bibfnamefont {E.~J.}\ \bibnamefont {Hinch}},\
  and\ \bibinfo {author} {\bibfnamefont {J.~R.}\ \bibnamefont {Lister}},\
  }\bibfield  {title} {\bibinfo {title} {\emph {{Self-Similar Capillary
  Pinchoff of an Inviscid Fluid}}},\ }\href
  {https://doi.org/10.1103/PhysRevLett.80.704} {\bibfield  {journal} {\bibinfo
  {journal} {\protect\JournalTitle{Physical Review Letters}}\ }\textbf
  {\bibinfo {volume} {80}},\ \bibinfo {pages} {704} (\bibinfo {year}
  {1998})}\BibitemShut {NoStop}%
\bibitem [{\citenamefont {Burton}\ \emph {et~al.}(2005)\citenamefont {Burton},
  \citenamefont {Waldrep},\ and\ \citenamefont {Taborek}}]{Burton2005}%
  \BibitemOpen
  \bibfield  {author} {\bibinfo {author} {\bibfnamefont {J.~C.}\ \bibnamefont
  {Burton}}, \bibinfo {author} {\bibfnamefont {R.}~\bibnamefont {Waldrep}},\
  and\ \bibinfo {author} {\bibfnamefont {P.}~\bibnamefont {Taborek}},\
  }\bibfield  {title} {\bibinfo {title} {\emph {{Scaling and instabilities in
  bubble pinch-off}}},\ }\href {https://doi.org/10.1103/PhysRevLett.94.184502}
  {\bibfield  {journal} {\bibinfo  {journal} {\protect\JournalTitle{Physical
  Review Letters}}\ }\textbf {\bibinfo {volume} {94}},\ \bibinfo {pages}
  {184502} (\bibinfo {year} {2005})}\BibitemShut {NoStop}%
\bibitem [{\citenamefont {Bergmann}\ \emph {et~al.}(2006)\citenamefont
  {Bergmann}, \citenamefont {van~der Meer}, \citenamefont {Stijnman},
  \citenamefont {Sandtke}, \citenamefont {Prosperetti},\ and\ \citenamefont
  {Lohse}}]{Bergmann2006}%
  \BibitemOpen
  \bibfield  {author} {\bibinfo {author} {\bibfnamefont {R.}~\bibnamefont
  {Bergmann}}, \bibinfo {author} {\bibfnamefont {D.}~\bibnamefont {van~der
  Meer}}, \bibinfo {author} {\bibfnamefont {M.}~\bibnamefont {Stijnman}},
  \bibinfo {author} {\bibfnamefont {M.}~\bibnamefont {Sandtke}}, \bibinfo
  {author} {\bibfnamefont {A.}~\bibnamefont {Prosperetti}},\ and\ \bibinfo
  {author} {\bibfnamefont {D.}~\bibnamefont {Lohse}},\ }\bibfield  {title}
  {\bibinfo {title} {\emph {{Giant Bubble Pinch-Off}}},\ }\href
  {https://doi.org/10.1103/PhysRevLett.96.154505} {\bibfield  {journal}
  {\bibinfo  {journal} {\protect\JournalTitle{Physical Review Letters}}\
  }\textbf {\bibinfo {volume} {96}},\ \bibinfo {pages} {154505} (\bibinfo
  {year} {2006})}\BibitemShut {NoStop}%
\bibitem [{\citenamefont {Castrej{\'{o}}n-Pita}\ \emph
  {et~al.}(2015)\citenamefont {Castrej{\'{o}}n-Pita}, \citenamefont
  {Castrej{\'{o}}n-Pita}, \citenamefont {Thete}, \citenamefont {Sambath},
  \citenamefont {Hutchings}, \citenamefont {Hinch}, \citenamefont {Lister},\
  and\ \citenamefont {Basaran}}]{Castrejon-Pita2015}%
  \BibitemOpen
  \bibfield  {author} {\bibinfo {author} {\bibfnamefont {J.~R.}\ \bibnamefont
  {Castrej{\'{o}}n-Pita}}, \bibinfo {author} {\bibfnamefont {A.~A.}\
  \bibnamefont {Castrej{\'{o}}n-Pita}}, \bibinfo {author} {\bibfnamefont
  {S.~S.}\ \bibnamefont {Thete}}, \bibinfo {author} {\bibfnamefont
  {K.}~\bibnamefont {Sambath}}, \bibinfo {author} {\bibfnamefont {I.~M.}\
  \bibnamefont {Hutchings}}, \bibinfo {author} {\bibfnamefont {J.}~\bibnamefont
  {Hinch}}, \bibinfo {author} {\bibfnamefont {J.~R.}\ \bibnamefont {Lister}},\
  and\ \bibinfo {author} {\bibfnamefont {O.~a.}\ \bibnamefont {Basaran}},\
  }\bibfield  {title} {\bibinfo {title} {\emph {{Plethora of transitions during
  breakup of liquid filaments}}},\ }\href
  {https://doi.org/10.1073/pnas.1418541112} {\bibfield  {journal} {\bibinfo
  {journal} {\protect\JournalTitle{Proceedings of the National Academy of
  Sciences}}\ }\textbf {\bibinfo {volume} {112}},\ \bibinfo {pages} {4582}
  (\bibinfo {year} {2015})}\BibitemShut {NoStop}%
\bibitem [{\citenamefont {Lagarde}\ \emph {et~al.}(2018)\citenamefont
  {Lagarde}, \citenamefont {Josserand},\ and\ \citenamefont
  {Proti{\`{e}}re}}]{Lagarde2018}%
  \BibitemOpen
  \bibfield  {author} {\bibinfo {author} {\bibfnamefont {A.}~\bibnamefont
  {Lagarde}}, \bibinfo {author} {\bibfnamefont {C.}~\bibnamefont {Josserand}},\
  and\ \bibinfo {author} {\bibfnamefont {S.}~\bibnamefont {Proti{\`{e}}re}},\
  }\bibfield  {title} {\bibinfo {title} {\emph {{Oscillating path between
  self-similarities in liquid pinch-off}}},\ }\href
  {https://doi.org/10.1073/pnas.1814242115} {\bibfield  {journal} {\bibinfo
  {journal} {\protect\JournalTitle{Proceedings of the National Academy of
  Sciences}}\ }\textbf {\bibinfo {volume} {115}},\ \bibinfo {pages} {12371}
  (\bibinfo {year} {2018})}\BibitemShut {NoStop}%
\bibitem [{\citenamefont {Ruth}\ \emph {et~al.}(2019)\citenamefont {Ruth},
  \citenamefont {Mostert}, \citenamefont {Perrard},\ and\ \citenamefont
  {Deike}}]{Ruth2019}%
  \BibitemOpen
  \bibfield  {author} {\bibinfo {author} {\bibfnamefont {D.~J.}\ \bibnamefont
  {Ruth}}, \bibinfo {author} {\bibfnamefont {W.}~\bibnamefont {Mostert}},
  \bibinfo {author} {\bibfnamefont {S.}~\bibnamefont {Perrard}},\ and\ \bibinfo
  {author} {\bibfnamefont {L.}~\bibnamefont {Deike}},\ }\bibfield  {title}
  {\bibinfo {title} {\emph {{Bubble pinch-off in turbulence}}},\ }\href
  {https://doi.org/10.1073/pnas.1909842116} {\bibfield  {journal} {\bibinfo
  {journal} {\protect\JournalTitle{Proceedings of the National Academy of
  Sciences}}\ }\textbf {\bibinfo {volume} {116}},\ \bibinfo {pages} {25412}
  (\bibinfo {year} {2019})}\BibitemShut {NoStop}%
\bibitem [{\citenamefont {Pahlavan}\ \emph {et~al.}(2019)\citenamefont
  {Pahlavan}, \citenamefont {Stone}, \citenamefont {McKinley},\ and\
  \citenamefont {Juanes}}]{Pahlavan2019}%
  \BibitemOpen
  \bibfield  {author} {\bibinfo {author} {\bibfnamefont {A.~A.}\ \bibnamefont
  {Pahlavan}}, \bibinfo {author} {\bibfnamefont {H.~A.}\ \bibnamefont {Stone}},
  \bibinfo {author} {\bibfnamefont {G.~H.}\ \bibnamefont {McKinley}},\ and\
  \bibinfo {author} {\bibfnamefont {R.}~\bibnamefont {Juanes}},\ }\bibfield
  {title} {\bibinfo {title} {\emph {{Restoring universality to the pinch-off of
  a bubble}}},\ }\href {https://doi.org/10.1073/pnas.1819744116} {\bibfield
  {journal} {\bibinfo  {journal} {\protect\JournalTitle{Proceedings of the
  National Academy of Sciences of the United States of America}}\ }\textbf
  {\bibinfo {volume} {116}},\ \bibinfo {pages} {13780} (\bibinfo {year}
  {2019})}\BibitemShut {NoStop}%
\bibitem [{\citenamefont {Paulsen}\ \emph {et~al.}(2012)\citenamefont
  {Paulsen}, \citenamefont {Burton}, \citenamefont {Nagel}, \citenamefont
  {Appathurai}, \citenamefont {Harris},\ and\ \citenamefont
  {Basaran}}]{Paulsen2012}%
  \BibitemOpen
  \bibfield  {author} {\bibinfo {author} {\bibfnamefont {J.~D.}\ \bibnamefont
  {Paulsen}}, \bibinfo {author} {\bibfnamefont {J.~C.}\ \bibnamefont {Burton}},
  \bibinfo {author} {\bibfnamefont {S.~R.}\ \bibnamefont {Nagel}}, \bibinfo
  {author} {\bibfnamefont {S.}~\bibnamefont {Appathurai}}, \bibinfo {author}
  {\bibfnamefont {M.~T.}\ \bibnamefont {Harris}},\ and\ \bibinfo {author}
  {\bibfnamefont {O.~A.}\ \bibnamefont {Basaran}},\ }\bibfield  {title}
  {\bibinfo {title} {\emph {{The inexorable resistance of inertia determines
  the initial regime of drop coalescence}}},\ }\href
  {https://doi.org/10.1073/pnas.1120775109} {\bibfield  {journal} {\bibinfo
  {journal} {\protect\JournalTitle{Proceedings of the National Academy of
  Sciences of the United States of America}}\ }\textbf {\bibinfo {volume}
  {109}},\ \bibinfo {pages} {6857} (\bibinfo {year} {2012})}\BibitemShut
  {NoStop}%
\bibitem [{\citenamefont {Hack}\ \emph {et~al.}(2020)\citenamefont {Hack},
  \citenamefont {Tewes}, \citenamefont {Xie}, \citenamefont {Datt},
  \citenamefont {Harth}, \citenamefont {Harting},\ and\ \citenamefont
  {Snoeijer}}]{Hack2020}%
  \BibitemOpen
  \bibfield  {author} {\bibinfo {author} {\bibfnamefont {M.~A.}\ \bibnamefont
  {Hack}}, \bibinfo {author} {\bibfnamefont {W.}~\bibnamefont {Tewes}},
  \bibinfo {author} {\bibfnamefont {Q.}~\bibnamefont {Xie}}, \bibinfo {author}
  {\bibfnamefont {C.}~\bibnamefont {Datt}}, \bibinfo {author} {\bibfnamefont
  {K.}~\bibnamefont {Harth}}, \bibinfo {author} {\bibfnamefont
  {J.}~\bibnamefont {Harting}},\ and\ \bibinfo {author} {\bibfnamefont {J.~H.}\
  \bibnamefont {Snoeijer}},\ }\bibfield  {title} {\bibinfo {title} {\emph
  {{Self-Similar Liquid Lens Coalescence}}},\ }\href
  {https://doi.org/10.1103/PhysRevLett.124.194502} {\bibfield  {journal}
  {\bibinfo  {journal} {\protect\JournalTitle{Physical Review Letters}}\
  }\textbf {\bibinfo {volume} {124}},\ \bibinfo {pages} {194502} (\bibinfo
  {year} {2020})}\BibitemShut {NoStop}%
\bibitem [{\citenamefont {Hogrefe}\ \emph {et~al.}(1998)\citenamefont
  {Hogrefe}, \citenamefont {Peffley}, \citenamefont {Goodridge}, \citenamefont
  {Shi}, \citenamefont {Hentschel},\ and\ \citenamefont
  {Lathrop}}]{Hogrefe1998}%
  \BibitemOpen
  \bibfield  {author} {\bibinfo {author} {\bibfnamefont {J.~E.}\ \bibnamefont
  {Hogrefe}}, \bibinfo {author} {\bibfnamefont {N.~L.}\ \bibnamefont
  {Peffley}}, \bibinfo {author} {\bibfnamefont {C.~L.}\ \bibnamefont
  {Goodridge}}, \bibinfo {author} {\bibfnamefont {W.~T.}\ \bibnamefont {Shi}},
  \bibinfo {author} {\bibfnamefont {H.~E.}\ \bibnamefont {Hentschel}},\ and\
  \bibinfo {author} {\bibfnamefont {D.~P.}\ \bibnamefont {Lathrop}},\
  }\bibfield  {title} {\bibinfo {title} {\emph {{Power-law singularities in
  gravity-capillary waves}}},\ }\href
  {https://doi.org/10.1016/S0167-2789(98)00120-1} {\bibfield  {journal}
  {\bibinfo  {journal} {\protect\JournalTitle{Physica D: Nonlinear Phenomena}}\
  }\textbf {\bibinfo {volume} {123}},\ \bibinfo {pages} {183} (\bibinfo {year}
  {1998})}\BibitemShut {NoStop}%
\bibitem [{\citenamefont {Brenner}(2000)}]{Brenner2000}%
  \BibitemOpen
  \bibfield  {author} {\bibinfo {author} {\bibfnamefont {M.~P.}\ \bibnamefont
  {Brenner}},\ }\bibfield  {title} {\bibinfo {title} {\emph {{Jets from a
  singular surface}}},\ }\href {https://doi.org/10.1038/35000330} {\bibfield
  {journal} {\bibinfo  {journal} {\protect\JournalTitle{Nature}}\ }\textbf
  {\bibinfo {volume} {403}},\ \bibinfo {pages} {377} (\bibinfo {year}
  {2000})}\BibitemShut {NoStop}%
\bibitem [{\citenamefont {{Krishna Raja}}\ \emph {et~al.}(2019)\citenamefont
  {{Krishna Raja}}, \citenamefont {Das},\ and\ \citenamefont
  {Hopfinger}}]{KrishnaRaja2019}%
  \BibitemOpen
  \bibfield  {author} {\bibinfo {author} {\bibfnamefont {D.}~\bibnamefont
  {{Krishna Raja}}}, \bibinfo {author} {\bibfnamefont {S.~P.}\ \bibnamefont
  {Das}},\ and\ \bibinfo {author} {\bibfnamefont {E.~J.}\ \bibnamefont
  {Hopfinger}},\ }\bibfield  {title} {\bibinfo {title} {\emph {{On standing
  gravity wave-depression cavity collapse and jetting}}},\ }\href
  {https://doi.org/10.1017/jfm.2019.86} {\bibfield  {journal} {\bibinfo
  {journal} {\protect\JournalTitle{Journal of Fluid Mechanics}}\ }\textbf
  {\bibinfo {volume} {866}},\ \bibinfo {pages} {112} (\bibinfo {year}
  {2019})}\BibitemShut {NoStop}%
\bibitem [{\citenamefont {Basak}\ \emph {et~al.}(2021)\citenamefont {Basak},
  \citenamefont {Farsoiya},\ and\ \citenamefont {Dasgupta}}]{Basak2021}%
  \BibitemOpen
  \bibfield  {author} {\bibinfo {author} {\bibfnamefont {S.}~\bibnamefont
  {Basak}}, \bibinfo {author} {\bibfnamefont {P.~K.}\ \bibnamefont
  {Farsoiya}},\ and\ \bibinfo {author} {\bibfnamefont {R.}~\bibnamefont
  {Dasgupta}},\ }\bibfield  {title} {\bibinfo {title} {\emph {{Jetting in
  finite-amplitude, free, capillary-gravity waves}}},\ }\href
  {https://doi.org/10.1017/jfm.2020.851} {\bibfield  {journal} {\bibinfo
  {journal} {\protect\JournalTitle{Journal of Fluid Mechanics}}\ }\textbf
  {\bibinfo {volume} {909}},\ \bibinfo {pages} {A3} (\bibinfo {year}
  {2021})}\BibitemShut {NoStop}%
\bibitem [{\citenamefont {Longuet-Higgins}(1983)}]{Longuet-Higgins1983}%
  \BibitemOpen
  \bibfield  {author} {\bibinfo {author} {\bibfnamefont {M.~S.}\ \bibnamefont
  {Longuet-Higgins}},\ }\bibfield  {title} {\bibinfo {title} {\emph {{Bubbles,
  breaking waves and hyperbolic jets at a free surface}}},\ }\href
  {https://doi.org/10.1017/S0022112083002645} {\bibfield  {journal} {\bibinfo
  {journal} {\protect\JournalTitle{Journal of Fluid Mechanics}}\ }\textbf
  {\bibinfo {volume} {127}},\ \bibinfo {pages} {103} (\bibinfo {year}
  {1983})}\BibitemShut {NoStop}%
\bibitem [{\citenamefont {Thoroddsen}\ \emph {et~al.}(2009)\citenamefont
  {Thoroddsen}, \citenamefont {Takehara}, \citenamefont {Etoh},\ and\
  \citenamefont {Ohl}}]{Thoroddsen2009}%
  \BibitemOpen
  \bibfield  {author} {\bibinfo {author} {\bibfnamefont {S.~T.}\ \bibnamefont
  {Thoroddsen}}, \bibinfo {author} {\bibfnamefont {K.}~\bibnamefont
  {Takehara}}, \bibinfo {author} {\bibfnamefont {T.~G.}\ \bibnamefont {Etoh}},\
  and\ \bibinfo {author} {\bibfnamefont {C.-D.}\ \bibnamefont {Ohl}},\
  }\bibfield  {title} {\bibinfo {title} {\emph {{Spray and microjets produced
  by focusing a laser pulse into a hemispherical drop}}},\ }\href
  {https://doi.org/10.1063/1.3253394} {\bibfield  {journal} {\bibinfo
  {journal} {\protect\JournalTitle{Physics of Fluids}}\ }\textbf {\bibinfo
  {volume} {21}},\ \bibinfo {pages} {112101} (\bibinfo {year}
  {2009})}\BibitemShut {NoStop}%
\bibitem [{\citenamefont {Reuter}\ and\ \citenamefont
  {Ohl}(2021)}]{Reuter2021}%
  \BibitemOpen
  \bibfield  {author} {\bibinfo {author} {\bibfnamefont {F.}~\bibnamefont
  {Reuter}}\ and\ \bibinfo {author} {\bibfnamefont {C.-d.}\ \bibnamefont
  {Ohl}},\ }\bibfield  {title} {\bibinfo {title} {\emph {{Supersonic needle-jet
  generation with single cavitation bubbles}}},\ }\href
  {https://doi.org/10.1063/5.0045705} {\bibfield  {journal} {\bibinfo
  {journal} {\protect\JournalTitle{Applied Physics Letters}}\ }\textbf
  {\bibinfo {volume} {118}},\ \bibinfo {pages} {134103} (\bibinfo {year}
  {2021})}\BibitemShut {NoStop}%
\bibitem [{\citenamefont {Ismail}\ \emph {et~al.}(2018)\citenamefont {Ismail},
  \citenamefont {Ga{\~{n}}{\'{a}}n-Calvo}, \citenamefont
  {Castrej{\'{o}}n-Pita}, \citenamefont {Herrada},\ and\ \citenamefont
  {Castrej{\'{o}}n-Pita}}]{Ismail2018}%
  \BibitemOpen
  \bibfield  {author} {\bibinfo {author} {\bibfnamefont {A.~S.}\ \bibnamefont
  {Ismail}}, \bibinfo {author} {\bibfnamefont {A.~M.}\ \bibnamefont
  {Ga{\~{n}}{\'{a}}n-Calvo}}, \bibinfo {author} {\bibfnamefont {J.~R.}\
  \bibnamefont {Castrej{\'{o}}n-Pita}}, \bibinfo {author} {\bibfnamefont
  {M.~A.}\ \bibnamefont {Herrada}},\ and\ \bibinfo {author} {\bibfnamefont
  {A.~A.}\ \bibnamefont {Castrej{\'{o}}n-Pita}},\ }\bibfield  {title} {\bibinfo
  {title} {\emph {{Controlled cavity collapse: Scaling laws of drop
  formation}}},\ }\href {https://doi.org/10.1039/c8sm00114f} {\bibfield
  {journal} {\bibinfo  {journal} {\protect\JournalTitle{Soft Matter}}\ }\textbf
  {\bibinfo {volume} {14}},\ \bibinfo {pages} {7671} (\bibinfo {year}
  {2018})}\BibitemShut {NoStop}%
\bibitem [{\citenamefont {{Van Rijn}}\ \emph {et~al.}(2021)\citenamefont {{Van
  Rijn}}, \citenamefont {Westerweel}, \citenamefont {{Van Brummen}},
  \citenamefont {Antkowiak},\ and\ \citenamefont {Bonn}}]{VanRijn2021}%
  \BibitemOpen
  \bibfield  {author} {\bibinfo {author} {\bibfnamefont {C.~J.}\ \bibnamefont
  {{Van Rijn}}}, \bibinfo {author} {\bibfnamefont {J.}~\bibnamefont
  {Westerweel}}, \bibinfo {author} {\bibfnamefont {B.}~\bibnamefont {{Van
  Brummen}}}, \bibinfo {author} {\bibfnamefont {A.}~\bibnamefont {Antkowiak}},\
  and\ \bibinfo {author} {\bibfnamefont {D.}~\bibnamefont {Bonn}},\ }\bibfield
  {title} {\bibinfo {title} {\emph {{Self-similar jet evolution after drop
  impact on a liquid surface}}},\ }\href
  {https://doi.org/10.1103/PhysRevFluids.6.034801} {\bibfield  {journal}
  {\bibinfo  {journal} {\protect\JournalTitle{Physical Review Fluids}}\
  }\textbf {\bibinfo {volume} {6}},\ \bibinfo {pages} {1} (\bibinfo {year}
  {2021})}\BibitemShut {NoStop}%
\bibitem [{\citenamefont {Blanco–Rodr{\'{i}}guez}\ and\ \citenamefont
  {Gordillo}(2021)}]{BlancoRodriguez2021}%
  \BibitemOpen
  \bibfield  {author} {\bibinfo {author} {\bibfnamefont {F.~J.}\ \bibnamefont
  {Blanco–Rodr{\'{i}}guez}}\ and\ \bibinfo {author} {\bibfnamefont {J.~M.}\
  \bibnamefont {Gordillo}},\ }\bibfield  {title} {\bibinfo {title} {\emph {{On
  the jets produced by drops impacting a deep liquid pool and by bursting
  bubbles}}},\ }\href {https://doi.org/10.1017/jfm.2021.207} {\bibfield
  {journal} {\bibinfo  {journal} {\protect\JournalTitle{Journal of Fluid
  Mechanics}}\ }\textbf {\bibinfo {volume} {916}},\ \bibinfo {pages} {A37}
  (\bibinfo {year} {2021})}\BibitemShut {NoStop}%
\bibitem [{\citenamefont {Lohse}\ \emph {et~al.}(2004)\citenamefont {Lohse},
  \citenamefont {Bergmann}, \citenamefont {Mikkelsen}, \citenamefont
  {Zeilstra}, \citenamefont {{Van Der Meer}}, \citenamefont {Versluis},
  \citenamefont {{Van Der Weele}}, \citenamefont {{Van Der Hoef}},\ and\
  \citenamefont {Kuipers}}]{Lohse2004}%
  \BibitemOpen
  \bibfield  {author} {\bibinfo {author} {\bibfnamefont {D.}~\bibnamefont
  {Lohse}}, \bibinfo {author} {\bibfnamefont {R.}~\bibnamefont {Bergmann}},
  \bibinfo {author} {\bibfnamefont {R.}~\bibnamefont {Mikkelsen}}, \bibinfo
  {author} {\bibfnamefont {C.}~\bibnamefont {Zeilstra}}, \bibinfo {author}
  {\bibfnamefont {D.}~\bibnamefont {{Van Der Meer}}}, \bibinfo {author}
  {\bibfnamefont {M.}~\bibnamefont {Versluis}}, \bibinfo {author}
  {\bibfnamefont {K.}~\bibnamefont {{Van Der Weele}}}, \bibinfo {author}
  {\bibfnamefont {M.}~\bibnamefont {{Van Der Hoef}}},\ and\ \bibinfo {author}
  {\bibfnamefont {H.}~\bibnamefont {Kuipers}},\ }\bibfield  {title} {\bibinfo
  {title} {\emph {{Impact on soft sand: Void collapse and jet formation}}},\
  }\href {https://doi.org/10.1103/PhysRevLett.93.198003} {\bibfield  {journal}
  {\bibinfo  {journal} {\protect\JournalTitle{Physical Review Letters}}\
  }\textbf {\bibinfo {volume} {93}},\ \bibinfo {pages} {198003} (\bibinfo
  {year} {2004})}\BibitemShut {NoStop}%
\bibitem [{\citenamefont {Gekle}\ \emph {et~al.}(2009)\citenamefont {Gekle},
  \citenamefont {Gordillo}, \citenamefont {van~der Meer},\ and\ \citenamefont
  {Lohse}}]{Gekle2009}%
  \BibitemOpen
  \bibfield  {author} {\bibinfo {author} {\bibfnamefont {S.}~\bibnamefont
  {Gekle}}, \bibinfo {author} {\bibfnamefont {J.~M.}\ \bibnamefont {Gordillo}},
  \bibinfo {author} {\bibfnamefont {D.}~\bibnamefont {van~der Meer}},\ and\
  \bibinfo {author} {\bibfnamefont {D.}~\bibnamefont {Lohse}},\ }\bibfield
  {title} {\bibinfo {title} {\emph {{High-Speed Jet Formation after Solid
  Object Impact}}},\ }\href {https://doi.org/10.1103/PhysRevLett.102.034502}
  {\bibfield  {journal} {\bibinfo  {journal} {\protect\JournalTitle{Physical
  Review Letters}}\ }\textbf {\bibinfo {volume} {102}},\ \bibinfo {pages}
  {034502} (\bibinfo {year} {2009})}\BibitemShut {NoStop}%
\bibitem [{\citenamefont {Truscott}\ \emph {et~al.}(2014)\citenamefont
  {Truscott}, \citenamefont {Epps},\ and\ \citenamefont
  {Belden}}]{Truscott2014}%
  \BibitemOpen
  \bibfield  {author} {\bibinfo {author} {\bibfnamefont {T.~T.}\ \bibnamefont
  {Truscott}}, \bibinfo {author} {\bibfnamefont {B.~P.}\ \bibnamefont {Epps}},\
  and\ \bibinfo {author} {\bibfnamefont {J.}~\bibnamefont {Belden}},\
  }\bibfield  {title} {\bibinfo {title} {\emph {{Water Entry of
  Projectiles}}},\ }\href {https://doi.org/10.1146/annurev-fluid-011212-140753}
  {\bibfield  {journal} {\bibinfo  {journal} {\protect\JournalTitle{Annual
  Review of Fluid Mechanics}}\ }\textbf {\bibinfo {volume} {46}},\ \bibinfo
  {pages} {355} (\bibinfo {year} {2014})}\BibitemShut {NoStop}%
\bibitem [{\citenamefont {Lee}\ \emph {et~al.}(2011)\citenamefont {Lee},
  \citenamefont {Weon}, \citenamefont {Park}, \citenamefont {Je}, \citenamefont
  {Fezzaa},\ and\ \citenamefont {Lee}}]{Lee2011}%
  \BibitemOpen
  \bibfield  {author} {\bibinfo {author} {\bibfnamefont {J.~S.}\ \bibnamefont
  {Lee}}, \bibinfo {author} {\bibfnamefont {B.~M.}\ \bibnamefont {Weon}},
  \bibinfo {author} {\bibfnamefont {S.~J.}\ \bibnamefont {Park}}, \bibinfo
  {author} {\bibfnamefont {J.~H.}\ \bibnamefont {Je}}, \bibinfo {author}
  {\bibfnamefont {K.}~\bibnamefont {Fezzaa}},\ and\ \bibinfo {author}
  {\bibfnamefont {W.-K.}\ \bibnamefont {Lee}},\ }\bibfield  {title} {\bibinfo
  {title} {\emph {{Size limits the formation of liquid jets during bubble
  bursting}}},\ }\href {https://doi.org/10.1038/ncomms1369} {\bibfield
  {journal} {\bibinfo  {journal} {\protect\JournalTitle{Nature
  Communications}}\ }\textbf {\bibinfo {volume} {2}},\ \bibinfo {pages} {367}
  (\bibinfo {year} {2011})}\BibitemShut {NoStop}%
\bibitem [{\citenamefont {Ghabache}\ \emph {et~al.}(2014)\citenamefont
  {Ghabache}, \citenamefont {Antkowiak}, \citenamefont {Josserand},\ and\
  \citenamefont {S{\'{e}}on}}]{Ghabache2014}%
  \BibitemOpen
  \bibfield  {author} {\bibinfo {author} {\bibfnamefont {E.}~\bibnamefont
  {Ghabache}}, \bibinfo {author} {\bibfnamefont {A.}~\bibnamefont {Antkowiak}},
  \bibinfo {author} {\bibfnamefont {C.}~\bibnamefont {Josserand}},\ and\
  \bibinfo {author} {\bibfnamefont {T.}~\bibnamefont {S{\'{e}}on}},\ }\bibfield
   {title} {\bibinfo {title} {\emph {{On the physics of fizziness: How bubble
  bursting controls droplets ejection}}},\ }\bibfield  {journal} {\bibinfo
  {journal} {\protect\JournalTitle{Physics of Fluids}}\ }\textbf {\bibinfo
  {volume} {26}},\ \href {https://doi.org/10.1063/1.4902820}
  {10.1063/1.4902820} (\bibinfo {year} {2014})\BibitemShut {NoStop}%
\bibitem [{\citenamefont {Krishnan}\ \emph {et~al.}(2017)\citenamefont
  {Krishnan}, \citenamefont {Hopfinger},\ and\ \citenamefont
  {Puthenveettil}}]{Krishnan2017}%
  \BibitemOpen
  \bibfield  {author} {\bibinfo {author} {\bibfnamefont {S.}~\bibnamefont
  {Krishnan}}, \bibinfo {author} {\bibfnamefont {E.~J.}\ \bibnamefont
  {Hopfinger}},\ and\ \bibinfo {author} {\bibfnamefont {B.~A.}\ \bibnamefont
  {Puthenveettil}},\ }\bibfield  {title} {\bibinfo {title} {\emph {{On the
  scaling of jetting from bubble collapse at a liquid surface}}},\ }\href
  {https://doi.org/10.1017/jfm.2017.214} {\bibfield  {journal} {\bibinfo
  {journal} {\protect\JournalTitle{Journal of Fluid Mechanics}}\ }\textbf
  {\bibinfo {volume} {822}},\ \bibinfo {pages} {791} (\bibinfo {year}
  {2017})}\BibitemShut {NoStop}%
\bibitem [{\citenamefont {Ga{\~{n}}{\'{a}}n-Calvo}(2017)}]{Ganan-Calvo2017}%
  \BibitemOpen
  \bibfield  {author} {\bibinfo {author} {\bibfnamefont {A.~M.}\ \bibnamefont
  {Ga{\~{n}}{\'{a}}n-Calvo}},\ }\bibfield  {title} {\bibinfo {title} {\emph
  {{Revision of Bubble Bursting: Universal Scaling Laws of Top Jet Drop Size
  and Speed}}},\ }\href {https://doi.org/10.1103/PhysRevLett.119.204502}
  {\bibfield  {journal} {\bibinfo  {journal} {\protect\JournalTitle{Physical
  Review Letters}}\ }\textbf {\bibinfo {volume} {119}},\ \bibinfo {pages}
  {204502} (\bibinfo {year} {2017})}\BibitemShut {NoStop}%
\bibitem [{\citenamefont {Ga{\~{n}}{\'{a}}n-Calvo}(2018)}]{Ganan-Calvo2018}%
  \BibitemOpen
  \bibfield  {author} {\bibinfo {author} {\bibfnamefont {A.~M.}\ \bibnamefont
  {Ga{\~{n}}{\'{a}}n-Calvo}},\ }\bibfield  {title} {\bibinfo {title} {\emph
  {{Scaling laws of top jet drop size and speed from bubble bursting including
  gravity and inviscid limit}}},\ }\href
  {https://doi.org/10.1103/PhysRevFluids.3.091601} {\bibfield  {journal}
  {\bibinfo  {journal} {\protect\JournalTitle{Physical Review Fluids}}\
  }\textbf {\bibinfo {volume} {3}},\ \bibinfo {pages} {091601(R)} (\bibinfo
  {year} {2018})}\BibitemShut {NoStop}%
\bibitem [{\citenamefont {Lai}\ \emph {et~al.}(2018)\citenamefont {Lai},
  \citenamefont {Eggers},\ and\ \citenamefont {Deike}}]{Lai2018}%
  \BibitemOpen
  \bibfield  {author} {\bibinfo {author} {\bibfnamefont {C.-Y.}\ \bibnamefont
  {Lai}}, \bibinfo {author} {\bibfnamefont {J.}~\bibnamefont {Eggers}},\ and\
  \bibinfo {author} {\bibfnamefont {L.}~\bibnamefont {Deike}},\ }\bibfield
  {title} {\bibinfo {title} {\emph {{Bubble Bursting: Universal Cavity and Jet
  Profiles}}},\ }\href {https://doi.org/10.1103/PhysRevLett.121.144501}
  {\bibfield  {journal} {\bibinfo  {journal} {\protect\JournalTitle{Physical
  Review Letters}}\ }\textbf {\bibinfo {volume} {121}},\ \bibinfo {pages}
  {144501} (\bibinfo {year} {2018})}\BibitemShut {NoStop}%
\bibitem [{\citenamefont {Brasz}\ \emph {et~al.}(2018)\citenamefont {Brasz},
  \citenamefont {Bartlett}, \citenamefont {Walls}, \citenamefont {Flynn},
  \citenamefont {Yu},\ and\ \citenamefont {Bird}}]{Brasz2018}%
  \BibitemOpen
  \bibfield  {author} {\bibinfo {author} {\bibfnamefont {C.~F.}\ \bibnamefont
  {Brasz}}, \bibinfo {author} {\bibfnamefont {C.~T.}\ \bibnamefont {Bartlett}},
  \bibinfo {author} {\bibfnamefont {P.~L.~L.}\ \bibnamefont {Walls}}, \bibinfo
  {author} {\bibfnamefont {E.~G.}\ \bibnamefont {Flynn}}, \bibinfo {author}
  {\bibfnamefont {Y.~E.}\ \bibnamefont {Yu}},\ and\ \bibinfo {author}
  {\bibfnamefont {J.~C.}\ \bibnamefont {Bird}},\ }\bibfield  {title} {\bibinfo
  {title} {\emph {{Minimum size for the top jet drop from a bursting
  bubble}}},\ }\href {https://doi.org/10.1103/PhysRevFluids.3.074001}
  {\bibfield  {journal} {\bibinfo  {journal} {\protect\JournalTitle{Physical
  Review Fluids}}\ }\textbf {\bibinfo {volume} {3}},\ \bibinfo {pages} {074001}
  (\bibinfo {year} {2018})}\BibitemShut {NoStop}%
\bibitem [{\citenamefont {Gordillo}\ and\ \citenamefont
  {Rodr{\'{i}}guez-Rodr{\'{i}}guez}(2019)}]{Gordillo2019}%
  \BibitemOpen
  \bibfield  {author} {\bibinfo {author} {\bibfnamefont {J.~M.}\ \bibnamefont
  {Gordillo}}\ and\ \bibinfo {author} {\bibfnamefont {J.}~\bibnamefont
  {Rodr{\'{i}}guez-Rodr{\'{i}}guez}},\ }\bibfield  {title} {\bibinfo {title}
  {\emph {{Capillary waves control the ejection of bubble bursting jets}}},\
  }\href {https://doi.org/10.1017/jfm.2019.161} {\bibfield  {journal} {\bibinfo
   {journal} {\protect\JournalTitle{Journal of Fluid Mechanics}}\ }\textbf
  {\bibinfo {volume} {867}},\ \bibinfo {pages} {556} (\bibinfo {year}
  {2019})}\BibitemShut {NoStop}%
\bibitem [{\citenamefont {Blanco–Rodr{\'{i}}guez}\ and\ \citenamefont
  {Gordillo}(2020)}]{BlancoRodriguez2020}%
  \BibitemOpen
  \bibfield  {author} {\bibinfo {author} {\bibfnamefont {F.~J.}\ \bibnamefont
  {Blanco–Rodr{\'{i}}guez}}\ and\ \bibinfo {author} {\bibfnamefont {J.~M.}\
  \bibnamefont {Gordillo}},\ }\bibfield  {title} {\bibinfo {title} {\emph {{On
  the sea spray aerosol originated from bubble bursting jets}}},\ }\href
  {https://doi.org/10.1017/jfm.2019.1061} {\bibfield  {journal} {\bibinfo
  {journal} {\protect\JournalTitle{Journal of Fluid Mechanics}}\ }\textbf
  {\bibinfo {volume} {886}},\ \bibinfo {pages} {R2} (\bibinfo {year}
  {2020})}\BibitemShut {NoStop}%
\bibitem [{\citenamefont {Ga{\~{n}}{\'{a}}n-Calvo}\ and\ \citenamefont
  {L{\'{o}}pez-Herrera}(2021)}]{Ganan-Calvo2021}%
  \BibitemOpen
  \bibfield  {author} {\bibinfo {author} {\bibfnamefont {A.~M.}\ \bibnamefont
  {Ga{\~{n}}{\'{a}}n-Calvo}}\ and\ \bibinfo {author} {\bibfnamefont {J.~M.}\
  \bibnamefont {L{\'{o}}pez-Herrera}},\ }\bibfield  {title} {\bibinfo {title}
  {\emph {{On the physics of transient ejection from bubble bursting}}},\
  }\href {https://doi.org/10.1017/jfm.2021.791} {\bibfield  {journal} {\bibinfo
   {journal} {\protect\JournalTitle{Journal of Fluid Mechanics}}\ }\textbf
  {\bibinfo {volume} {929}},\ \bibinfo {pages} {A12} (\bibinfo {year}
  {2021})}\BibitemShut {NoStop}%
\bibitem [{\citenamefont {Dhuper}\ \emph {et~al.}(2021)\citenamefont {Dhuper},
  \citenamefont {Guleria},\ and\ \citenamefont {Kumar}}]{Dhuper2021}%
  \BibitemOpen
  \bibfield  {author} {\bibinfo {author} {\bibfnamefont {K.}~\bibnamefont
  {Dhuper}}, \bibinfo {author} {\bibfnamefont {S.~D.}\ \bibnamefont
  {Guleria}},\ and\ \bibinfo {author} {\bibfnamefont {P.}~\bibnamefont
  {Kumar}},\ }\bibfield  {title} {\bibinfo {title} {\emph {{Interface dynamics
  at the impact of a drop onto a deep pool of immiscible liquid}}},\ }\href
  {https://doi.org/10.1016/j.ces.2021.116541} {\bibfield  {journal} {\bibinfo
  {journal} {\protect\JournalTitle{Chemical Engineering Science}}\ }\textbf
  {\bibinfo {volume} {237}},\ \bibinfo {pages} {116541} (\bibinfo {year}
  {2021})}\BibitemShut {NoStop}%
\bibitem [{\citenamefont {Thoroddsen}\ \emph {et~al.}(2005)\citenamefont
  {Thoroddsen}, \citenamefont {Etoh}, \citenamefont {Takehara}, \citenamefont
  {Ootsuka},\ and\ \citenamefont {Hatsuki}}]{Thoroddsen2005}%
  \BibitemOpen
  \bibfield  {author} {\bibinfo {author} {\bibfnamefont {S.~T.}\ \bibnamefont
  {Thoroddsen}}, \bibinfo {author} {\bibfnamefont {T.~G.}\ \bibnamefont
  {Etoh}}, \bibinfo {author} {\bibfnamefont {K.}~\bibnamefont {Takehara}},
  \bibinfo {author} {\bibfnamefont {N.}~\bibnamefont {Ootsuka}},\ and\ \bibinfo
  {author} {\bibfnamefont {Y.}~\bibnamefont {Hatsuki}},\ }\bibfield  {title}
  {\bibinfo {title} {\emph {{The air bubble entrapped under a drop impacting on
  a solid surface}}},\ }\href {https://doi.org/10.1017/S0022112005006919}
  {\bibfield  {journal} {\bibinfo  {journal} {\protect\JournalTitle{Journal of
  Fluid Mechanics}}\ }\textbf {\bibinfo {volume} {545}},\ \bibinfo {pages}
  {203} (\bibinfo {year} {2005})}\BibitemShut {NoStop}%
\bibitem [{\citenamefont {Li}\ and\ \citenamefont {Thoroddsen}(2015)}]{Li2015}%
  \BibitemOpen
  \bibfield  {author} {\bibinfo {author} {\bibfnamefont {E.~Q.}\ \bibnamefont
  {Li}}\ and\ \bibinfo {author} {\bibfnamefont {S.~T.}\ \bibnamefont
  {Thoroddsen}},\ }\bibfield  {title} {\bibinfo {title} {\emph {{Time-resolved
  imaging of a compressible air disc under a drop impacting on a solid
  surface}}},\ }\href {https://doi.org/10.1017/jfm.2015.466} {\bibfield
  {journal} {\bibinfo  {journal} {\protect\JournalTitle{Journal of Fluid
  Mechanics}}\ }\textbf {\bibinfo {volume} {780}},\ \bibinfo {pages} {636}
  (\bibinfo {year} {2015})}\BibitemShut {NoStop}%
\bibitem [{\citenamefont {Ghabache}\ \emph {et~al.}(2016)\citenamefont
  {Ghabache}, \citenamefont {Liger-Belair}, \citenamefont {Antkowiak},\ and\
  \citenamefont {S{\'{e}}on}}]{Ghabache2016}%
  \BibitemOpen
  \bibfield  {author} {\bibinfo {author} {\bibfnamefont {E.}~\bibnamefont
  {Ghabache}}, \bibinfo {author} {\bibfnamefont {G.}~\bibnamefont
  {Liger-Belair}}, \bibinfo {author} {\bibfnamefont {A.}~\bibnamefont
  {Antkowiak}},\ and\ \bibinfo {author} {\bibfnamefont {T.}~\bibnamefont
  {S{\'{e}}on}},\ }\bibfield  {title} {\bibinfo {title} {\emph {{Evaporation of
  droplets in a Champagne wine aerosol}}},\ }\href
  {https://doi.org/10.1038/srep25148} {\bibfield  {journal} {\bibinfo
  {journal} {\protect\JournalTitle{Scientific Reports}}\ }\textbf {\bibinfo
  {volume} {6}},\ \bibinfo {pages} {25148} (\bibinfo {year}
  {2016})}\BibitemShut {NoStop}%
\bibitem [{\citenamefont {Spiel}(1995)}]{Spiel1995}%
  \BibitemOpen
  \bibfield  {author} {\bibinfo {author} {\bibfnamefont {D.~E.}\ \bibnamefont
  {Spiel}},\ }\bibfield  {title} {\bibinfo {title} {\emph {{On the births of
  jet drops from bubbles bursting on water surfaces}}},\ }\href
  {https://doi.org/10.1029/94JC03055} {\bibfield  {journal} {\bibinfo
  {journal} {\protect\JournalTitle{Journal of Geophysical Research}}\ }\textbf
  {\bibinfo {volume} {100}},\ \bibinfo {pages} {4995} (\bibinfo {year}
  {1995})}\BibitemShut {NoStop}%
\bibitem [{\citenamefont {Jain}\ \emph {et~al.}(2019)\citenamefont {Jain},
  \citenamefont {Jalaal}, \citenamefont {Lohse},\ and\ \citenamefont {van~der
  Meer}}]{Jain2019}%
  \BibitemOpen
  \bibfield  {author} {\bibinfo {author} {\bibfnamefont {U.}~\bibnamefont
  {Jain}}, \bibinfo {author} {\bibfnamefont {M.}~\bibnamefont {Jalaal}},
  \bibinfo {author} {\bibfnamefont {D.}~\bibnamefont {Lohse}},\ and\ \bibinfo
  {author} {\bibfnamefont {D.}~\bibnamefont {van~der Meer}},\ }\bibfield
  {title} {\bibinfo {title} {\emph {{Deep pool water-impacts of viscous oil
  droplets}}},\ }\href {https://doi.org/10.1039/C9SM00318E} {\bibfield
  {journal} {\bibinfo  {journal} {\protect\JournalTitle{Soft Matter}}\ }\textbf
  {\bibinfo {volume} {15}},\ \bibinfo {pages} {4629} (\bibinfo {year}
  {2019})}\BibitemShut {NoStop}%
\bibitem [{\citenamefont {Popinet}(2021)}]{Popinet2021}%
  \BibitemOpen
  \bibfield  {author} {\bibinfo {author} {\bibfnamefont {S.}~\bibnamefont
  {Popinet}},\ }\bibfield  {title} {\bibinfo {title} {\emph {{Basilisk flow
  solver}}},\ }\href {http://basilisk.fr/} {\bibfield  {journal} {\bibinfo
  {journal} {\protect\JournalTitle{http://basilisk.fr/}}\ } (\bibinfo {year}
  {2021})}\BibitemShut {NoStop}%
\bibitem [{\citenamefont {Popinet}(2003)}]{Popinet2003}%
  \BibitemOpen
  \bibfield  {author} {\bibinfo {author} {\bibfnamefont {S.}~\bibnamefont
  {Popinet}},\ }\bibfield  {title} {\bibinfo {title} {\emph {{Gerris: a
  tree-based adaptive solver for the incompressible Euler equations in complex
  geometries}}},\ }\href {https://doi.org/10.1016/S0021-9991(03)00298-5}
  {\bibfield  {journal} {\bibinfo  {journal} {\protect\JournalTitle{Journal of
  Computational Physics}}\ }\textbf {\bibinfo {volume} {190}},\ \bibinfo
  {pages} {572} (\bibinfo {year} {2003})}\BibitemShut {NoStop}%
\bibitem [{\citenamefont {Popinet}(2009)}]{Popinet2009}%
  \BibitemOpen
  \bibfield  {author} {\bibinfo {author} {\bibfnamefont {S.}~\bibnamefont
  {Popinet}},\ }\bibfield  {title} {\bibinfo {title} {\emph {{An accurate
  adaptive solver for surface-tension-driven interfacial flows}}},\ }\href
  {https://doi.org/10.1016/j.jcp.2009.04.042} {\bibfield  {journal} {\bibinfo
  {journal} {\protect\JournalTitle{Journal of Computational Physics}}\ }\textbf
  {\bibinfo {volume} {228}},\ \bibinfo {pages} {5838} (\bibinfo {year}
  {2009})}\BibitemShut {NoStop}%
\bibitem [{\citenamefont {Popinet}(2018)}]{Popinet2018}%
  \BibitemOpen
  \bibfield  {author} {\bibinfo {author} {\bibfnamefont {S.}~\bibnamefont
  {Popinet}},\ }\bibfield  {title} {\bibinfo {title} {\emph {{Numerical Models
  of Surface Tension}}},\ }\href
  {https://doi.org/10.1146/annurev-fluid-122316-045034} {\bibfield  {journal}
  {\bibinfo  {journal} {\protect\JournalTitle{Annual Review of Fluid
  Mechanics}}\ }\textbf {\bibinfo {volume} {50}},\ \bibinfo {pages} {49}
  (\bibinfo {year} {2018})}\BibitemShut {NoStop}%
\bibitem [{\citenamefont {Plesset}\ and\ \citenamefont
  {Prosperetti}(1977)}]{Plesset1977}%
  \BibitemOpen
  \bibfield  {author} {\bibinfo {author} {\bibfnamefont {M.~S.}\ \bibnamefont
  {Plesset}}\ and\ \bibinfo {author} {\bibfnamefont {A.}~\bibnamefont
  {Prosperetti}},\ }\bibfield  {title} {\bibinfo {title} {\emph {{Bubble
  Dynamics and Cavitation}}},\ }\href
  {https://doi.org/10.1146/annurev.fl.09.010177.001045} {\bibfield  {journal}
  {\bibinfo  {journal} {\protect\JournalTitle{Annual Review of Fluid
  Mechanics}}\ }\textbf {\bibinfo {volume} {9}},\ \bibinfo {pages} {145}
  (\bibinfo {year} {1977})}\BibitemShut {NoStop}%
\bibitem [{\citenamefont {Thoroddsen}\ \emph {et~al.}(2007)\citenamefont
  {Thoroddsen}, \citenamefont {Etoh},\ and\ \citenamefont
  {Takehara}}]{Thoroddsen2007}%
  \BibitemOpen
  \bibfield  {author} {\bibinfo {author} {\bibfnamefont {S.~T.}\ \bibnamefont
  {Thoroddsen}}, \bibinfo {author} {\bibfnamefont {T.~G.}\ \bibnamefont
  {Etoh}},\ and\ \bibinfo {author} {\bibfnamefont {K.}~\bibnamefont
  {Takehara}},\ }\bibfield  {title} {\bibinfo {title} {\emph {{Experiments on
  bubble pinch-off}}},\ }\href {https://doi.org/10.1063/1.2710269} {\bibfield
  {journal} {\bibinfo  {journal} {\protect\JournalTitle{Physics of Fluids}}\
  }\textbf {\bibinfo {volume} {19}},\ \bibinfo {pages} {042101} (\bibinfo
  {year} {2007})}\BibitemShut {NoStop}%
\bibitem [{\citenamefont {Eggers}\ \emph {et~al.}(2007)\citenamefont {Eggers},
  \citenamefont {Fontelos}, \citenamefont {Leppinen},\ and\ \citenamefont
  {Snoeijer}}]{Eggers2007}%
  \BibitemOpen
  \bibfield  {author} {\bibinfo {author} {\bibfnamefont {J.}~\bibnamefont
  {Eggers}}, \bibinfo {author} {\bibfnamefont {M.~A.}\ \bibnamefont
  {Fontelos}}, \bibinfo {author} {\bibfnamefont {D.}~\bibnamefont {Leppinen}},\
  and\ \bibinfo {author} {\bibfnamefont {J.~H.}\ \bibnamefont {Snoeijer}},\
  }\bibfield  {title} {\bibinfo {title} {\emph {{Theory of the collapsing
  axisymmetric cavity}}},\ }\href
  {https://doi.org/10.1103/PhysRevLett.98.094502} {\bibfield  {journal}
  {\bibinfo  {journal} {\protect\JournalTitle{Physical Review Letters}}\
  }\textbf {\bibinfo {volume} {98}},\ \bibinfo {pages} {094502} (\bibinfo
  {year} {2007})}\BibitemShut {NoStop}%
\bibitem [{\citenamefont {Gordillo}\ and\ \citenamefont
  {Fontelos}(2007)}]{Gordillo2007}%
  \BibitemOpen
  \bibfield  {author} {\bibinfo {author} {\bibfnamefont {J.~M.}\ \bibnamefont
  {Gordillo}}\ and\ \bibinfo {author} {\bibfnamefont {M.~A.}\ \bibnamefont
  {Fontelos}},\ }\bibfield  {title} {\bibinfo {title} {\emph {{Satellites in
  the Inviscid Breakup of Bubbles}}},\ }\href
  {https://doi.org/10.1103/PhysRevLett.98.144503} {\bibfield  {journal}
  {\bibinfo  {journal} {\protect\JournalTitle{Physical Review Letters}}\
  }\textbf {\bibinfo {volume} {98}},\ \bibinfo {pages} {144503} (\bibinfo
  {year} {2007})}\BibitemShut {NoStop}%
\bibitem [{\citenamefont {Keller}\ and\ \citenamefont
  {Miksis}(1983)}]{Keller1983}%
  \BibitemOpen
  \bibfield  {author} {\bibinfo {author} {\bibfnamefont {J.~B.}\ \bibnamefont
  {Keller}}\ and\ \bibinfo {author} {\bibfnamefont {M.~J.}\ \bibnamefont
  {Miksis}},\ }\bibfield  {title} {\bibinfo {title} {\emph {{Surface Tension
  Driven Flows}}},\ }\href {https://doi.org/10.1137/0143018} {\bibfield
  {journal} {\bibinfo  {journal} {\protect\JournalTitle{SIAM Journal on Applied
  Mathematics}}\ }\textbf {\bibinfo {volume} {43}},\ \bibinfo {pages} {268}
  (\bibinfo {year} {1983})}\BibitemShut {NoStop}%
\bibitem [{\citenamefont {Ram{\'{i}}rez-Soto}\ \emph
  {et~al.}(2020)\citenamefont {Ram{\'{i}}rez-Soto}, \citenamefont {Sanjay},
  \citenamefont {Lohse}, \citenamefont {Pham},\ and\ \citenamefont
  {Vollmer}}]{Ramirez-Soto2020}%
  \BibitemOpen
  \bibfield  {author} {\bibinfo {author} {\bibfnamefont {O.}~\bibnamefont
  {Ram{\'{i}}rez-Soto}}, \bibinfo {author} {\bibfnamefont {V.}~\bibnamefont
  {Sanjay}}, \bibinfo {author} {\bibfnamefont {D.}~\bibnamefont {Lohse}},
  \bibinfo {author} {\bibfnamefont {J.~T.}\ \bibnamefont {Pham}},\ and\
  \bibinfo {author} {\bibfnamefont {D.}~\bibnamefont {Vollmer}},\ }\bibfield
  {title} {\bibinfo {title} {\emph {{Lifting a sessile oil drop from a
  superamphiphobic surface with an impacting one}}},\ }\href
  {https://doi.org/10.1126/sciadv.aba4330} {\bibfield  {journal} {\bibinfo
  {journal} {\protect\JournalTitle{Science Advances}}\ }\textbf {\bibinfo
  {volume} {6}},\ \bibinfo {pages} {eaba4330} (\bibinfo {year}
  {2020})}\BibitemShut {NoStop}%
\bibitem [{\citenamefont {Wei}\ and\ \citenamefont {Thoraval}(2021)}]{Wei2021}%
  \BibitemOpen
  \bibfield  {author} {\bibinfo {author} {\bibfnamefont {Y.}~\bibnamefont
  {Wei}}\ and\ \bibinfo {author} {\bibfnamefont {M.-J.}\ \bibnamefont
  {Thoraval}},\ }\bibfield  {title} {\bibinfo {title} {\emph {{Maximum
  spreading of an impacting air-in-liquid compound drop}}},\ }\href
  {https://doi.org/10.1063/5.0053384} {\bibfield  {journal} {\bibinfo
  {journal} {\protect\JournalTitle{Physics of Fluids}}\ }\textbf {\bibinfo
  {volume} {33}},\ \bibinfo {pages} {061703} (\bibinfo {year}
  {2021})}\BibitemShut {NoStop}%
\end{thebibliography}%


\begin{thebibliography}{6}%
\makeatletter
\providecommand \@ifxundefined [1]{%
 \@ifx{#1\undefined}
}%
\providecommand \@ifnum [1]{%
 \ifnum #1\expandafter \@firstoftwo
 \else \expandafter \@secondoftwo
 \fi
}%
\providecommand \@ifx [1]{%
 \ifx #1\expandafter \@firstoftwo
 \else \expandafter \@secondoftwo
 \fi
}%
\providecommand \natexlab [1]{#1}%
\providecommand \enquote  [1]{``#1''}%
\providecommand \bibnamefont  [1]{#1}%
\providecommand \bibfnamefont [1]{#1}%
\providecommand \citenamefont [1]{#1}%
\providecommand \href@noop [0]{\@secondoftwo}%
\providecommand \href [0]{\begingroup \@sanitize@url \@href}%
\providecommand \@href[1]{\@@startlink{#1}\@@href}%
\providecommand \@@href[1]{\endgroup#1\@@endlink}%
\providecommand \@sanitize@url [0]{\catcode `\\12\catcode `\$12\catcode
  `\&12\catcode `\#12\catcode `\^12\catcode `\_12\catcode `\%12\relax}%
\providecommand \@@startlink[1]{}%
\providecommand \@@endlink[0]{}%
\providecommand \url  [0]{\begingroup\@sanitize@url \@url }%
\providecommand \@url [1]{\endgroup\@href {#1}{\urlprefix }}%
\providecommand \urlprefix  [0]{URL }%
\providecommand \Eprint [0]{\href }%
\providecommand \doibase [0]{https://doi.org/}%
\providecommand \selectlanguage [0]{\@gobble}%
\providecommand \bibinfo  [0]{\@secondoftwo}%
\providecommand \bibfield  [0]{\@secondoftwo}%
\providecommand \translation [1]{[#1]}%
\providecommand \BibitemOpen [0]{}%
\providecommand \bibitemStop [0]{}%
\providecommand \bibitemNoStop [0]{.\EOS\space}%
\providecommand \EOS [0]{\spacefactor3000\relax}%
\providecommand \BibitemShut  [1]{\csname bibitem#1\endcsname}%
\let\auto@bib@innerbib\@empty
\bibitem [{\citenamefont {Blanken}\ \emph {et~al.}(2020)\citenamefont
  {Blanken}, \citenamefont {Saleem}, \citenamefont {Antonini},\ and\
  \citenamefont {Thoraval}}]{Blanken2020}%
  \BibitemOpen
  \bibfield  {author} {\bibinfo {author} {\bibfnamefont {N.}~\bibnamefont
  {Blanken}}, \bibinfo {author} {\bibfnamefont {M.~S.}\ \bibnamefont {Saleem}},
  \bibinfo {author} {\bibfnamefont {C.}~\bibnamefont {Antonini}},\ and\
  \bibinfo {author} {\bibfnamefont {M.-J.}\ \bibnamefont {Thoraval}},\
  }\bibfield  {title} {\bibinfo {title} {\emph {{Rebound of self-lubricating
  compound drops}}},\ }\href {https://doi.org/10.1126/sciadv.aay3499}
  {\bibfield  {journal} {\bibinfo  {journal} {\protect\JournalTitle{Science
  Advances}}\ }\textbf {\bibinfo {volume} {6}},\ \bibinfo {pages} {eaay3499}
  (\bibinfo {year} {2020})}\BibitemShut {NoStop}%
\bibitem [{\citenamefont {Bartolo}\ \emph {et~al.}(2006)\citenamefont
  {Bartolo}, \citenamefont {Josserand},\ and\ \citenamefont
  {Bonn}}]{Bartolo2006}%
  \BibitemOpen
  \bibfield  {author} {\bibinfo {author} {\bibfnamefont {D.}~\bibnamefont
  {Bartolo}}, \bibinfo {author} {\bibfnamefont {C.}~\bibnamefont {Josserand}},\
  and\ \bibinfo {author} {\bibfnamefont {D.}~\bibnamefont {Bonn}},\ }\bibfield
  {title} {\bibinfo {title} {\emph {{Singular Jets and Bubbles in Drop
  Impact}}},\ }\href {https://doi.org/10.1103/PhysRevLett.96.124501} {\bibfield
   {journal} {\bibinfo  {journal} {\protect\JournalTitle{Physical Review
  Letters}}\ }\textbf {\bibinfo {volume} {96}},\ \bibinfo {pages} {124501}
  (\bibinfo {year} {2006})}\BibitemShut {NoStop}%
\bibitem [{\citenamefont {Wildeman}\ \emph {et~al.}(2016)\citenamefont
  {Wildeman}, \citenamefont {Visser}, \citenamefont {Sun},\ and\ \citenamefont
  {Lohse}}]{Wildeman2016}%
  \BibitemOpen
  \bibfield  {author} {\bibinfo {author} {\bibfnamefont {S.}~\bibnamefont
  {Wildeman}}, \bibinfo {author} {\bibfnamefont {C.~W.}\ \bibnamefont
  {Visser}}, \bibinfo {author} {\bibfnamefont {C.}~\bibnamefont {Sun}},\ and\
  \bibinfo {author} {\bibfnamefont {D.}~\bibnamefont {Lohse}},\ }\bibfield
  {title} {\bibinfo {title} {\emph {{On the spreading of impacting drops}}},\
  }\href {https://doi.org/10.1017/jfm.2016.584} {\bibfield  {journal} {\bibinfo
   {journal} {\protect\JournalTitle{Journal of Fluid Mechanics}}\ }\textbf
  {\bibinfo {volume} {805}},\ \bibinfo {pages} {636} (\bibinfo {year}
  {2016})}\BibitemShut {NoStop}%
\bibitem [{\citenamefont {Thoroddsen}\ \emph {et~al.}(2018)\citenamefont
  {Thoroddsen}, \citenamefont {Takehara}, \citenamefont {Nguyen},\ and\
  \citenamefont {Etoh}}]{Thoroddsen2018}%
  \BibitemOpen
  \bibfield  {author} {\bibinfo {author} {\bibfnamefont {S.~T.}\ \bibnamefont
  {Thoroddsen}}, \bibinfo {author} {\bibfnamefont {K.}~\bibnamefont
  {Takehara}}, \bibinfo {author} {\bibfnamefont {H.~D.}\ \bibnamefont
  {Nguyen}},\ and\ \bibinfo {author} {\bibfnamefont {T.~G.}\ \bibnamefont
  {Etoh}},\ }\bibfield  {title} {\bibinfo {title} {\emph {{Singular jets during
  the collapse of drop-impact craters}}},\ }\href
  {https://doi.org/10.1017/jfm.2018.435} {\bibfield  {journal} {\bibinfo
  {journal} {\protect\JournalTitle{Journal of Fluid Mechanics}}\ }\textbf
  {\bibinfo {volume} {848}},\ \bibinfo {pages} {R3} (\bibinfo {year}
  {2018})}\BibitemShut {NoStop}%
\bibitem [{\citenamefont {Yang}\ \emph {et~al.}(2020)\citenamefont {Yang},
  \citenamefont {Tian},\ and\ \citenamefont {Thoroddsen}}]{Yang2020}%
  \BibitemOpen
  \bibfield  {author} {\bibinfo {author} {\bibfnamefont {Z.~Q.}\ \bibnamefont
  {Yang}}, \bibinfo {author} {\bibfnamefont {Y.~S.}\ \bibnamefont {Tian}},\
  and\ \bibinfo {author} {\bibfnamefont {S.~T.}\ \bibnamefont {Thoroddsen}},\
  }\bibfield  {title} {\bibinfo {title} {\emph {{Multitude of dimple shapes can
  produce singular jets during the collapse of immiscible drop-impact
  craters}}},\ }\href {https://doi.org/10.1017/jfm.2020.694} {\bibfield
  {journal} {\bibinfo  {journal} {\protect\JournalTitle{Journal of Fluid
  Mechanics}}\ }\textbf {\bibinfo {volume} {904}},\ \bibinfo {pages} {A19}
  (\bibinfo {year} {2020})}\BibitemShut {NoStop}%
\bibitem [{\citenamefont {Blanco–Rodr{\'{i}}guez}\ and\ \citenamefont
  {Gordillo}(2021)}]{BlancoRodriguez2021}%
  \BibitemOpen
  \bibfield  {author} {\bibinfo {author} {\bibfnamefont {F.~J.}\ \bibnamefont
  {Blanco–Rodr{\'{i}}guez}}\ and\ \bibinfo {author} {\bibfnamefont {J.~M.}\
  \bibnamefont {Gordillo}},\ }\bibfield  {title} {\bibinfo {title} {\emph {{On
  the jets produced by drops impacting a deep liquid pool and by bursting
  bubbles}}},\ }\href {https://doi.org/10.1017/jfm.2021.207} {\bibfield
  {journal} {\bibinfo  {journal} {\protect\JournalTitle{Journal of Fluid
  Mechanics}}\ }\textbf {\bibinfo {volume} {916}},\ \bibinfo {pages} {A37}
  (\bibinfo {year} {2021})}\BibitemShut {NoStop}%
\end{thebibliography}%

\end{document}


\title{Supplementary Information for: Singular jets in compound drop impact}

\author{Zeyang Mou}
\affiliation{
State Key Laboratory for Strength and Vibration of Mechanical Structures,
International Center for Applied Mechanics, School of Aerospace,
Xi'an Jiaotong University, Xi'an 710049, P. R. China
}
\author{Zheng Zheng}
\affiliation{
State Key Laboratory for Strength and Vibration of Mechanical Structures,
International Center for Applied Mechanics, School of Aerospace,
Xi'an Jiaotong University, Xi'an 710049, P. R. China
}
\author{Zhen Jian}
\affiliation{
State Key Laboratory for Strength and Vibration of Mechanical Structures,
International Center for Applied Mechanics, School of Aerospace,
Xi'an Jiaotong University, Xi'an 710049, P. R. China
}
\author{Carlo Antonini}
\affiliation{
Department of Materials Science, University of Milano-Bicocca, Via R. Cozzi 55, 20125, Milano, Italy
}
\author{Christophe Josserand}
\affiliation{Laboratoire d’Hydrodynamique (LadHyX), UMR 7646 CNRS-Ecole Polytechnique, IP Paris, F-91128 Palaiseau CEDEX, France}
\author{Marie-Jean Thoraval}
\email[]{mjthoraval@xjtu.edu.cn}
\affiliation{
State Key Laboratory for Strength and Vibration of Mechanical Structures,
International Center for Applied Mechanics, School of Aerospace,
Xi'an Jiaotong University, Xi'an 710049, P. R. China
}

\pacs{}

\maketitle

\onecolumngrid

\renewcommand{\thesection}{S\arabic{section}}
\renewcommand{\thesubsection}{\thesection\arabic{subsection}.}
\renewcommand{\thefigure}{S\arabic{figure}}
\renewcommand{\thetable}{S\arabic{table}}
\renewcommand{\theequation}{S\arabic{equation}}
\renewcommand{\thepage}{S\arabic{page}}

\section{Upper limit}

During spreading, the levitated oil lamella splashes into small droplets and leaves saw-toothed instability at the contact line, see \cref{figS:Rupture:A}.
As the water core spreads more close to the irregular periphery of oil shell at high $V_\textnormal{i}$, the shape of water core is deformed. The the focusing process is disturbed due to the asymmetric retraction. 
The oil film beneath the water core remains intact, so the cavity and the following jet is produced anyway, but with lower emitting velocity.

As discussed in previous work \cite{Blanken2020} and shown in our main text in Fig.~1 \textit{C} and \textit{D} (black cross), the oil film beneath the water core breaks during spreading when the core sinks to the bottom of the shell. From top-view experiments, the rupture of oil film is distinguishable by the contact line of water and substrate, as indicated in \cref{figS:Rupture:B}.

\begin{figure}[hb]
    \centering
    \includegraphics[width=\textwidth]{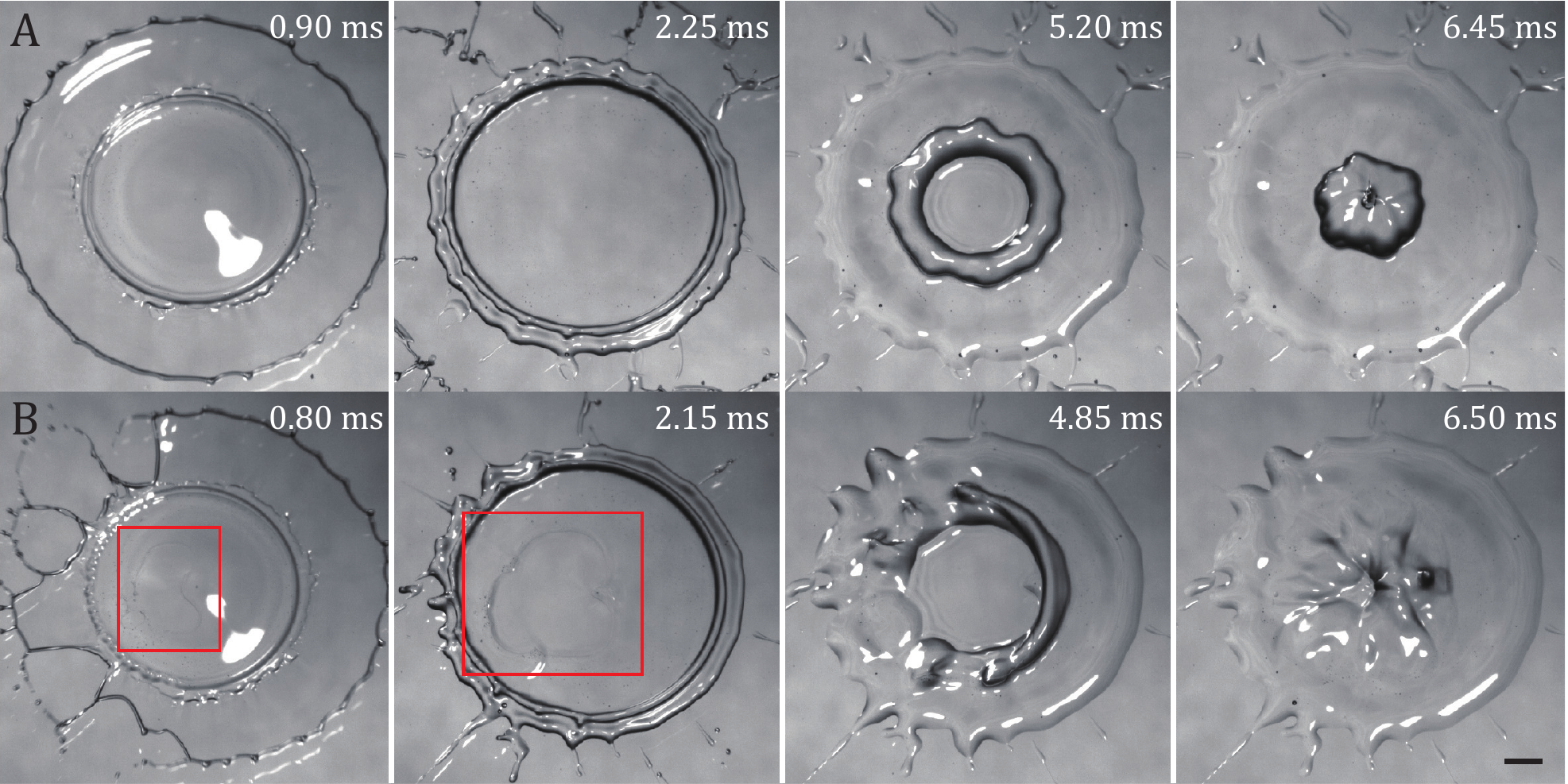}
    \phantomsubfloat{figS:Rupture:A}
    \phantomsubfloat{figS:Rupture:B}
    \vspace{-2\baselineskip}
    \vspace{0.5\baselineskip}
    \caption{Top-view snapshots for $\alpha = 0.3$ near the upper limitation. 
    (\textit{A}) $V_{\textnormal{i}} = \SI{2.80}{\m\per\s}$, $We_\textnormal{w}=324$.
    From $t= \SI{0.90}{ms}$ to $\SI{2.25}{ms}$, The prompt splash of oil shell influence the shape of contact line. The trailing lamella (water) retracts with an irregular shape (shown as $\SI{5.20}{ms}$ and $\SI{6.45}{ms}$) due to the close distance to the preceding one (oil) ($\SI{2.25}{ms})$).
    (\textit{B}) $V_{\textnormal{i}} = \SI{3.16}{\m\per\s}$, $We_\textnormal{w}=412$.
    The water-substrate contact is distinguishable from a gradually expanding edge with an irregular shape (marked in the red square).
    The direct contact with glass slide jeopardize the recoiling process and form an asymmetric rim, thus destroy the indispensable cavity to emit a high-speed jet.
    The scale bar in the images corresponds to \SI{1}{\mm}.}
	\label{figS:Rupture}
    \vspace{-15pt}
\end{figure}

\clearpage
\section{Size of jet and entrapped bubble}

\begin{figure}[htb]
    \centering
    \includegraphics[width=0.6\textwidth]{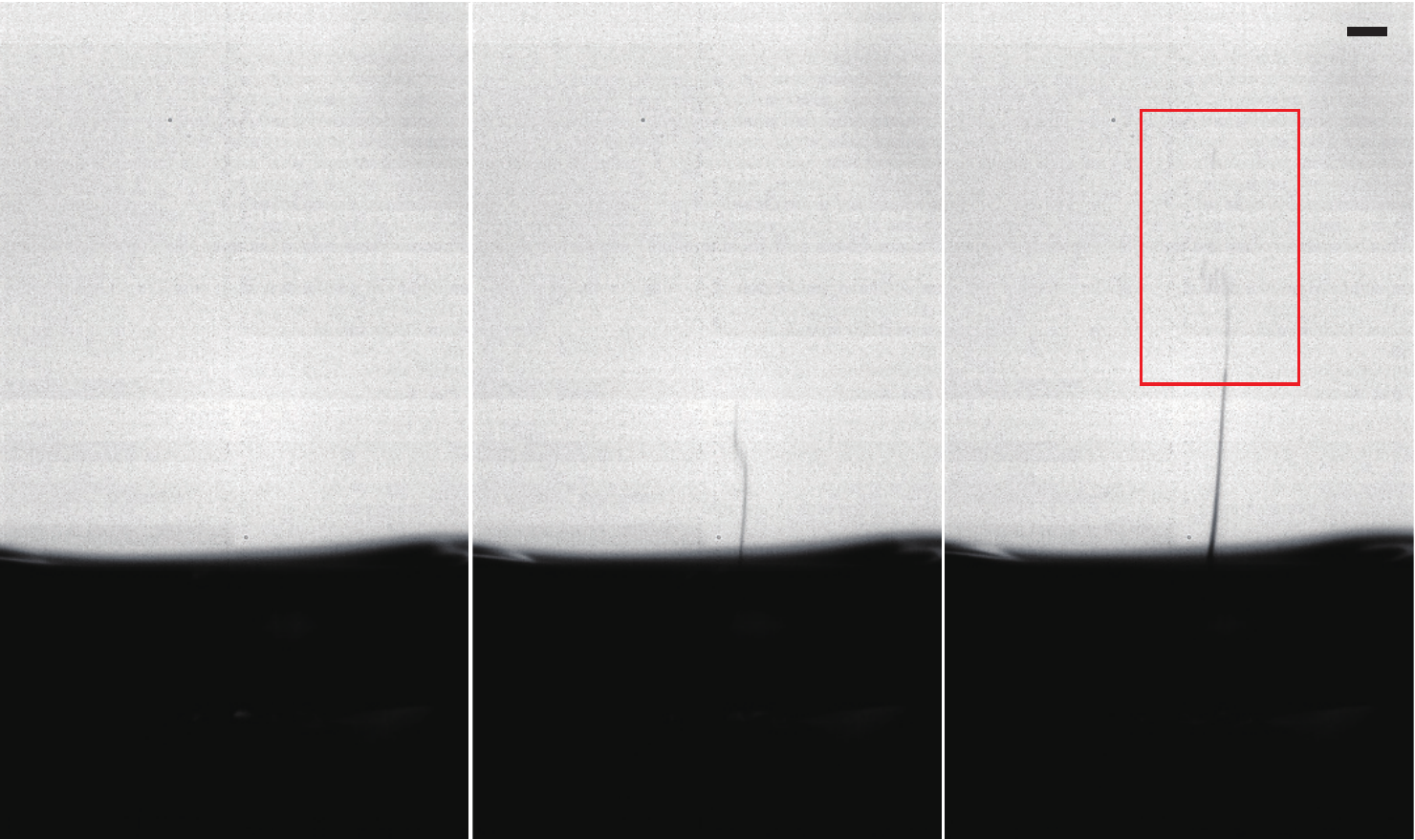}
    \caption{The magnified snapshots of jet when $\alpha = 0.3$, $V_{\textnormal{i}} = \SI{2.59}{\m\per\s}$ as the second velocity peak in Fig.~1\textit{C} (Movie S14).
    The time interval between them is \SI[parse-numbers = false]{1/80 000}{s}. 
    The jet is so thin that it is bent over by the air drag force. 
    So the jet velocity is underestimated under this certain initial condition.
    The black scale bar represents \SI{100}{\um}.}
	\label{figS:JetMag}
\end{figure}

The velocity of the singular jet could be underestimated because of the limitations when we track the singular jet in Fig.~1\textit{D}, as shown in \cref{figS:JetMag}.
The minimal diameter of the jet and drops that pinch-off from it are less than one pixel (\SI{3.21}{\um}). 
Beyond the jet tip that we track from Matlab codes, there is a gray and blur tracks left to show the approximate location of the droplets and the jet tip (marked in the red square). 
These threads corresponds to liquid droplets which moves too fast to be captured within the certain shutter time (\SI[parse-numbers = false]{1/1680000}{s}). 
The gray threads shows the motion of the droplet for approximately $10$ pixels.

\begin{figure}[hbt]
    \centering
    \includegraphics[width=0.6\textwidth]{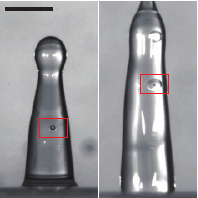}
    \caption{Experimental snapshots to show entrapped bubbles at two entrapment regime when $V_{\textnormal{i}} = \SI{2.12}{\m\per\s}$ and $V_{\textnormal{i}} = \SI{2.8}{\m\per\s}$ in Fig.~1\textit{C}. The bubble size (of diameter $\sim$ \SI{40}{\um}) reduces significantly in the second regime compared to the first one (of diameter $\sim$ \SI{130}{\um}).
    The bubble size is close to numerical results, where the maximal value of equivalent bubble diameter is \SI{140}{\um} and \SI{50}{\um}.
    The black scale bar represents \SI{1}{\mm}.}
	\label{figS:BubbleVolume}
\end{figure}

In \cref{figS:BubbleVolume}, two typical images are shown to visualize the bubble entrapped inside the drop. The bubble diameter at the second bubble entrainment regime ($V_{\textnormal{i}} = \SI{2.8}{\m\per\s}$) are so small that it reaches the limitation of resolution, which may explain the absence of bubbles merges from smaller toroidal bubbles before the second velocity peak in experiments (Fig.~1\textit{C}).

\clearpage
\section{Experimental results for various volumetric ratios \texorpdfstring{$\alpha$}{α}}

\begin{figure}[htbp]
    \centering
    \includegraphics[width=\textwidth]{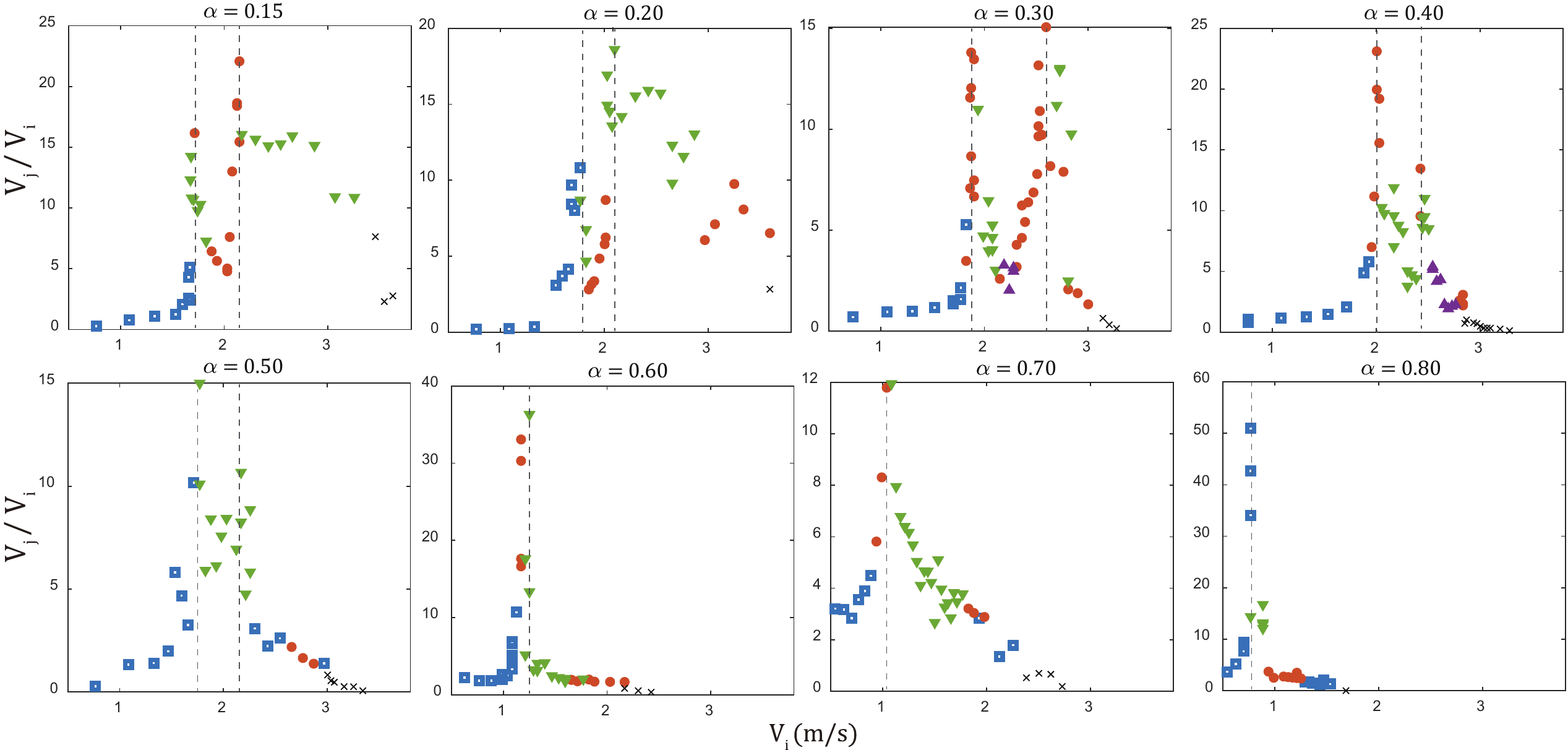}
    \caption{Nondimensional jetting velocity from experiments of various volumetric ratio $\alpha$.
    As summarized in our phase diagram Fig.~1\textit{D}, there are two velocity peaks when $\alpha \leq 0.5$, but only one peak for cases that have larger proportion of water. 
    The vertical dashed lines indicate the jet velocity peaks.}
	\label{figS:Experiments}
\end{figure}

\cref{figS:Experiments} shows the experimental results for varied volumetric ratio $\alpha = 0.15$-$0.8$.
The upper limits for impact velocity (black crosses) are controlled by the condition when the core drop sinks to the bottom, which is easier to be satisfied for larger volumetric proportion drops.

\begin{figure}[htbp]
    \centering
    \includegraphics[width=\textwidth]{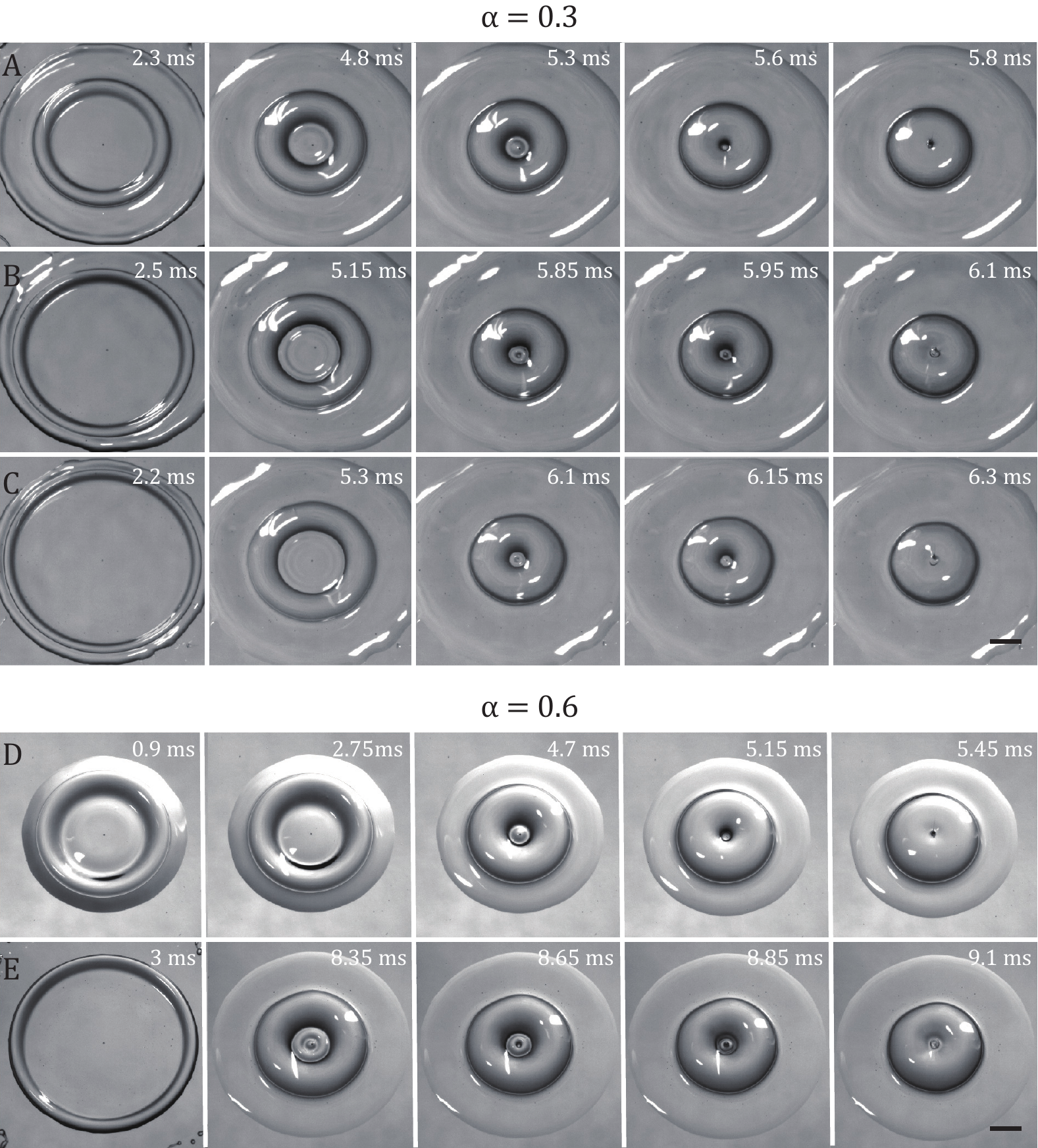}
    \vspace*{0.5em}
    \phantomsubfloat{figS:Top:A}
    \phantomsubfloat{figS:Top:B}
    \phantomsubfloat{figS:Top:C}
    \phantomsubfloat{figS:Top:D}
    \phantomsubfloat{figS:Top:E}
    \vspace{-2\baselineskip}
    \caption{Top-view snapshots from experiments for $\alpha = 0.3, 0.6$ . 
    (\textit{A}) $V_{\textnormal{i}} = \SI{1.87}{\m\per\s}$, $We_{\textnormal{w}} = 144.5$ at the first velocity peak in \cref{figS:Experiments}. The capillary waves accumulate in the center and oscillate downward before collapse;
    (\textit{B}) $V_{\textnormal{i}} = \SI{2.30}{\m\per\s}$, $We_{\textnormal{w}} = 218.6$ at the valley value of jet velocity, where the waves for a protrusion at the base of the cavity;
    (\textit{C}) $V_{\textnormal{i}} = \SI{2.43}{\m\per\s}$, $We_{\textnormal{w}} = 244$ at the second peak, where the film remains nearly flat with small propagating waves until final collapse;
    (\textit{D}) $V_{\textnormal{i}} = \SI{1.13}{\m\per\s}$, $We_{\textnormal{w}} = 60.9$ at the first velocity peak;
    (\textit{E}) $V_{\textnormal{i}} = \SI{2.03}{\m\per\s}$, $We_{\textnormal{w}} = 196.6$, after the single peak and near the upper limit to rupture.
    The scale bar in the images corresponds to \SI{1}{\mm}.}
	\label{figS:Top}
\end{figure}

\clearpage
\cref{figS:Top} shows sequences of experimental results to verify the relationship between geometry and jetting velocity when $\alpha=0.3$ and $0.6$.
If we compare the process between \cref{figS:Top:A} and \cref{figS:Top:D}, \cref{figS:Top:B} and \cref{figS:Top:E}, they are quite akin to each other.
\cref{figS:Top:A} and \cref{figS:Top:D} correspond to the first jetting velocity peak thanks to the downward and slender cavity that is easily recognized by the bright zone located at the axis of symmetry, while \cref{figS:Top:B} and \cref{figS:Top:E} are corresponding to low emission speed because of the convex in center. 

The phenomena in \cref{figS:Top:C} could be supplemented by our numerical results from which the decoupled effect of two interfaces is clearly shown, see the last column in Fig.~2.
Here we observe the capillary waves with shorter wave length and smaller amplitude propagating along the nearly flat, thin film.
As it is shown from the consistent high pixel value (bright) of the film, those waves fail to deform the base to the similar length scale as the rim.
The absence of the second jetting velocity peak for cases with larger volumetric ratio is a result of constant geometric shape in the course of contraction (\cref{figS:Top:E}).
Thinner oil layer on top has little effect in suppressing waves, even when close to the upper limit of $V_\textnormal{i}$ where the most amount of oil is accumulated. 
Note that the violent oscillations imply more water participating in forming central protrusion, whose height may exceed the thickness of the rim. So that the jet is not ejected by the closure of the rim but the central oscillation itself, see similar discussions in single drop impact as the reason for the decline stage after the second peak (without bubble entrapment) \cite{Bartolo2006}.

\begin{figure}[htbp] 
    \centering
    \includegraphics[width=\textwidth]{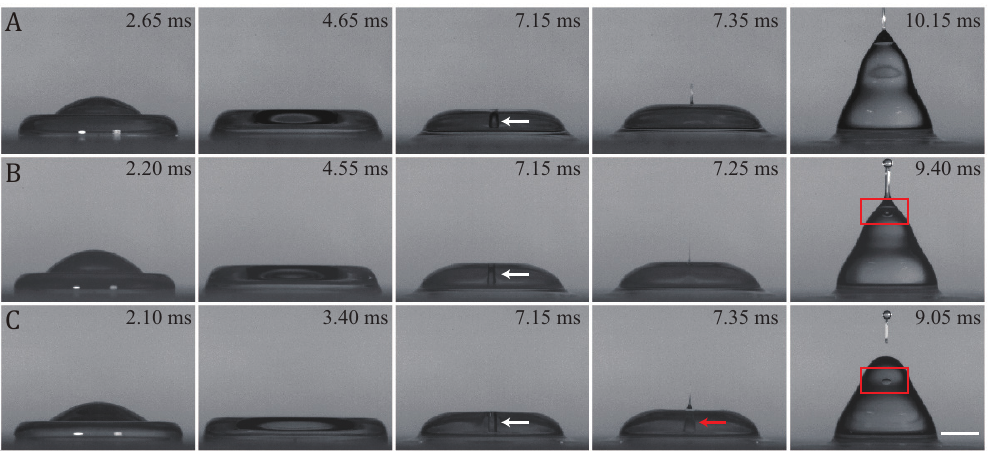}
    \phantomsubfloat{figS:alpha0.8:A}
    \phantomsubfloat{figS:alpha0.8:B}
    \phantomsubfloat{figS:alpha0.8:C}
    \vspace{-2\baselineskip}
    \caption{Snapshots from side-view image when $\alpha=0.8$.
    The large proportion of water means thin oil shell wrapping outside the core, so the deformation of two interfaces are highly synchronous.
    (\textit{A}) $V_{\textnormal{i}} = \SI{0.70}{\m\per\s}$.
    The white arrow refers to the cylindrical air cavity leading to the first singular jet.
    At $t=\SI{10.15}{ms}$, neither bubble nor oil is trapped after jet emergence.
    (\textit{B}) $V_{\textnormal{i}} = \SI{0.77}{\m\per\s}$, corresponding to the peak velocity in \cref{figS:Experiments}.
    Bubble-in-oil entrapment is clearly shown in the red box.
    (\textit{C}) $V_{\textnormal{i}} = \SI{0.89}{\m\per\s}$.
    The red arrow points to the water-oil interface that will close on top to entrap an oil drop, as circled in the red square.
    The white scale bar in the images corresponds to \SI{1}{\mm}.
    }
	\label{figS:alpha0.8}
\end{figure}

\clearpage
When the volume ratio reaches close to $1$, the compound drop acts more similar to a pure drop. In \cref{figS:alpha0.8}, the base of cavity is not formed by the oscillation of central waves induced ahead of rim (as shown in Fig.~2), but by the downward spire on the top of droplets directly. Even though there are no distinct stairs to build a pyramidal structure of the compound drop, the formation of cavity is quite akin to the pure drops \cite{Bartolo2006}.

\clearpage
\section{Conditions for the second singularity}

In practical experiments and numerical reproductions (Fig.~3 and Fig.~4), the two variables $V_{\textnormal{i}}$ and $d^*$ are coupled together, so in this section, we keep the same impact velocity ($V_{\textnormal{i}} = \SI{2.59}{\m\per\s}$) and vary $d^*$ only to investigate the effect brought by relative position.
\cref{figS:RelativePosition:A} illustrates the variation of jet velocity and its dependence on cavity geometry,
where the nondimensional jet velocity first decreases by moving the core drop down from $d^* = 1$ to $d^* = -0.65$.
When the water core is nearer to the bottom of compound drop ($-0.8 > d^* > -1$), jet velocity increases significantly, which is reminiscent of the ascend to the second peak in Fig.~3 \textit{A}, where the base oscillation is inhibited by the thick oil layer on top.
Not surprisingly, the power law switches from $0.55$ to $2/3$ when reaching to the second bubble entrained band (\cref{figS:PowerLaw2.59}).

In \cref{figS:RelativePosition:D}, the maximal spreading ratio of the water core, which is only a function of $We$ for single drops, rises as the water core sinks to the bottom and more oil accumulates on top of water core to facilitate spreading.
This result clearly exceeds the observation of pure drop expansion on hydrophobic surface \cite{Wildeman2016} and is also due to the water drop sliding on the substrate because of the presence of the lubrication oil layer between the substrate and the water core.

\begin{figure}[htbp]
	\centering
	\includegraphics[width=0.5\textwidth]{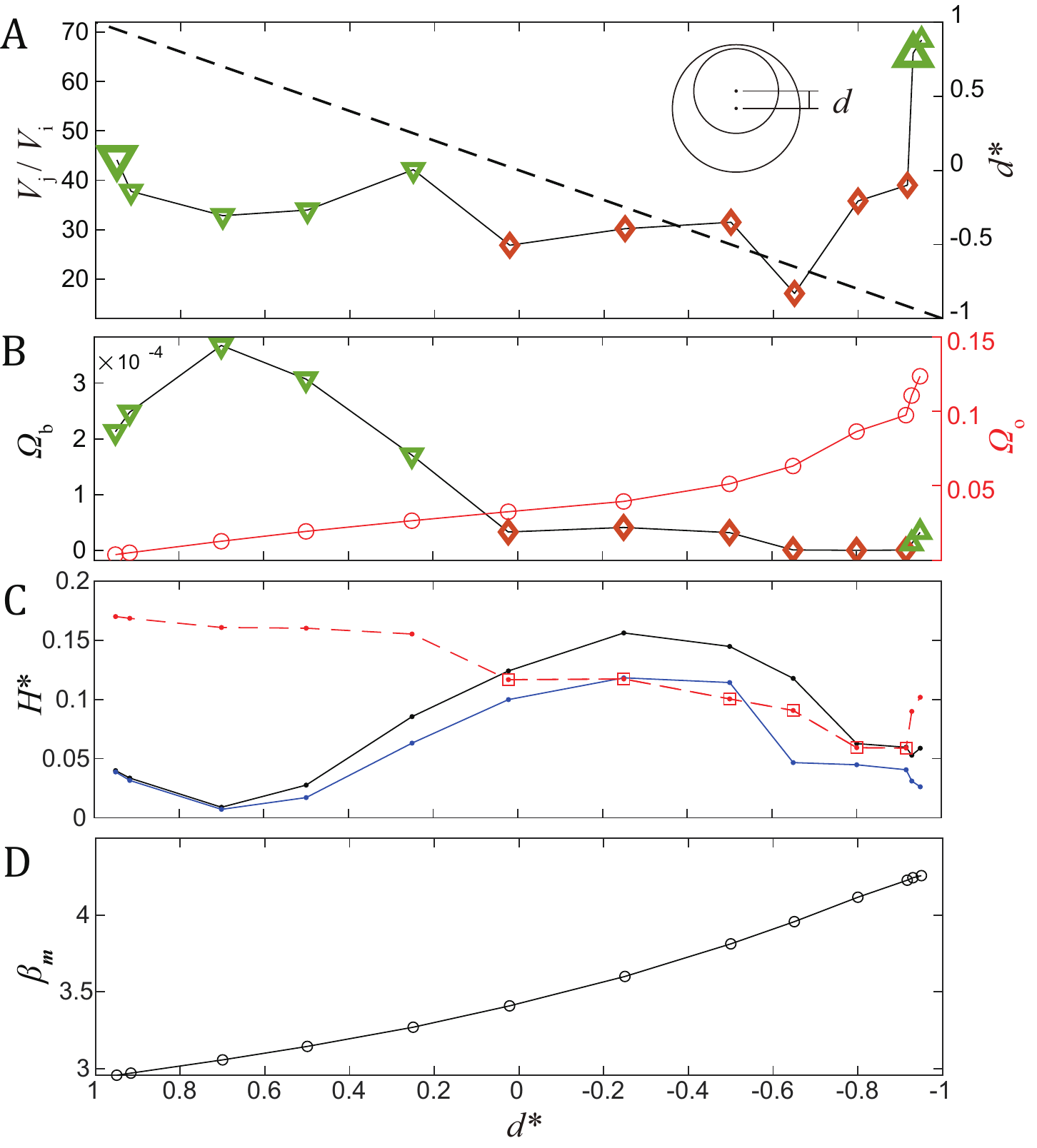}
	\phantomsubfloat{figS:RelativePosition:A}
    \phantomsubfloat{figS:RelativePosition:B}
    \phantomsubfloat{figS:RelativePosition:C}
    \phantomsubfloat{figS:RelativePosition:D}
    \vspace{-2\baselineskip}
    \vspace{0.5\baselineskip}
	\caption{Numerical results to investigate the effects of eccentricity $d^*$ at the identical $V_\textnormal{i}$ represented as the magnified orange diamond in Fig.~3 \textit{A}.
	(\textit{A}) Nondimensional jet velocity versus offset distance $d^*$ at $V_{\textnormal{i}} = \SI{2.59}{\m\per\s}$. dashed line: eccentricity $d^*$. When $d^*>0$, the water core is in the upper section of the whole drop.
	(\textit{B}) The corresponding volume of air encapsulation $\Omega_\textnormal{b}$ and oil on top of water core $\Omega_\textnormal{o}$ calculated at the instant of jet emergence based on the same definition in Fig.~3 \textit{B}.
	(\textit{C}) Tip height for two interfaces and the pinch-off location before jet initiation, with the same definition in Fig.~3\textit{C}.
	(\textit{D}) Maximum spreading ratio $\beta_{\textnormal{m}}$ of water core.}
	\vspace{-15pt}
	\label{figS:RelativePosition}
\end{figure}

\begin{figure}[htbp]
	\centering
	\includegraphics[width=0.5\textwidth]{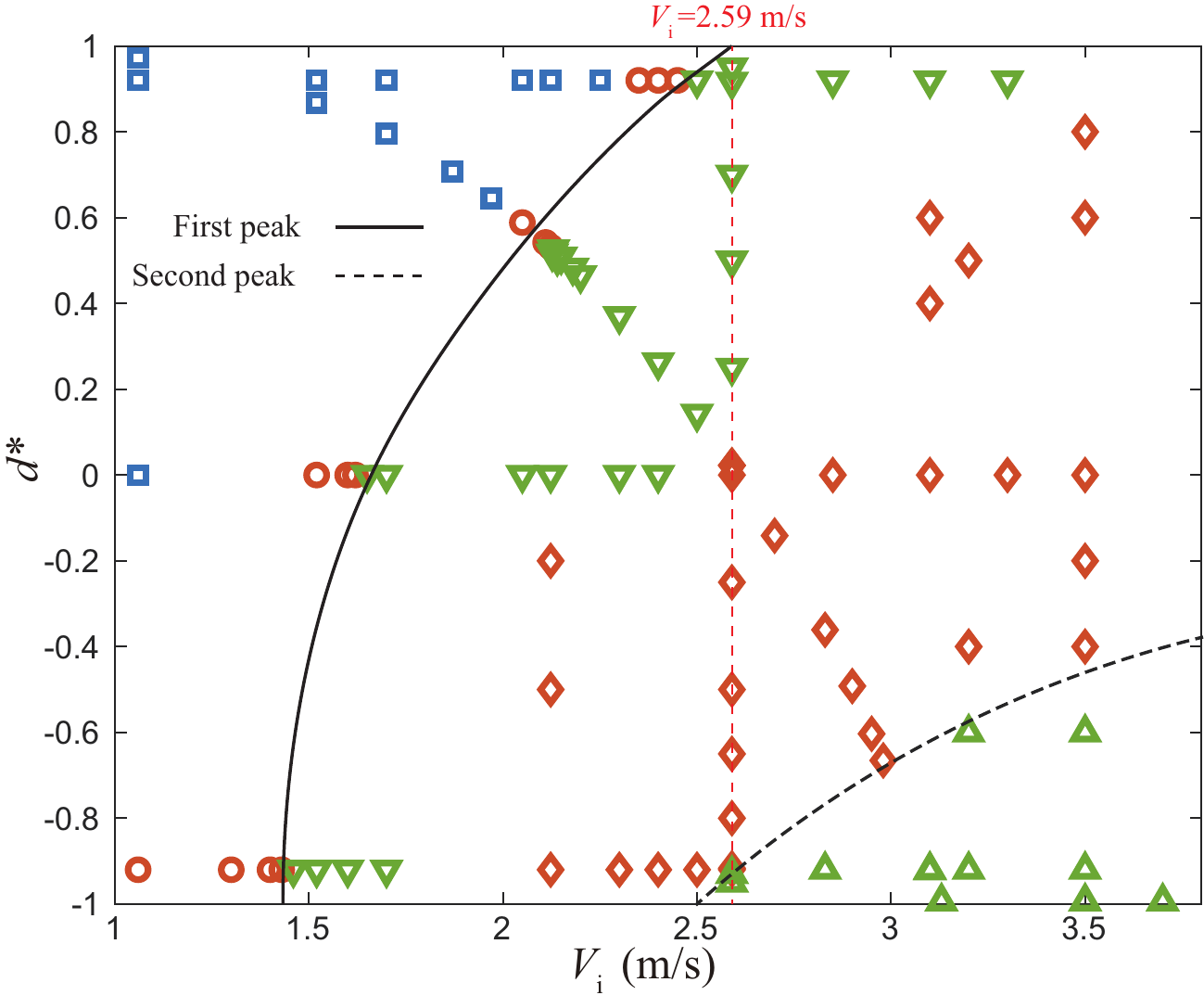}
	\caption{Phase diagram for encapsulation in terms of $V_\textnormal{i}$ and $d^*$ at $\alpha=0.3$ from simulations.
	The black solid and dashed line represent the approximate condition of topological transition, so they correspond to the first and the second jetting velocity peak.
	The red vertical dashed line represents the result when $V_\textnormal{i}=\SI{2.59}{m \per s}$ and $d^*$ is the only variable, as illustrate in \cref{figS:RelativePosition}.}
	\label{figS:SimulationPhase}
\end{figure}

\clearpage
We investigate the encapsulation types for various $V_\textnormal{i}$ and $d^*$ when $\alpha=0.3$ in \cref{figS:SimulationPhase}. 
According to the observation that the singularity occurs at the threshold of topological change, we mark the condition of two singularities by the solid and dashed curves. 
Apparently, neither impact velocity nor the eccentricity is the single variable determines the occurrence of singularity. 
Within this certain range of impact velocity, the second type of singularity only happens when $V_\textnormal{i}$ and is sufficiently large and water core is close to the bottom of the compound drop.

\clearpage
\section{Power law}

\begin{figure*}[htp]
	\centering	
	\setlength{\abovecaptionskip}{0.cm}
	\includegraphics[width=0.8\linewidth]{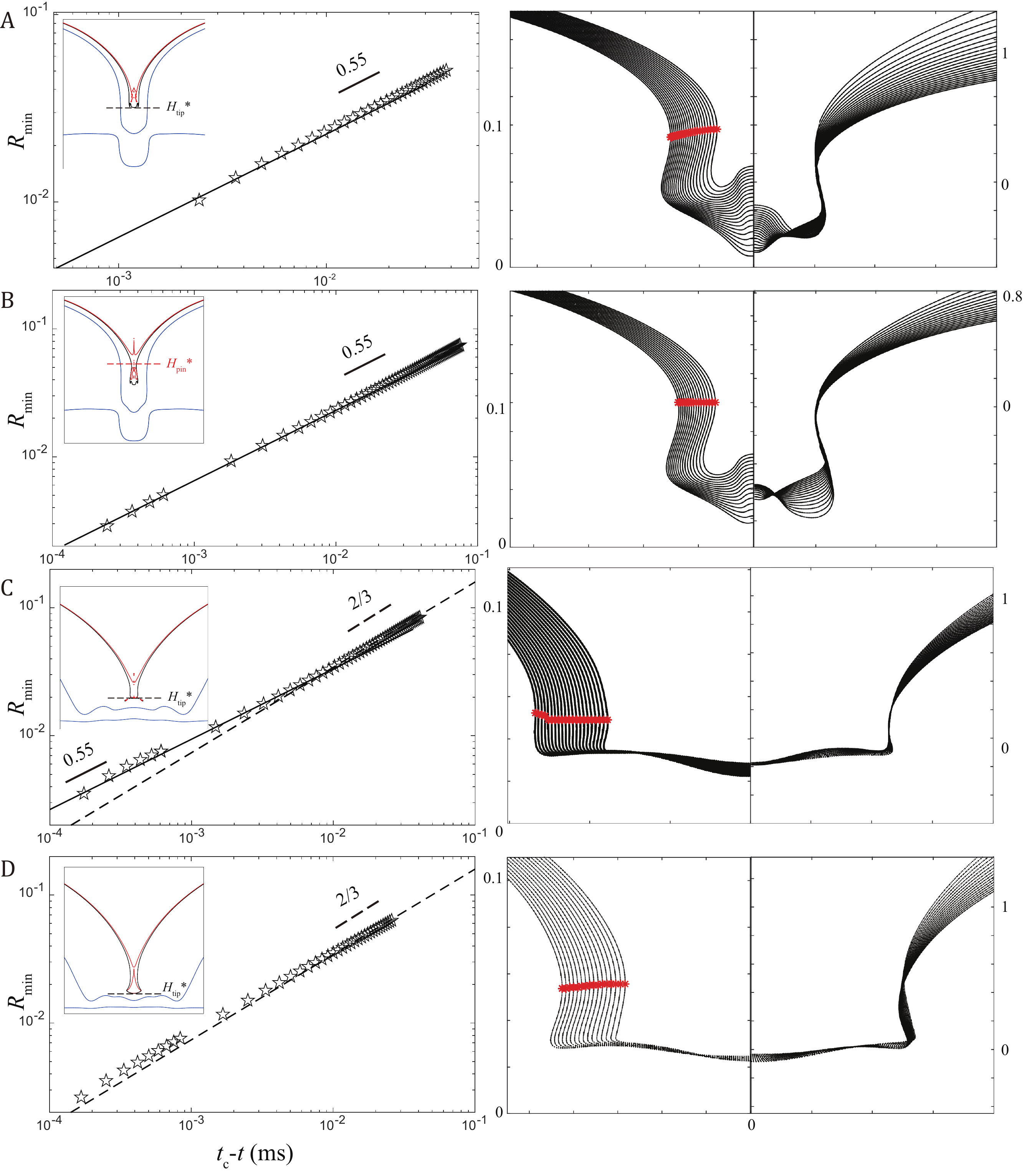}
	\vspace{\spaceBelowFigure}
	\phantomsubfloat{figS:PowerLawSI:A}
    \phantomsubfloat{figS:PowerLawSI:B}
    \phantomsubfloat{figS:PowerLawSI:C}
    \phantomsubfloat{figS:PowerLawSI:D}
	\caption{Power law analysis at the impact conditions corresponding to the second to the fifth column in Fig.~2.
	(\textit{A}) $V_{\textnormal{i}} = \SI{2.123}{\m\per\s}$. 
	(\textit{B}) $V_{\textnormal{i}} = \SI{2.13}{\m\per\s}$, the same as Fig.~4\textit{A}.
	Solid line: $R_{\textnormal{min}} = 0.29(t_{\textnormal{c}}-t)^{0.55}$.
	The profile is plotted from $t_{\textnormal{c}}-t=\SI{0.0261}{ms}$ to $\SI{0.0079}{ms}$.
	(\textit{C}) $V_{\textnormal{i}} = \SI{2.98}{\m\per\s}$.
	(\textit{D}) $V_{\textnormal{i}} = \SI{3.1}{\m\per\s}$.
	Dash line: $R_{\textnormal{min}} = 0.74(t_{\textnormal{c}}-t)^{0.667}$.
	The profile is plotted from $t_{\textnormal{c}}-t=\SI{0.0259}{ms}$ to $\SI{0.0134}{ms}$.}
	\label{figS:PowerLawSI}
\end{figure*}

\cref{figS:PowerLawSI} depicts the power law fitting to illustrate the collapse dynamics in the vicinity of two jet velocity peaks.
Remarkably, we observe a cross-over power law between $2/3$ and $1/2$ when the impact velocity is not sufficient to steepen the interface and trap a bubble in \cref{figS:PowerLawSI:C}, where the retracting velocity follows $2/3$ rule firstly but accelerates near the singularity, and yields to purely inertial $1/2$ power law \cite{Thoroddsen2018, Yang2020} for the more slender and cylindrical cavity shape.
This cross-over power law has been discussed recently for other configurations \cite{Yang2020,BlancoRodriguez2021} on account of local geometrical transition.
Note that the cross-over solution is not seen in the second band of bubble entrainment anymore. For instance, a simple and consistent $2/3$ power law is found in \cref{figS:PowerLawSI:D} and Fig.~2\textit{B}. 

\begin{figure*}[htp]
	\centering
	\includegraphics[width=0.8\linewidth]{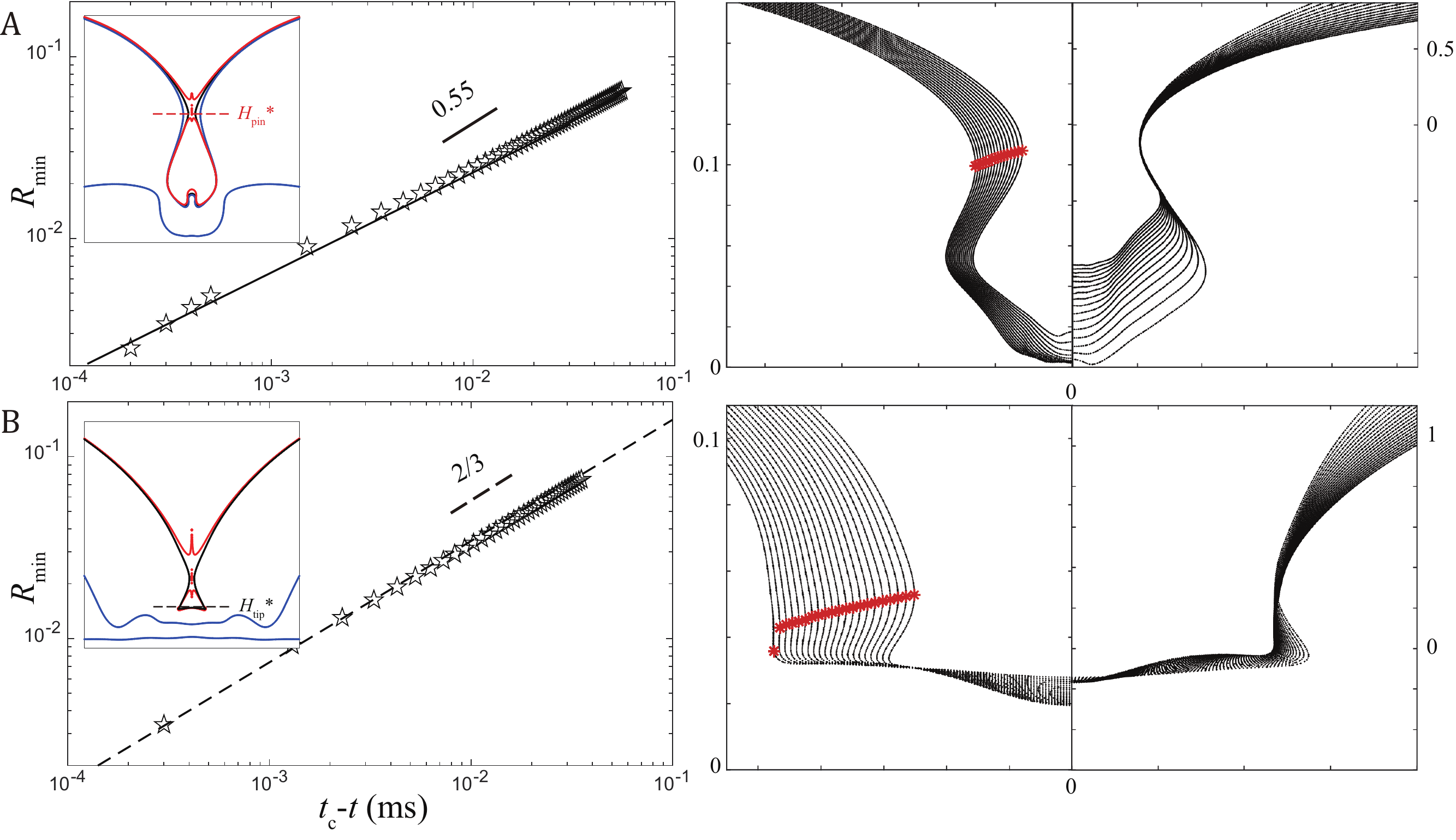}
	\vspace{\spaceBelowFigure}
	\phantomsubfloat{figS:PowerLaw2.59:A}
    \phantomsubfloat{figS:PowerLaw2.59:B}
	\caption{Scaling for the cavity collapse dynamics in two bubble entrainment regimes when $V_{\textnormal{i}} = \SI{2.59}{\m\per\s}$, marked as the two magnified triangles in \cref{figS:RelativePosition:A}.
	(\textit{A}) $d^* = 0.95$.
	Solid line: $R_{\textnormal{min}} = 0.30(t_{\textnormal{c}}-t)^{0.55}$.
	The profile is plotted from $t_{\textnormal{c}}-t=\SI{0.034}{\m\per\s}$ to $\SI{0.0014}{ms}$
	(\textit{B}) $d^* = -0.93$.
	Dash line: $R_{\textnormal{min}} = 0.74(t_{\textnormal{c}}-t)^{2/3}$.
	The profile is plotted from $t_{\textnormal{c}}-t=\SI{0.032}{\m\per\s}$ to $\SI{0.0012}{ms}$.}
	\label{figS:PowerLaw2.59}
\end{figure*}

\clearpage
To decouple the influences brought by $V_\textnormal{i}$ and eccentricity $d^*$, we fix the impact velocity and change the geometrical configuration only in \cref{figS:RelativePosition}.
The reproduction of two bubble entrainment bands encourages us to plot the power law of the two typical conditions representing the two separated regimes, as indicated in \cref{figS:PowerLaw2.59}.

\clearpage
\section{Mesh refinement and simulation convergence}

\begin{figure}[htp]
	\centering
	\includegraphics[width=0.8\linewidth]{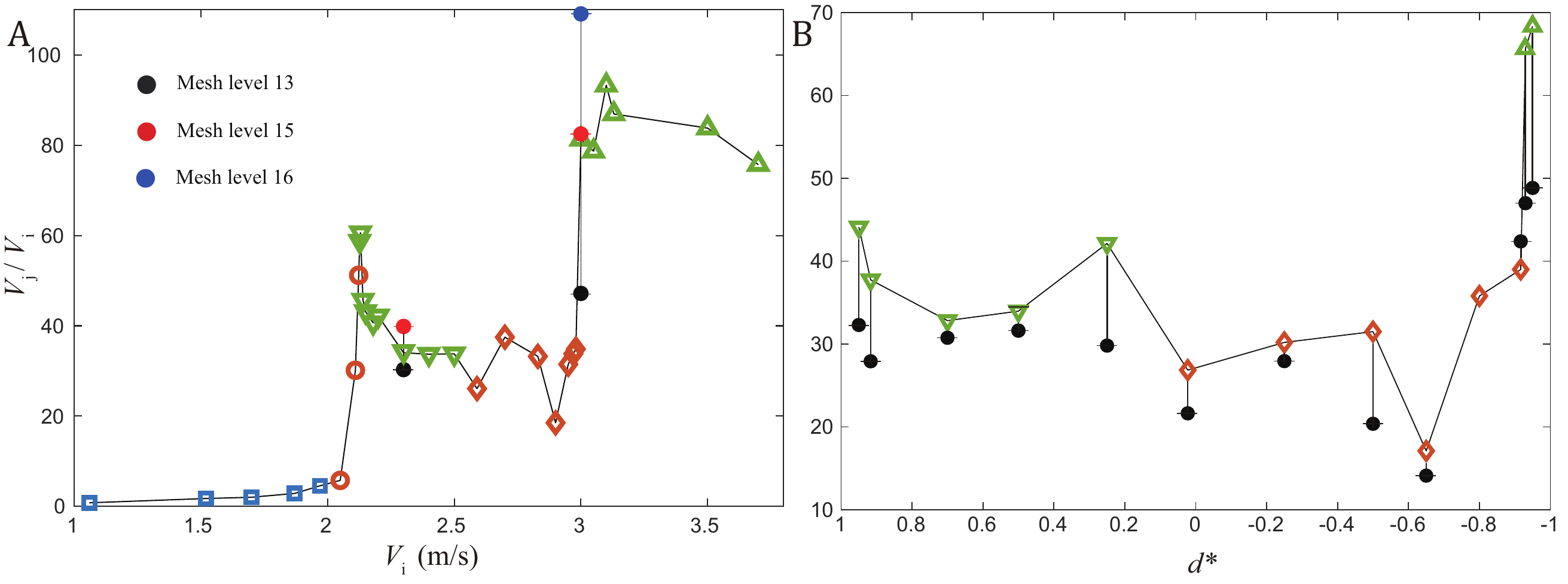}
	\vspace{\spaceBelowFigure}
	\phantomsubfloat{MeshRefinementVJet:A}
    \phantomsubfloat{MeshRefinementVJet:B}
    \phantomsubfloat{MeshRefinementVJet:C}
    \phantomsubfloat{MeshRefinementVJet:D}
	\caption{Mesh independence test in terms of jet velocity.
	(\textit{A}) Jet velocity at different mesh level to test mesh independence as Fig.~3\textit{A}.
	(\textit{B}) Expansion of \cref{figS:RelativePosition:A} when mesh level is at $13$ and $14$.
	}
	\label{figS:MeshRefinementVjet}
\end{figure}

The maximal mesh refinement level is chosen to be $12$ during the spreading and retraction stages, then $14$ just before and after the jet emergence,
corresponding to the smallest mesh size $\delta$ being $1061$ ($D_{\textnormal{o}} / \delta = 2^{12}$) or $4244$ ($D_{\textnormal{o}} / \delta = 2^{14}$) times smaller than the outer drop diameter $D_{\textnormal{o}}$ respectively.
The oil film beneath the water core and the deformed water core at the axis of symmetry never breaks until jet emergence in our simulation. 
When $V_\textnormal{i}=\SI{2.30}{m \per s}$, 
the normalized jet speed $V_\textnormal{j}/V_\textnormal{i}$ increases from $30$ to $34$ when enhancing the mesh level from $13$ to $14$. The jet velocity $V_\textnormal{j}/V_\textnormal{i}$ further rises to $39$ when mesh level is $15$.
Higher resolution captures more details of jet geometry so that faster shooting speed is measured.
This conclusion is also valid for the second bubble entrapment regime when $V_\textnormal{i}=\SI{3.0}{m \per s}$, $V_\textnormal{j}/V_\textnormal{i}$ increases from $47$ to $81$ as mesh level changes from $13$ to $14$.
By further increasing the mesh level to $15$, the jet is easier to break and $V_\textnormal{j}/V_\textnormal{i}$ rises slightly to $83$. This velocity can even increase to $109$ when using smaller mesh size (level $16$). 
Here the jet breaks to micro droplets even when the time interval is $10^{-5}$, so we remove the small droplets pinches-off from the jet and track the tip of the jet connected to the body drop.
Therefore, we set the maximal mesh level to $14$ to cover the entire parameter space.
We also prove that the starting time to enhance mesh level to $14$ has little influence on the velocity of jet and the time of its occurrence.
\cref{figS:MeshRefinementVjet} shows the influence of mesh refinement on the jet velocity of Fig.~3\textit{A} and \cref{figS:RelativePosition:A}. Even though the higher resolution of mesh leads to higher jetting velocity, the dependence between jet velocity and local geometry (shown as entrapment types) is still valid and the trend of the curve is consistent at different mesh levels.

\begin{figure}[htp]
	\centering
	\includegraphics[width=0.8\linewidth]{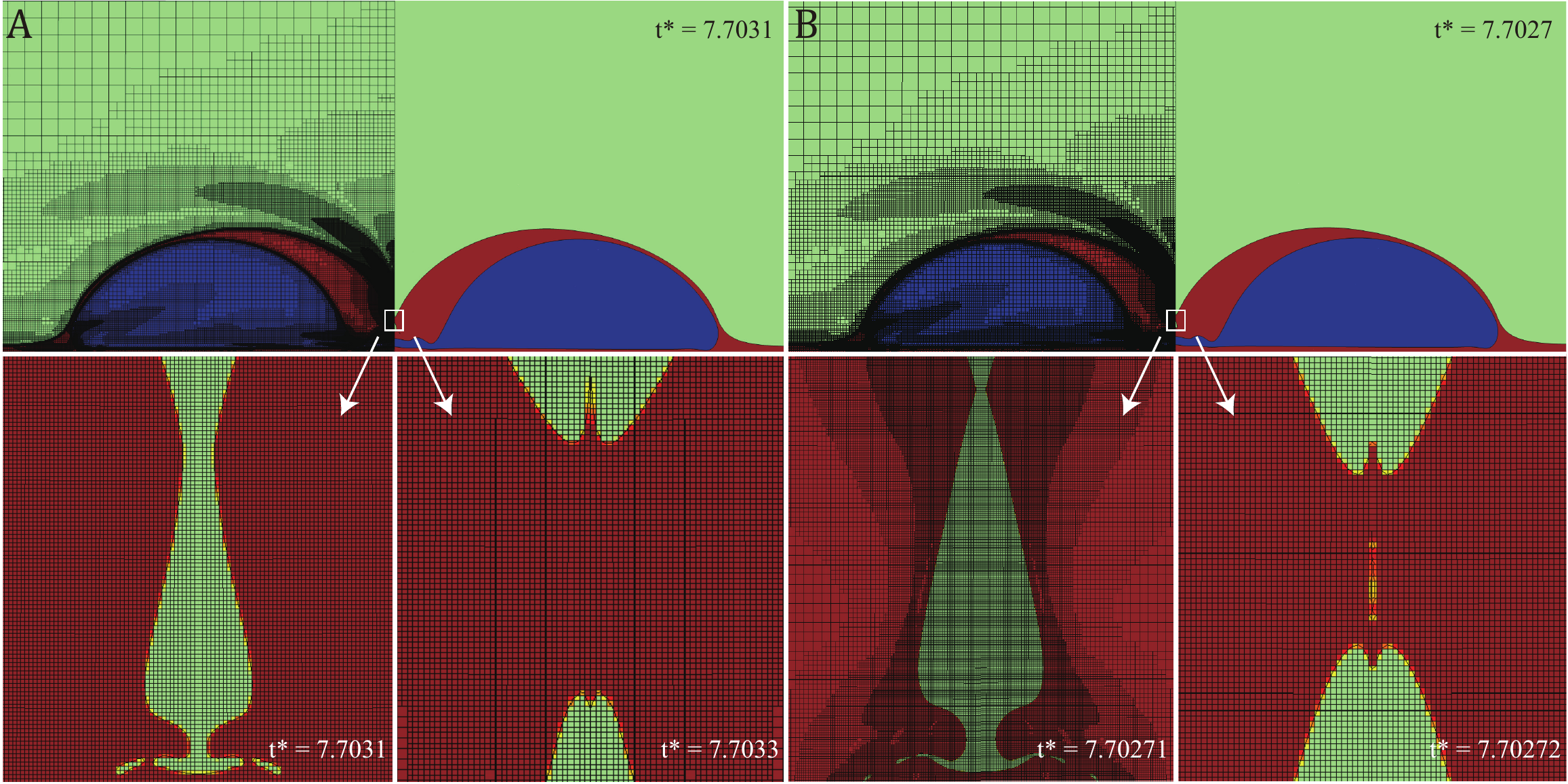}
	\vspace{-0.5em}
	\phantomsubfloat{MeshRefinement1:A}
    \phantomsubfloat{MeshRefinement1:B}
	\caption{Mesh refinement of the singular jet at different mesh level when $V_{\textnormal{i}} = \SI{3.0}{\m\per\s}$ in Fig.~3\textit{A}.
	(\textit{A}) Mesh level $L=14$. Top: the assembly of numerical snapshots with and without mesh just before the cavity collapse, where the most refined meshes are concentrated on the solid surface, the two interfaces and the region near to the axis of symmetry where the interfaces is most curved.
	Bottom: A zoom just before the cavity collapse, where the thinnest radius of cavity contains four cells. The minimal cell is of the width of $\SI{610}{nm}$.
	(\textit{B}) Mesh level $L=15$. The size of the smallest mesh is only $\SI{305}{nm}$.
	}
	\label{figS:MeshRefinement}
    \vspace{-15pt}
\end{figure}

\clearpage
\section{Supplementary videos}

\movie{Experimental results in Fig.~1\textit{B} when $V_{\textnormal{i}} = \SI{1.87}{\m\per\s}$, $We_\textnormal{w}=144$, from side-view.}
\movie{Experimental results in Fig.~1\textit{B} when $V_{\textnormal{i}} = \SI{1.87}{\m\per\s}$, $We_\textnormal{w}=144$, from top-view.}
\movie{Numerical results in \cref{figS:SimulationPhase} when $V_{\textnormal{i}} = \SI{3.50}{\m\per\s}$, $We_\textnormal{w}=506$, $d^* = 0.80$. The double layer emulsion (water-in-oil) is observed in this case when the central protrusion pitches off and is engulfed by the reverse-curvature water-oil interface.}
\movie{Experimental results in Fig.~1\textit{E} when $V_{\textnormal{i}} = \SI{1.06}{\m\per\s}$, $We_\textnormal{w}=46$, no entrapment.}
\movie{Experimental results in Fig.~1\textit{E} when $V_{\textnormal{i}} = \SI{2.07}{\m\per\s}$, $We_\textnormal{w}=177$, bubble entrapment.}
\movie{Experimental results in Fig.~1\textit{E} when $V_{\textnormal{i}} = \SI{2.24}{\m\per\s}$, $We_\textnormal{w}=207$, water-in-oil entrapment.}
\movie{Experimental results in Fig.~1\textit{E} when $V_{\textnormal{i}} = \SI{2.31}{\m\per\s}$, $We_\textnormal{w}=220$, oil entrapment.}
\movie{Experimental results in Fig.~1\textit{E} when $V_{\textnormal{i}} = \SI{3.14}{\m\per\s}$, $We_\textnormal{w}=407$, oil film break.}
\movie{Numerical results in Fig.~2 when $V_{\textnormal{i}} = \SI{1.87}{\m\per\s}$, $We_\textnormal{w}=144$, $d^* = 0.71$.}
\movie{Numerical results in Fig.~2 when $V_{\textnormal{i}} = \SI{2.123}{\m\per\s}$, $We_\textnormal{w}=186$, $d^* = 0.53$.}
\movie{Numerical results in Fig.~2 when $V_{\textnormal{i}} = \SI{2.40}{\m\per\s}$, $We_\textnormal{w}=238$, $d^* = 0.26$.}
\movie{Numerical results in Fig.~2 when $V_{\textnormal{i}} = \SI{2.59}{\m\per\s}$, $We_\textnormal{w}=277$, $d^* = 0.02$.}
\movie{Numerical results in Fig.~2 when $V_{\textnormal{i}} = \SI{3.10}{\m\per\s}$, $We_\textnormal{w}=397$, $d^* = -0.92$.}
\movie{Experimental results in \cref{figS:JetMag} to show the details of the jet when $V_{\textnormal{i}} = \SI{2.59}{\m\per\s}$, $We_\textnormal{w}=277$.}


\par All numerical codes used to produce the results in this paper are available at https://github.com/zhengzhengfr/PNAS-Compound-drop-code.git. Note that the open-source software Basilisk (basilisk.fr) should be installed before running these codes.

\bibliographystyle{jabbrv_apsrev4-2}
\bibliography{References}